\documentclass[usenatbib]{mnras}
\usepackage{graphicx}
\usepackage{amsmath}
\usepackage{times}
\usepackage{xspace}
\usepackage{color}
\usepackage{breqn}

\usepackage[normalem]{ulem}
\usepackage{float}
\usepackage{epsfig}
\usepackage{multirow}
\usepackage{epstopdf}
\usepackage{verbatim}
\usepackage{nicefrac}
\usepackage{url}
\usepackage{subfig}
\usepackage{bm}
\usepackage[table,usenames,dvipsnames]{xcolor}
\usepackage{amssymb}
\usepackage{booktabs}
\usepackage{threeparttable}
\usepackage{amsbsy}
\usepackage{amsfonts}
\usepackage{pgfplots}
\usepackage{upgreek}

\usepackage{tabularx}
\newcolumntype{R}{>{\raggedleft\arraybackslash}X}

\def\mnras{MNRAS}
\def\ssr{Space Science Reviews}
\def\apj{ApJ}
\def\prd{PRD}
\def\aap{A\&A}
\def\jcap{JCAP}

\newcommand{\sjaddress}{\url{https://github.com/sjoudaki/CosmoLSS}\xspace}
\newcommand{\kidsaddress}{\url{http://kids.strw.leidenuniv.nl}\xspace}
\newcommand{\twodflensaddress}{\url{http://2dflens.swin.edu.au}\xspace}

\newcommand{\be}{\begin{equation}}
\newcommand{\ee}{\end{equation}}
\newcommand{\bea}{\begin{eqnarray}}
\newcommand{\eea}{\end{eqnarray}}

\newcolumntype{P}[1]{>{\centering\arraybackslash}p{#1}}

\newcommand{\cfhtlens}{CFHTLenS\xspace}

\newcommand{\halofit}{\textsc{halofit}\xspace}
\newcommand{\hmcode}{\textsc{hmcode}\xspace}
\newcommand{\theli}{\textsc{theli}\xspace}
\newcommand{\astrowise}{\textsc{Astro-WISE}\xspace}
\newcommand{\isitgr}{\textsc{ISiTGR}\xspace}

\newcommand{\athena}{\textsc{athena}\xspace}
\newcommand{\bpz}{\textsc{bpz}\xspace}
\newcommand{\cosmomc}{\textsc{CosmoMC}\xspace}
\newcommand{\cosmolss}{\textsc{CosmoLSS}\xspace}
\newcommand{\plik}{\textsc{Plik}\xspace}
\newcommand{\simlow}{\textsc{SimLow}\xspace}
\newcommand{\camb}{\textsc{CAMB}\xspace}

\begin{document}
\voffset=-1.0cm
\hoffset=0.1cm

\title[KiDS+2dFLenS combined analysis]{KiDS-450 + 2dFLenS: Cosmological parameter constraints from weak gravitational lensing tomography and overlapping redshift-space galaxy clustering}

\author[Joudaki et al.]{\parbox[t]{\textwidth}{Shahab Joudaki$^{1,2,3}$\thanks{E-mail: shahab.joudaki@physics.ox.ac.uk},
    Chris Blake$^{1,2}$, Andrew Johnson$^{1,2}$, Alexandra Amon$^4$, \\ Marika Asgari$^4$, Ami Choi$^5$, Thomas Erben$^6$, Karl Glazebrook$^1$, \\ Joachim Harnois-D\'{e}raps$^4$, Catherine Heymans$^4$, Hendrik Hildebrandt$^6$, \\ Henk Hoekstra$^7$, Dominik Klaes$^6$, Konrad Kuijken$^7$, Chris Lidman$^8$, Alexander Mead$^9$, Lance Miller$^3$, David Parkinson$^{10}$, Gregory B. Poole$^1$, Peter Schneider$^6$, \\ Massimo Viola$^7$, Christian Wolf$^{11}$} \\ \\ $^1$ Centre for Astrophysics \&  Supercomputing, Swinburne University of Technology, P.O.\ Box 218,  Hawthorn, VIC 3122, Australia \\ $^2$ ARC Centre of Excellence for All-sky Astrophysics (CAASTRO)  \\ $^3$ Department of Physics, University of Oxford, Denys Wilkinson Building, Keble Road, Oxford OX1 3RH, U.K. \\ $^4$ Scottish Universities Physics Alliance, 
    Institute for Astronomy, University of Edinburgh, Royal Observatory, Blackford
  Hill, Edinburgh, EH9 3HJ, U.K. \\ $^5$ Center for Cosmology and AstroParticle Physics, The Ohio State University, 191 West Woodruff Avenue, Columbus, OH 43210, USA \\ $^6$ Argelander Institute for Astronomy, University of Bonn, Auf dem Hugel 71, 53121 Bonn, Germany \\ $^7$ Leiden Observatory, Leiden University, Niels Bohrweg 2, 2333 CA Leiden, the Netherlands \\ $^8$ Australian Astronomical Observatory, North Ryde, NSW 2113, Australia \\ $^9$ Department of Physics and Astronomy, The University of British Columbia, 6224 Agricultural Road, Vancouver, B.C., V6T 1Z1, Canada \\ $^{10}$ School of Mathematics and Physics, University of Queensland, Brisbane, QLD 4072, Australia \\ $^{11}$ Research School of Astronomy and Astrophysics, Australian National University, Canberra, ACT 2611, Australia
  }

\pubyear{2017}
\date{\today}

\maketitle

\begin{abstract}
We perform a combined analysis of cosmic shear tomography, galaxy-galaxy lensing tomography, and redshift-space multipole power spectra (monopole and quadrupole) using 450~deg$^2$ of imaging data by the Kilo Degree Survey (KiDS-450) overlapping with two spectroscopic surveys: the 2-degree Field Lensing Survey (2dFLenS) and the Baryon Oscillation Spectroscopic Survey (BOSS). We restrict the galaxy-galaxy lensing and multipole power spectrum measurements to the overlapping regions with KiDS, and self-consistently compute the full covariance between the different observables using a large suite of $N$-body simulations. We methodically analyze different combinations of the observables, finding that galaxy-galaxy lensing measurements are particularly useful in improving the constraint on the intrinsic alignment amplitude (by 30\%, positive at $3.5\sigma$ in the fiducial data analysis), while the multipole power spectra are useful in tightening the constraints along the lensing degeneracy direction (e.g.~factor of two improvement in the matter density constraint in the fiducial analysis). The fully combined constraint on $S_8 \equiv \sigma_8 \sqrt{\Omega_{\rm m}/0.3} = 0.742 \pm 0.035$, which is an improvement by 20\% compared to KiDS alone, corresponds to a $2.6\sigma$ discordance with Planck, and is not significantly affected by fitting to a more conservative set of scales. Given the tightening of the parameter space, we are unable to resolve the discordance with an extended cosmology that is simultaneously favored in a model selection sense, including the sum of neutrino masses, curvature, evolving dark energy (both constant and time-varying equations of state), and modified gravity. The complementarity of our observables allows for constraints on modified gravity degrees of freedom that are not simultaneously bounded with either probe alone, and up to a factor of three improvement in the $S_8$ constraint in the extended cosmology compared to KiDS alone. We make our measurements and fitting pipeline public at \sjaddress.
\end{abstract}

\begin{keywords}
surveys -- large-scale structure of universe -- cosmology: observations
\vspace{-0.993cm}
\end{keywords}

\section{Introduction}
\label{Introduction}
\setcounter{footnote}{0}
\renewcommand{\thefootnote}{\arabic{footnote}}

The promise of future cosmological surveys rests not only in their increased statistical precision, but also in the combined analysis of the cosmological observables that they enable (e.g.~\citealt{hj04,JB10,JK12,deputter13, fr14}). In this regard, the complementarity between imaging and spectroscopic surveys is particularly fruitful, as it allows for an improved calibration of astrophysical systematics in observations of weak gravitational lensing and galaxy clustering (e.g.~arising from photometric redshift uncertainties, intrinsic galaxy alignments, baryonic 
effects in the nonlinear matter power spectrum, and galaxy bias), and the breaking of degeneracies with cosmologically desired quantities such as neutrino mass and dark energy (e.g.~\citealt{Joachimi11,cb12, hmts13,mw13,deputter14,fr14,eg15}). 

The complementarity between imaging and spectroscopic surveys is moreover crucial in testing gravity on cosmic scales. This is achieved by probing the relationship between the gravitational potentials $\psi$ and $\phi$ describing temporal and spatial perturbations to the spacetime metric, respectively (e.g.~\citealt{zhang07,jz08,guzik10,song11}). While the two potentials are equal in General Relativity (GR; in the absence of anisotropic stresses), their equality is generally broken in distinct modified gravity scenarios, in a way than can depend on both length scale and time (e.g.~\citealt{bz08,baker13,ps16}).

As weak gravitational lensing mainly probes the sum of the metric potentials $\psi+\phi$ modifying the relativistic deflection of light, and redshift-space distortions (RSD) probe the potential $\psi$ modifying the growth of large-scale structure, their combination enables simultaneous constraints on each of the potentials. This complementarity has notably been used in \citet{simpson13} to constrain deviations from General Relativity with the Canada-France-Hawaii Telescope Lensing Survey (CFHTLenS; \citealt{heymans12,hildebrandt12,erben13, miller13}), WiggleZ Dark Energy Survey \citep{blake11,blake12}, and the Six-degree-Field Galaxy Survey (6dFGS;~\citealt{beutler11,beutler12}). For overlapping imaging and spectroscopic surveys, the relationship between the metric potentials can also be tested with the gravitational slip statistic $E_G$ obtained from the ratio of the galaxy-shear and galaxy-velocity cross-spectra (\citealt{zhang07}, also see e.g.~\citealt{reyes10,blake16eg}).

In our analysis, we consider the full set of primary observables of weak gravitational lensing and redshift-space galaxy clustering that can be extracted from overlapping imaging and spectroscopic surveys. Employing tomography of the source distributions, we measure the two-point shear correlation functions, galaxy-galaxy lensing angular cross-correlation, and multipole power spectra using 450 deg$^2$ of imaging data from the Kilo Degree Survey (KiDS-450; \citealt{kuijken15}) overlapping with two distinct spectroscopic surveys: the Baryon Oscillation Spectroscopic Survey (BOSS; \citealt{dawson13}), and the 2-degree Field Lensing Survey (2dFLenS; \citealt{blake16}). As the first wide-area spectroscopic survey to specifically target the footprint of a deep-imaging lensing survey, 2dFLenS was designed to overlap on the sky with KiDS to optimize the science that can be achieved from a joint analysis of weak lensing and redshift-space galaxy clustering.

We restrict the multipole power spectrum measurements from 2dFLenS and BOSS to the overlapping regions with KiDS. Although these surveys extend beyond the KiDS-450 area, this paper focuses on creating and applying an analysis pipeline for overlapping lensing and spectroscopic datasets, paying particular attention to generating a self-consistent covariance between different statistics and scales using a large suite of $N$-body simulations. Our restriction of the multipole power spectra to the overlapping regions with KiDS implies that our cosmological constraints from the power spectra are not as strong as the constraints achievable from the full datasets. As the KiDS area expands, the overlap increases with 2dFLenS in particular. We leave the cosmological analysis using the full spectroscopic datasets to future work.

To extract cosmological information from the observables, we have developed a self-consistent Markov Chain Monte Carlo (MCMC) fitting pipeline in \cosmomc \citep{Lewis:2002ah}, which is an extension of the pipeline used in the cosmic shear analyses of \cfhtlens \citep{joudaki16} and KiDS-450 \citep{Hildebrandt16, joudaki17}. The pipeline allows for simultaneous variation of all the key astrophysical systematics, including intrinsic galaxy alignments (IA), baryonic feedback affecting the nonlinear matter power spectrum, photometric redshift uncertainties, galaxy bias, pairwise velocity dispersion, and non-Poissonian shot noise. We methodically consider distinct and full combinations of the observables, and constrain both $\Lambda$CDM and extended cosmological models, including neutrino mass, curvature, evolving dark energy, and modified gravity.

A particular aim of this work is to examine the level of concordance of our combined lensing and RSD measurements with the cosmic microwave background (CMB) temperature measurements of the Planck satellite. This comparison has garnered particular interest given the previously reported $\sim2\sigma$ discordance between both \cfhtlens and KiDS-450 with Planck \citep{planck13, maccrann15, planck15, kohlinger15, kohlinger17, Hildebrandt16, joudaki16, joudaki17}. While \cfhtlens and KiDS are concordant with pre-Planck CMB surveys \citep{calabrese13,calabrese17,wmap9}, cosmic shear measurements by the Deep Lens Survey (DLS, \citealt{jee2016}) and the Dark Energy Survey (DES, \citealt{dessv}) show greater degrees of concordance with Planck. Given the inability to bring about concordance between KiDS and Planck through any known systematic uncertainty~\citep{Hildebrandt16, joudaki17}, we examined the discordance in the context of extended cosmological models in \citet{joudaki17}, which revealed evolving dark energy as a viable candidate that is simultaneously favored in a model selection sense. We will examine to what extent this picture holds when including galaxy-galaxy lensing and redshift-space galaxy clustering measurements.

A parallel KiDS analysis that is similar in nature, in which KiDS-450 cosmic shear measurements are combined with galaxy-galaxy lensing and angular clustering from the Galaxy And Mass Assembly survey (GAMA;~\citealt{liske15}), has been simultaneously released by \citet{vu17}.

In Section~\ref{theorysec}, we describe the underlying theory of our observables. In Section~\ref{measlab}, we present our measurements of cosmic shear, galaxy-galaxy lensing, and multipole power spectra, along with their full covariance. In Section~\ref{cosmomcsec}, we describe our new cosmology fitting pipeline, and the statistics used to assess the relative preference between models and concordance between datasets. In Section~\ref{Results}, we present cosmological constraints in $\Lambda$CDM for distinct combinations of the observables. We moreover explore the impact on our constraints from an extended treatment of the astrophysical systematics, and from different selections of physical and angular scales used in the analysis. In Section~\ref{Resultsextended}, we allow for an expansion of the underlying cosmology, in the form of massive neutrinos, curvature, evolving dark energy, and modified gravity. We conclude with a discussion of our results in Section~\ref{conclusionslab}.
\nocite{spergel15}

\section{Theory}
\label{theorysec}

In our cosmological analysis, we consider five statistics: the two-point shear correlation functions ($\xi_+, \xi_-$), the galaxy-galaxy lensing angular cross-correlation $\gamma_{\rm t}$, and the redshift space galaxy clustering spectra in the form of the monopole and quadrupole ($P_0, P_2$). The underlying theory of these observables is described below. 

\subsection{Cosmic shear ($\xi_+, \xi_-$)}
\label{cslab}

The observed two-point shear correlation functions receive contributions not only from the shear, but also from the intrinsic alignment of galaxies (reviewed in e.g.~\citealt{joachimi15}). As the two components are additive at the one-point level, the observed two-point functions for a given set of angular scales $\theta$ and tomographic bin combination $\{i,j\}$ are composed of three distinct pieces: shear-shear (GG), intrinsic-intrinsic (II), and shear-intrinsic (GI), such that
\begin{equation}
\xi_{\pm}^{ij}(\theta)_{\rm obs} = \xi_{\pm}^{ij}(\theta)_{\rm GG} + \xi_{\pm}^{ij}(\theta)_{\rm II} + \xi_{\pm}^{ij}(\theta)_{\rm GI}.
\label{eqn:xipmtotal} 
\end{equation}
To compute these distinct correlation functions $(\rm GG, II, GI)$, we follow the same  procedure laid out in earlier work (e.g.~\citealt{joudaki16}, and references therein), where
\begin{equation}
\xi_{\pm}^{ij}(\theta)_{\{{\rm GG, II, GI}\}} = \frac{1}{2\pi}\int d\ell \,\ell \,C_{\{{\rm GG, II, GI}\}}^{ij}(\ell) \, J_{0,4}(\ell \theta),
\label{eqn:xipmgg}
\end{equation}
such that $C_{\rm GG}^{ij}(\ell)$ is the shear power spectrum at angular wavenumber $\ell$, and $J_{0,4}$ are the zeroth and fourth order Bessel functions of the first kind which correspond to the `+' and `-' correlation functions, respectively. Employing the extended Limber and flat-sky approximations (\citealt{Limber,LA08, Kilbinger17}), the shear power spectrum is then expressed as
\begin{equation}
C_{\rm GG}^{ij}(\ell) = \int_0^{\chi_{\rm H}} d\chi \, 
\frac{q_i(\chi)q_j(\chi)}{[f_K(\chi)]^2} \, P_{\delta\delta} \left(\frac{\ell+1/2}{f_K(\chi)},\chi \right),
\label{eqn:ckk}
\end{equation}
where $\chi$ is the comoving distance, to the horizon in $\chi_{\rm H}$, and $f_K(\chi)$ is the comoving angular diameter distance. The matter power spectrum is denoted by $P_{\delta\delta}$, and the lensing kernel for tomographic bin $i$ is given by
\begin{equation}
q_i(\chi) = \frac{3 \Omega_{\mathrm m} H_0^2}{2c^2} \frac{f_K(\chi)}{a(\chi)}\int_\chi^{\chi_{\rm H}}\, d\chi'\ n_i(\chi') \frac{f_K(\chi'-\chi)}{f_K(\chi')},
\label{eqn:qi} 
\end{equation}
where $\Omega_{\mathrm m}$ is the present matter density, $H_0$ is the Hubble constant, $c$ is the speed of light, $a(\chi)$ is the scale factor, and $n_i(\chi)$ is the normalized source galaxy distribution for a given tomographic bin, the integral over which is unity. Furthermore, the intrinsic-intrinsic and shear-intrinsic power spectra \citep{HS04,BK07} are respectively given by 
\begin{equation}
C_{\rm II}^{ij}(\ell) = \int_0^{\chi_{\rm H}} d\chi \, 
\frac{n_i(\chi)n_j(\chi)F_i(\chi)F_j(\chi)}{[f_K(\chi)]^2} \, P_{\delta\delta} \left( \frac{\ell+1/2}{f_K(\chi)},\chi \right),
\label{eqn:xipmii}
\end{equation}
and
\begin{dmath}
C_{\rm GI}^{ij}(\ell) = \int_0^{\chi_{\rm H}} d\chi \frac{q_i(\chi)n_j(\chi)F_j(\chi)}{[f_K(\chi)]^2}P_{\delta\delta} \left( \frac{\ell+1/2}{f_K(\chi)},\chi \right) + 
\int_0^{\chi_{\rm H}} d\chi \frac{n_i(\chi)F_i(\chi)q_j(\chi) }{[f_K(\chi)]^2} P_{\delta\delta} \left({\frac{\ell+1/2}{f_K(\chi)},\chi}\right),
\label{eqn:xipmgi}
\end{dmath}
where the intrinsic alignments can vary with amplitude ($A_{\rm IA}$), luminosity ($L$ via the power $\beta_{\rm IA}$), and redshift ($z$ via the power $\eta_{\rm IA}$), such that 
\begin{equation}
F_i(\chi) = - A_{\rm IA} C_1 \rho_{\rm cr} \frac{\Omega_{\mathrm m}}{D(\chi)} \left({1+z(\chi)} \over {1+z_0}\right)^{\eta_{\rm IA}} \left({L_i \over L_0}\right)^{\beta_{\rm IA}} .
\label{eqn:fz}
\end{equation}
Here, the normalization constant $C_1 = 5 \times 10^{-14} \, h^{-2} M_\odot^{-1} {\rm Mpc}^3$, $\rho_{\rm cr}$ is the present critical density, $D(\chi)$ is the linear growth factor normalized to unity at present, $z_0 = 0.3$ is the arbitrary pivot redshift, and the pivot luminosity $L_0$ corresponds to an absolute r-band magnitude of $-22$ (e.g. \citealt{Joachimi11}). This standard parameterization of the intrinsic alignments does not include a scale dependence, and assumes the validity of the `nonlinear linear alignment' model over the scales probed by our measurements (\citealt{HS04,BK07}; also see \citealt{singh15}). In the cosmological analysis of KiDS-450 data, we do not account for any luminosity dependence (i.e. $\beta_{\rm IA} = 0$) as the mean luminosity is effectively the same across tomographic bins \citep{Hildebrandt16}.

\subsection{Galaxy-galaxy lensing ($\gamma_{\rm t}$)}
\label{ggllab}

We consider the galaxy-galaxy~lensing angular cross-correlation $\gamma_{\rm t}$ between KiDS and two spectroscopic surveys that substantially overlap on the sky: 2dFLenS and BOSS. Both of these surveys are further sub-divided such that we have a total of four lens samples, covering $0.15 < z < 0.43$ (2dFLOZ, LOWZ) and $0.43 < z < 0.70$ (2dFHIZ, CMASS). Incorporating intrinsic galaxy alignments, for a given lens sample and tomographic bin $j$, the observed cross-correlation takes the form
\begin{equation}
\gamma_{\rm t}^{j}(\theta)_{\rm obs} = \gamma_{\rm t}^{j}(\theta)_{\rm gG} + \gamma_{\rm t}^{j}(\theta)_{\rm gI},
\label{eqn:gttotal} 
\end{equation}
where the galaxy-shear ($\rm gG$) and galaxy-intrinsic ($\rm gI$) components are given by (e.g.~\citealt{hj04})
\begin{equation}
\gamma_{\rm t}^{j}(\theta)_{\{\rm gG, gI\}} = \frac{1}{2\pi} \int d\ell \,\ell \,C_{\{\rm gG, gI\}}^{j}(\ell) \, J_2(\ell \theta) .
\label{eqn:gt}
\end{equation}
Here, $J_2$ is the second-order Bessel function of the first kind, and the $\{\rm gG, gI\}$ power spectra are respectively given by (e.g.~\citealt{hj04,JB10,JK12}):
\begin{equation}
C_{\rm gG}^{j}(\ell) = \int_0^{\chi_{\rm H}} d\chi \, 
\frac{{\tilde n}(\chi)b(k,\chi)q_j(\chi)}{[f_K(\chi)]^2} \, P_{\delta\delta} \left( \frac{\ell+1/2}{f_K(\chi)},\chi \right),
\label{eqn:gtgG}
\end{equation}
and
\begin{equation}
C_{\rm gI}^{j}(\ell) = \int_0^{\chi_{\rm H}} d\chi \, 
\frac{{\tilde n}(\chi)b(k,\chi) n_j(\chi)F_j(\chi)}{[f_K(\chi)]^2} \, P_{\delta\delta} \left( \frac{\ell+1/2}{f_K(\chi)},\chi \right),
\label{eqn:gtgi}
\end{equation}
where ${\tilde n}(\chi)$ refers to the lens galaxy distribution, normalized to integrate to unity, and $b(k,\chi)$ is the bias factor relating the galaxy and matter density contrasts (such that $\delta_{\rm g} = b \delta$ to linear order). Given sufficiently large-scale cuts to our data (Section~\ref{secgt}), we assume that the galaxy bias is a constant for each of the lens samples, such that there are four additional parameters that are independently varied in our analysis. As a result, the bias can be seen as simply modifying the amplitude of the $\gamma_{\rm t}$ measurements, and is therefore strongly correlated with cosmological parameters such as $\sigma_8$ and $\Omega_{\rm m}$.

\subsection{Redshift-space multipole power spectra ($P_0, P_2$)}
\label{rsdlab}

We also consider the redshift-space multipole power spectra in the overlapping regions with KiDS, which enter as the coefficients of a Legendre expansion of the redshift-space galaxy power spectrum (e.g.~\citealt{th96,ballinger96,beutler14,johnson15}):
\begin{equation}
P_{\rm gg}^{\rm s}(k,\mu) = \sum_{{\rm even} \, m} P_m(k) L_m(\mu) ,
\label{eqn:p1}
\end{equation}
where $s$ denotes redshift space, $L_m$ are the Legendre polynomials of order $m$, and $\mu$ encapsulates the cosine of the angle between the Fourier mode $\rm{\bf k}$ and the line of sight. The redshift dependence of the equation is implicit, and the power spectra are evaluated at the effective redshift of each galaxy sample (detailed in Section~\ref{measlab}). In the linear regime, the only non-vanishing orders are $m = \{0, 2, 4\}$ \citep{kaiser87}. Incorporating the Alcock-Paczynski effect \citep{ap79}, which causes distortions in the galaxy clustering measurements due to the need to assume a fiducial cosmology in converting from angles and redshifts to distances, the multipole power spectra can be expressed as:
\begin{equation} 
P_m(k) = \frac{2m + 1}{2\alpha_\perp^2\alpha_\parallel} \int_{-1}^1 d\mu \, P_{\rm gg}^{\rm s}(k',\mu') L_m(\mu) ,
\label{eqn:p2}
\end{equation}
where the scaling factors are given by $\alpha_\parallel = {\hat H}(z)/H(z)$ and $\alpha_\perp = D_{\mathrm A}(z)/{\hat D}_{\mathrm A}(z)$, such that ^ refers to the quantities at the assumed fiducial cosmology (at the effective redshift of each sample). Moreover, the apparent wavenumbers and angles are $k' = \sqrt{k^{'2}_\parallel + k^{'2}_\perp}$ and $\mu' = k'_\parallel/k'$, related to the true wavenumbers and angles via the scaling factors: $k'_\parallel = k_\parallel/\alpha_\parallel$ and $k'_\perp = k_\perp/\alpha_\perp$.
The apparent redshift-space galaxy power spectrum is then 
\begin{equation}
P_{\rm gg}^{\rm s}(k',\mu') = \left[P_{\rm gg}(k') - 2 \mu^{\prime2} P_{{\rm g}\theta}(k') + \mu^{\prime4} P_{\theta\theta}(k')\right]\mathcal{D}(k',\mu') ,
\label{eqn:p3}
\end{equation}
where $P_{\rm gg}$ is the galaxy power spectrum, $P_{\theta\theta}$ is the peculiar velocity power spectrum, $P_{{\rm g}\theta}$ is the galaxy-velocity cross spectrum, and the small-scale damping term is 
\begin{equation} 
\mathcal{D}(k',\mu') = \exp\left[{-\left({k' \mu' \sigma_{\rm v}}\right)^2}\right] .
\label{eqn:p7}
\end{equation}
Here, the pairwise velocity dispersion $\sigma_{\rm v}$ is a free parameter that is varied in our analysis (for each sample, hence four additional parameters). Relating the peculiar velocity and density fields through the continuity equation (such that $\theta = -f\delta$, where $f$ is the growth rate), the galaxy and density fields through a linear bias $b$ (treated as a free parameter, independently for each sample) and assuming there is no peculiar velocity bias (i.e.~$b_{\rm v} = 1$ for all samples), the redshift-space spectrum simplifies to
\begin{equation} 
P_{\rm gg}^{\rm s}(k',\mu') = b^2 \left({P_{\delta\delta}(k') + N_{\rm shot}}\right)\left({1 + f(k')\mu^2/b}\right)^2 \mathcal{D}(k',\mu') .
\label{eqn:p5}
\end{equation}
We have further included the shot noise contribution $N_{\rm shot}$ as a free parameter (independently for each sample). We note that $N_{\rm shot}$ is quantifying any residual non-Poissonian contribution, as we directly subtract a Poisson shot noise from the measurements. Our conclusions are qualitatively unchanged by different implementations of the shot noise nuisance parameter~\citep{beutler14}.

In comparing the theory with the measurements, we finally need to convolve the multipole power spectra in equation~(\ref{eqn:p2}) with the survey selection function (e.g.~\citealt{blake16}). We do so by first arranging the multipole power spectra into an array $P_{\rm mod} = \left\{P_0(k_1), P_0(k_2),..., P_2(k_1), P_2(k_2), ..., P_4(k_1), P_4(k_2), ...\right\}$, where $\Delta k = 0.05~h~{\rm Mpc}^{-1}$ between $0<k<0.50~h~{\rm Mpc}^{-1}$. The theoretically estimated power spectra, which are compared to the observations, can then be expressed as
\begin{equation}
P_{\rm obs}(i) = \sum_j M_{ij} \, P_{\rm mod}(j) ,
\label{eqpkconv}
\end{equation}
where $M_{ij}$ is the convolution matrix that is computed in accordance with \citet{blake16}. As we consider 10 $k$-values for each of the statistics ($P_0, P_2, P_4$), $\bf M$ is a $30 \times 30$ matrix. Given the low signal-to-noise of the hexadecapole power spectrum measurements \citep{blake16}, we restrict our cosmological analysis to the $m=\{0, 2\}$ elements of the array $P_{\rm obs}$ at the measured $k$ values (detailed in Section~\ref{measlab}).

\begin{figure*}
\hspace{-1.4em}
\vspace{-0.5em}
\resizebox{18.0cm}{!}{\includegraphics{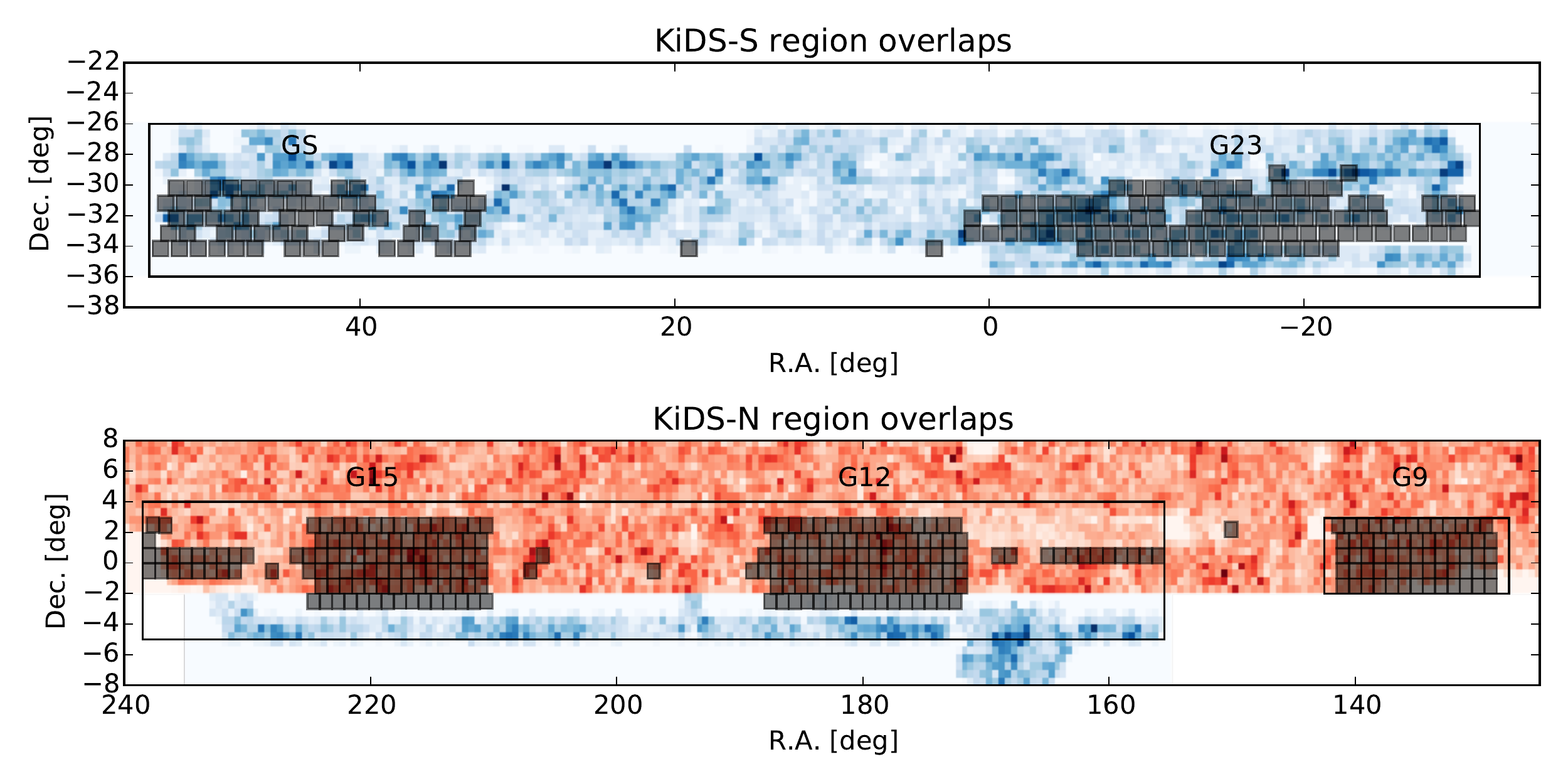}}
\caption{Overlapping imaging and spectroscopic surveys: dark squares are KiDS-450 pointings, and the fluctuating background is the gridded number density of 2dFLenS (blue) and BOSS (red) galaxies. The solid rectangles outline the footprint of the full KiDS survey. 
}
\label{figoverlap}
\end{figure*}

\begin{table}
\vspace{-0.11cm}
\caption{\label{taboverlap} Number of lenses and overlapping area of the imaging and spectroscopic surveys. Each KiDS field overlaps with either 2dFLenS or BOSS.
} 
\begin{tabular}{p{1.6cm}>{\raggedright}p{1.55cm}>{\raggedleft}p{1.5cm}>{\raggedleft\arraybackslash}p{1.5cm}}
\toprule
KiDS field & Spec.~overlap & Area (deg$^2$) & $N_{\rm lens}$\\
\midrule
${\rm G}9$     & ${\rm LOWZ}$ & $9.7$ & $414$ \\
${\rm G}9$     & ${\rm CMASS}$ & $44.0$ & $4272$ \\
${\rm G}12$     & ${\rm LOWZ}$ & $27.9$ & $849$ \\
${\rm G}12$     & ${\rm CMASS}$ & $90.3$ & $7451$ \\
${\rm G}15$     & ${\rm LOWZ}$ & $87.4$ & $3781$ \\
${\rm G}15$     & ${\rm CMASS}$ & $87.4$ & $8753$ \\
${\rm G}23$     & ${\rm 2dFLOZ}$ & $72.9$ & $1491$ \\
${\rm G}23$     & ${\rm 2dFHIZ}$ & $72.9$ & $2494$ \\
${\rm GS}$     & ${\rm 2dFLOZ}$ & $49.5$ & $723$ \\
${\rm GS}$     & ${\rm 2dFHIZ}$ & $49.5$ & $1182$ \\
\bottomrule
\end{tabular}
\end{table}

\begin{figure*}
\vspace{1em}
\begin{center}
\resizebox{17.4cm}{!}{\rotatebox{270}{\includegraphics{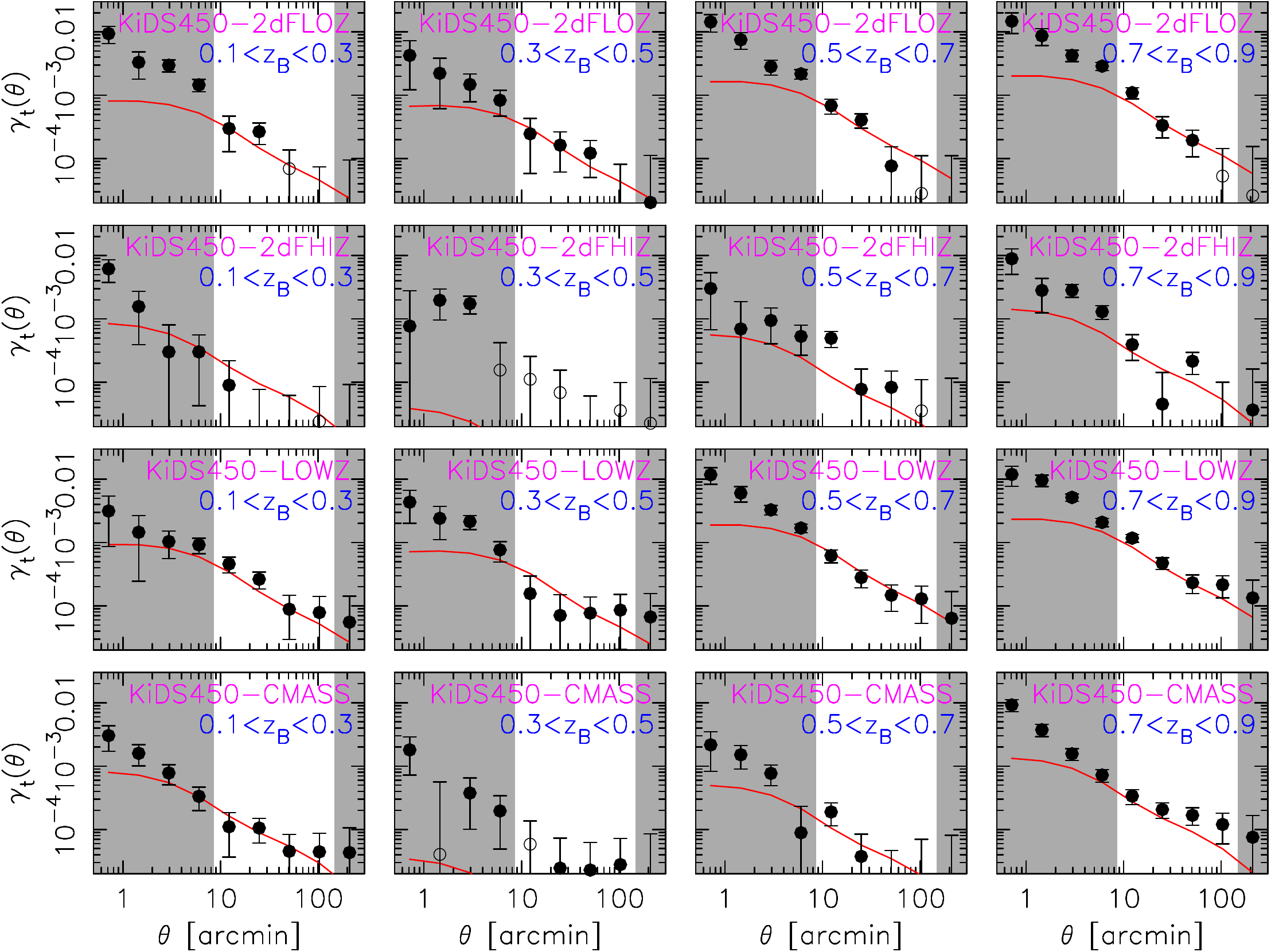}}}
\end{center}
\caption{Measurements of the galaxy-galaxy lensing angular cross-correlation ($\gamma_{\rm t}$) for KiDS overlapping with 2dFLenS and BOSS against angular scale~($\theta$) in arcminutes. The grey regions denote angular scales that were removed from the cosmological analysis when employing fiducial cuts to the data (with conservative cuts, the measurements at 12 arcminutes were also removed for all tomographic bins). The open circles indicate negative values, and we have included best-fit theory lines in red (solid) for comparison. 
}
\label{figgt}
\end{figure*}

\begin{figure*}
\vspace{0.4em}
\begin{center}
\resizebox{17.7cm}{!}{\rotatebox{270}{\includegraphics{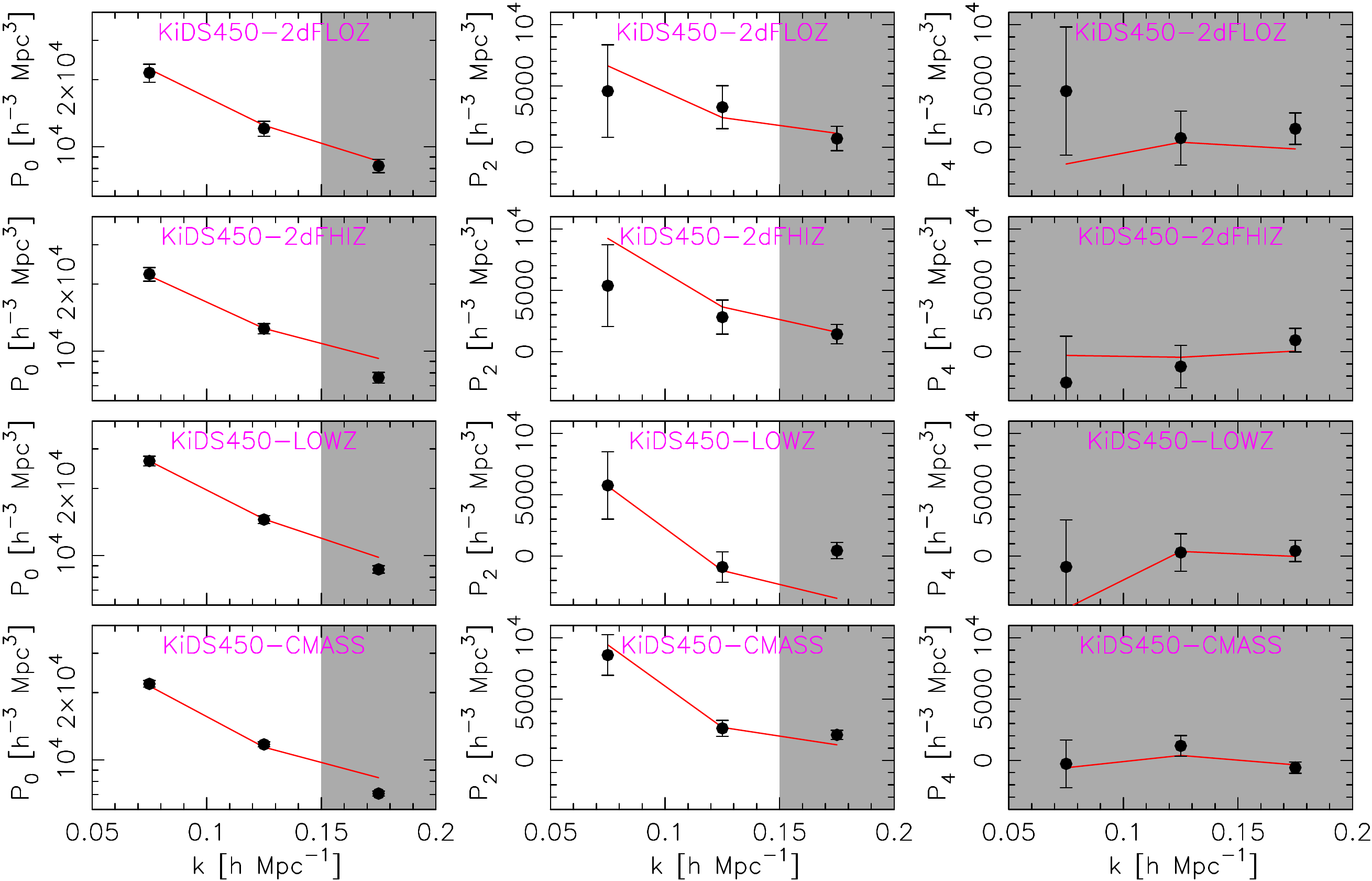}}}
\end{center}
\caption{Measurements of the redshift-space multipole power spectra $\{P_0, P_2, P_4\}$ for 2dFLenS and BOSS in the overlap regions with KiDS at the bin mid-points $k = \{0.075, 0.125, 0.175\}~h~{\rm Mpc}^{-1}$. The grey regions denote physical scales that were removed from the cosmological analysis when employing fiducial cuts to the data (with conservative cuts, the measurements at $k = 0.125~h~{\rm Mpc}^{-1}$ were also removed for all galaxy samples). We have included best-fit theory lines in red (solid) for comparison.
}
\label{figpl}
\end{figure*}

\section{Measurements and covariance}
\label{measlab}

\subsection{Datasets}

\subsubsection{KiDS}

Our galaxy shape measurements are obtained from the KiDS-450 dataset
\citep{kuijken15, Hildebrandt16, dejong17}, which covers an effective area of 360~deg$^2$ on the sky.
Its median redshift is $z_{\mathrm m} = 0.53$, and it contains an effective number density of $n_{\rm eff} = 8.5$ galaxies arcmin$^{-2}$. While \theli \citep{erben13} and \astrowise \citep{begeman13,dejong15} were used to process the raw pixel data, the shears were measured using ${\it{lens}}$fit \citep{miller13} and calibrated using a dedicated large suite of image simulations \citep{fc16}. We divided the dataset into 4 tomographic bins in the range $0.1 < z_{\mathrm B} < 0.9$ (equal widths of $\Delta z_{\mathrm B} = 0.2$), where $z_{\mathrm B}$ is the best-fitting redshift output by the Bayesian photometric redshift code \bpz \citep{benitez2000}, estimated from four-band ugri photometry. The photometric redshift distributions were calibrated using the `weighted direct calibration' (DIR) method described in \citet{Hildebrandt16}.

\subsubsection{BOSS}

BOSS \citep{eisenstein11} is a completed spectroscopic follow-up of the SDSS III imaging survey (SDSS denotes the Sloan Digital Sky Survey), which used the Sloan Telescope to obtain redshifts for over a million galaxies covering $10{,}000$ deg$^2$ on the sky. Color and magnitude cuts were used to select two classes of galaxies in BOSS. The classes consist of the `LOWZ' sample, which contains red galaxies for $z < 0.43$, and the higher-redshift `CMASS' sample that is designed to be approximately stellar-mass limited for $z > 0.43$. We used the data catalogues of the SDSS 10th Data Release (DR10; \citealt{dawson13, anderson14}), and note that the completed BOSS DR12 dataset does not include further observations that overlap with the KiDS regions. Following standard practice \citep{anderson14}, we cut the LOWZ and CMASS datasets to encompass $0.15 < z < 0.43$ and $0.43 < z < 0.70$, respectively, to create homogeneous galaxy samples. Lastly, we used the completeness weights assigned to the BOSS galaxies to correct for the effects of redshift failures, fibre collisions, and other known systematics affecting the angular completeness.

\subsubsection{2dFLenS}

2dFLenS \citep{blake16} is a completed spectroscopic survey conducted by the Anglo-Australian Telescope, covering an area of 731 deg$^2$ principally located in the KiDS regions, with the aim of expanding the
overlap area between galaxy redshift samples and gravitational lensing imaging surveys.  The 2dFLenS spectroscopic dataset contains two main target classes: approximately $40{,}000$ Luminous Red Galaxies (LRGs) across a range of redshifts $z < 0.9$, selected by SDSS-inspired cuts \citep{dawson13}, and a magnitude-limited sample of approximately $30{,}000$ objects between $17 < r < 19.5$ to assist with direct photometric redshift calibration \citep{wolf17}.  In our study, we analyzed the 2dFLenS LRG sample, splitting it into the redshift ranges $0.15 < z < 0.43$ (`2dFLOZ') and $0.43 < z < 0.70$ (`2dFHIZ') to mirror the division of the BOSS dataset. We refer the reader to \citet{blake16} for a full description of the construction of the 2dFLenS selection function and random catalogues.

\subsubsection{Overlapping regions}

KiDS-450 has been divided into five approximately contiguous regions for analysis. The three regions in KiDS-N (G9, G12, G15) overlap with the BOSS dataset, and the two regions in KiDS-S (G23, GS) overlap with the 2dFLenS dataset.  For each region, we restricted both the shape and density samples to the subsets lying within the areas of overlap. As detailed in Table~\ref{taboverlap}, the \{G9, G12, G15\} regions have overlap area $\{9.7, 27.9, 87.4\}$ deg$^2$ with LOWZ and $\{44.0, 90.3, 87.4\}$ deg$^2$ with CMASS. The \{G23, GS\} regions have overlap area $\{72.9, 49.5\}$ deg$^2$ with 2dFLenS. The number of lenses overlapping with the \{G9, G12, G15\} regions is $\{414, 849, 3781\}$ for LOWZ and $\{4272, 7451, 8753\}$ for CMASS, while the number of lenses overlapping with the \{G23, GS\} regions is $\{1491, 723\}$ for 2dFLOZ and $\{2494, 1182\}$ for 2dFHIZ. These statistics will continue to improve with future releases of KiDS.

\subsubsection{Planck}

In our analysis of the KiDS, 2dFLenS, and BOSS datasets, we examine their concordance with the cosmic microwave background measurements of Planck~\citep{planck15,planck15like}. To this end, we consider Planck CMB temperature and polarization information on large angular scales, including multipoles $\ell \leq 29$ (via the low-$\ell$ TEB likelihood), along with CMB temperature information on smaller angular scales (via the \plik TT likelihood). We denote these `TT+lowP' measurements `Planck~2015'. Conservatively, we do not include Planck polarization data on small angular scales, and we also do not include Planck CMB lensing measurements (the two would slightly decrease and increase the discordance with the KiDS-450 cosmic shear measurements, respectively, as noted in \citealt{joudaki17}). However, in Appendix~\ref{tausec}, we further consider the impact of the updated Planck measurement of the optical depth in \citet{plancktau}, illustrating that it does not significantly affect our results.

\begin{figure}
\vspace{0.6em}
\hspace{-1.0em}
\resizebox{8.7cm}{!}{\includegraphics{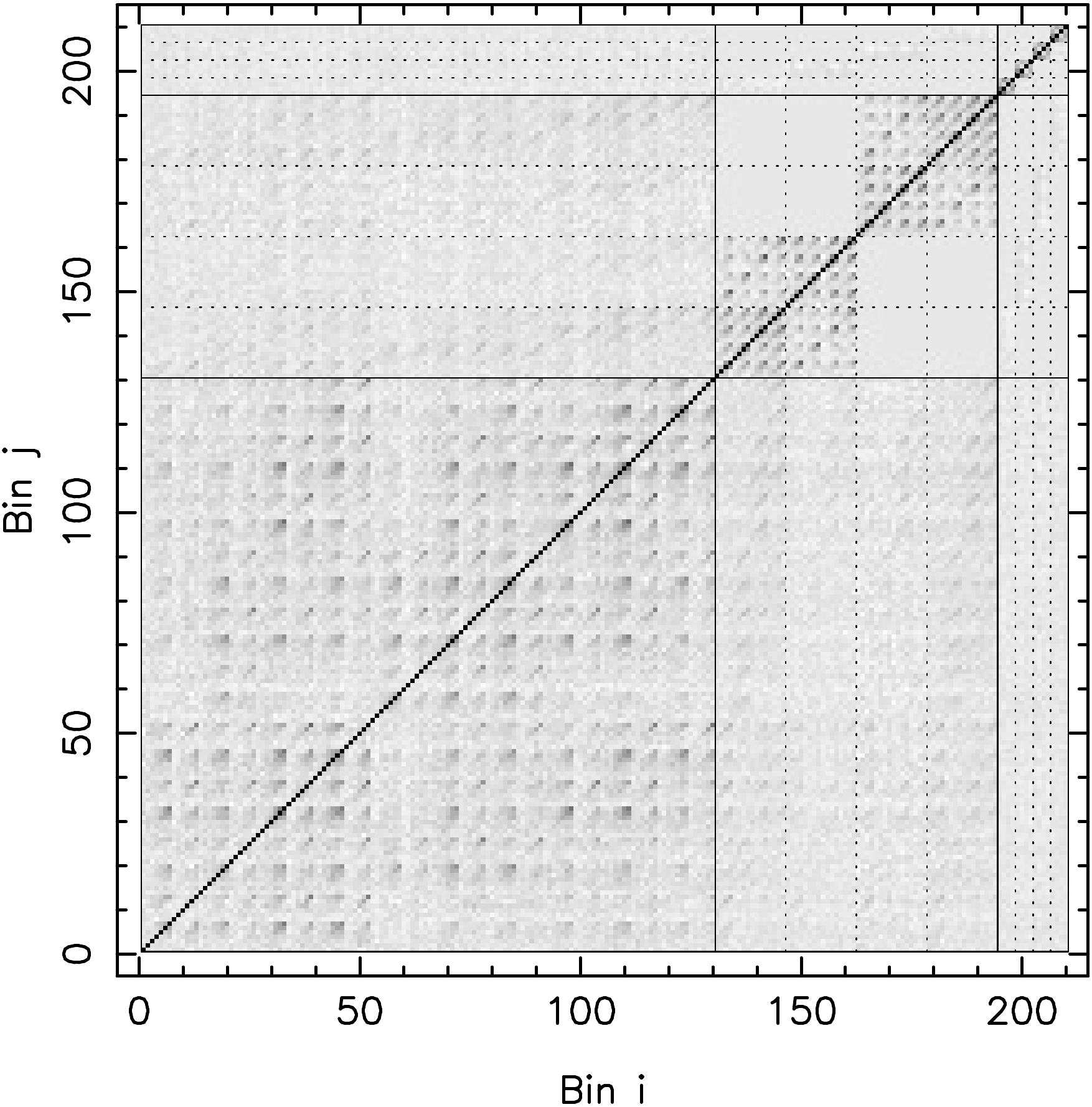}}
\vspace{-1.5em}
\caption{Correlation coefficients $r$ of the covariance matrix of the full data vector of cosmic shear, galaxy-galaxy lensing, and multipole power spectrum measurements for KiDS overlapping with 2dFLenS and BOSS (coefficients defined in equation~\ref{rcceqn}). We show the elements of the $\{\xi_{\pm}, \gamma_{\rm t}, P_{0/2}\}$ data vector that were employed in the fiducial analysis (with selections detailed in Table~\ref{tabcuts}). There are 130 elements of $\xi_{\pm}$, 64 elements of $\gamma_{\rm t}$, and 16 elements of $P_{0/2}$, delineated by thin solid lines. The $\gamma_{\rm t}$ and $P_{0/2}$ measurements are further delineated by thin dotted lines indicating the divisions between 2dFLOZ, 2dFHIZ, CMASS, and LOWZ. The ordering of the $\xi_{\pm}$ elements is the same as in our previous cosmic shear analyses (e.g.~\citealt{Heymans13, joudaki16, Hildebrandt16}), where for 4 tomographic and 9 angular bins it follows $\{ \xi^{11}_+(\theta_1), \xi^{11}_+(\theta_2), ..., \xi^{11}_-(\theta_8), \xi^{11}_-(\theta_9), \xi^{12}_+(\theta_1), ..., \xi^{44}_-(\theta_9) \}$. We use a greyscale where white represents $r=-0.1$ and black represents $r=1$.
}
\label{figcov}
\end{figure}

\subsection{Measurements}

\subsubsection{Cosmic shear measurements}
\label{cssec}

Our lensing observables are given by the tomographic two-point shear correlation functions $\xi_{\pm}^{ij}(\theta)$ for an angular range of 0.5 to 300 arcmin (as detailed in Section~\ref{cslab}). We follow \citet{Hildebrandt16} in using 7 angular bins in $\xi_{+}^{ij}(\theta)$ between 0.5 to 72 arcmin, and 6 angular bins in $\xi_{-}^{ij}(\theta)$ between 4.2 to 300 arcmin. Put differently, considering the nine angular bin mid-points at $[0.713, 1.45, 2.96, 6.01, 12.2, 24.9, 50.7, 103, 210 ]$ arcmin\footnote{These angular scales do not account for the galaxy weights, which causes a marginal $0.3\sigma$ increase in the relative discordance of KiDS with Planck.}, we retain the first 7 bins for $\xi_{+}^{ij}(\theta)$, and the last 6 bins for $\xi_{-}^{ij}(\theta)$. Given our four tomographic bins, the cosmic shear data vector consists of 130 elements. We applied multiplicative shear bias corrections to the cosmic shear measurements using the method described by \citet{Hildebrandt16}, and combined measurements in different regions through weighting by the effective pair number. We do not show the cosmic shear measurements as these were presented in \citet{Hildebrandt16}.

\subsubsection{Galaxy-galaxy lensing measurements}
\label{secgt}

We measured the galaxy-galaxy lensing signal $\gamma_{\rm t}^j(\theta)$ between each lens sample (LOWZ, CMASS, 2dFLOZ, 2dFHIZ) and the KiDS-450 tomographic bins labelled by $j$ (see Section~\ref{ggllab} for the theoretical description). These measurements fiducially cover the 4 angular bins with central values at $[12.2, 24.9, 50.7, 103]$ arcmin. We do not include the measurements at 210 arcmin due to low signal-to-noise, and we remove the measurements below 12.2 arcmin given the aim to avoid nonlinear galaxy bias. We also consider a conservative case where we remove all $\gamma_{\rm t}$ measurements below 24.9 arcmin, and a `large-scale' case where we remove all $\gamma_{\rm t}$ measurements below 50.7 arcmin, 
as detailed in Table~\ref{tabcuts}. 

Our cuts to $\gamma_{\rm t}$ are motivated by the scale of the nonlinear galaxy bias as roughly the 1-halo to 2-halo transition scale, which is at $r \simeq 2~h^{-1}~{\rm Mpc}$ for luminous red galaxies (e.g.~\citealt{parejko13,more15}). While the effect must also depend on galaxy type, i.e.~increase with halo mass (lower for 2dFLOZ and LOWZ compared to 2dFHIZ and CMASS), we employ the same angular cuts to all of our galaxy samples. For reference, the smallest angular scale of 12 arcmin used in the galaxy-galaxy lensing analysis corresponds to $\sim3~h^{-1}{\rm Mpc}$ at $z_{\rm eff} = 0.32$ (2dFLOZ, LOWZ) and $\sim5~h^{-1}{\rm Mpc}$ at $z_{\rm eff} = 0.57$ (2dFHIZ, CMASS). Our fiducial cuts with 4 angular bins were also verified to yield consistent results when discarding further angular scales in the conservative and large-scale cases (as discussed in Section~\ref{Results}).

We corrected for any additive shear bias by subtracting the correlation between the shear sample and a random lens catalogue, and applied multiplicative shear bias corrections as above. The random-catalogue correction also suppresses the sample variance error \citep{singh16}. We combined measurements (for each lens sample) in different regions through weighting by the effective pair number, and present our galaxy-galaxy lensing measurements in Fig.~\ref{figgt}. For these measurements, we do not discard $\gamma_{\rm t}$ obtained from source bins at lower redshift than lens bins (for instance, the correlation between tomographic bin 1 where $0.1 < z_{\rm B} \leq 0.3$, and 2dFHIZ which covers $0.43 < z_{\rm 2dFHIZ} \leq 0.7$) given the width of the source distributions for each tomographic bin (nonzero up to $z=3.5$ for all bins). We find that the choice between keeping or discarding these specific $\gamma_{\rm t}$ measurements does not particularly impact our cosmological parameter constraints.

We note that galaxies from the source sample that are physically associated with the lenses will not be lensed, and may bias the tangential shear measurements.  We tested for this effect by measuring the overdensity of source galaxies around lenses, showing that the resulting `boost factor' was significant on small scales, but not important for the range of scales used in our fits (at most 2\%, and always consistent with 1.0 within the errors; also see e.g.~\citealt{amon17,dvornik17}).  We therefore did not apply this correction.

\subsubsection{Multipole power spectrum measurements}

We estimated the multipole power spectra $\{P_0(k), P_2(k), P_4(k)\}$ of the different lens samples, within the boundaries of each KiDS-450 region, using the direct Fast Fourier Transform method presented by \citet{bianchi15}, following the procedure described in Section 7.3 of \citet{blake16}.  Motivated by the relatively small overlap volumes, we adopted relatively wide Fourier bins of width $\Delta k = 0.05 \, h$ Mpc$^{-1}$. The lack of available modes in the first bin, with centre $k = 0.025 \, h$ Mpc$^{-1}$, necessitated us excluding this bin from the analysis and utilizing the remaining bins with centres $k = \{ 0.075, 0.125, ... \} \, h$ Mpc$^{-1}$. 

As detailed in \citet{blake16}, we constructed a data vector $\{ P_0(k_1), P_0(k_2), ..., P_2(k_1), P_2(k_2), ..., P_4(k_1), P_4(k_2), ... \}$ for each lens sample, and derived a convolution matrix that enabled us to generate an equivalent model power spectrum allowing for the survey window function. We excluded the hexadecapole $(P_4)$ terms from our final fits, as they contained no significant signal, and combined measurements (for each lens sample) in different regions through weighting by their area. These measurements are presented in Fig.~\ref{figpl}, where the statistical significance of $P_0$ is higher than $P_2$, and the BOSS measurements are currently stronger than those from 2dFLenS (in the overlapping regions with KiDS). In our cosmological analysis, to avoid nonlinearities in the matter power spectrum and galaxy bias, we only retain the measurements at $k = \{0.075, 0.125\} \, h$ Mpc$^{-1}$ in a `fiducial' case, and the measurements at $k = 0.075~h~{\rm Mpc}^{-1}$ in a `conservative' case (as detailed in Table~\ref{tabcuts}).

\subsection{Covariance}
\label{covlab}

We computed the full covariance between the different observables, scales, and samples using a large suite of $N$-body simulations\footnote{We note that the cosmic shear $\xi_{\pm}$ part of the covariance is also
constructed from $N$-body simulations, as compared to the analytic covariance used in \citealt{Hildebrandt16} and \citealt{joudaki17}.}. Our mocks are built from the SLICS (Scinet LIght Cone Simulations) series
\citep{hdw15}, which consists of 930 independent dark matter only simulations in which $1536^3$ particles inside a $3072^3$ grid are evolved within a box-size $L = 505 \, h^{-1}$ Mpc with the high-performance {\small CUBEP$^3$M} $N$-body code \citep{hd13}. The projected density field and full halo catalogues were stored at 18 snapshots in the range $z < 3$.

The gravitational lensing shear within the simulations is computed at these multiple lens planes using the flat-sky Born approximation, and a survey cone spanning 60 deg$^2$ is constructed.  We constructed mock source catalogues by populating each cone using a source redshift distribution and an effective source density matching KiDS-450, by Monte-Carlo sampling sources from the density field. We applied shape noise to the two-component shears, drawing the noise components from a Gaussian distribution matching that of the lensing survey.  We also produced mock lens catalogues within the simulations, by populating the dark matter haloes with a Halo Occupation Distribution (HOD) model tuned to match the large-scale clustering amplitude and number density of the lens samples. We refer the reader to \citet{blake16} for a full description of our HOD approach.

We measured the cosmic shear, galaxy-galaxy lensing, and multipole power spectrum statistics of each of the 930 mocks (using the \athena software of \citealt{2014ascl.soft02026K} for $\xi_{\pm}$ and $\gamma_{\rm t}$, and using direct Fast Fourier Transforms as described for $P_{0/2}$), and constructed the joint covariance by scaling each piece with the appropriate overlap area $A_{\rm overlap}$ (i.e., by multiplying the covariance by
$60~{{\rm deg}^2}/A_{\rm overlap}$). In the case of the shear, galaxy-galaxy lensing and multipole pieces, the overlap area corresponds to the masked lensing area, the subset of that area overlapping with the lens distribution, and the full area of each field, respectively. We propagated the error in the multiplicative shear bias correction into the cosmic shear and galaxy-galaxy lensing pieces of the covariance. Due to finite box effects and neglecting super-sample variance, we slightly underestimate the variance on the largest scales ($\sim10\%$ on the largest scale of $\xi_+$, other statistics not affected;~\citealt{hdw15}).

In Fig.~\ref{figcov}, we show the covariance between the measurements via the respective correlation coefficients,
\begin{equation}
\label{rcceqn}
r(i,j) = {{\rm Cov}(i,j)} / \sqrt{{\rm Cov}(i,i) \, {\rm Cov}(j,j)},
\end{equation}
where `Cov' corresponds to the covariance between the measurement pairs $\{i, j\}$. For fiducial cuts, the correlation matrix contains 210 elements on each side, corresponding to the post-masked elements of the $\{\xi_+, \xi_-, \gamma_{\rm t}, P_0, P_2\}$ data vector. As expected, the correlation coefficients are larger between the elements of the same class of observables, and between elements of $\{\xi_\pm, \gamma_{\rm t}\}$ as compared to $\{\xi_\pm, P_{0/2}\}$ and $\{\gamma_{\rm t}, P_{0/2}\}$. The covariance is nonzero between the different elements with the exception of a zero covariance between the $\gamma_{\rm t}$ and $P_{0/2}$ elements of different lens samples, between the $\gamma_{\rm t}$ elements of 2dFLenS and BOSS (aside from a minor nonzero contribution by the propagation of the uncertainty in the multiplicative shear bias correction), and between the $P_{0/2}$ elements of different samples. 

Lastly, instead of correcting for the inverse of our numerically estimated covariance matrix with the approach of 
\citet{kaufman67} and \citet{hartlap07}, previously used in e.g.~\citet{Heymans13} and \citet{joudaki16}, we avoid biasing our cosmological parameter constraints by employing the \citet{sh15} correction to the likelihood in our MCMC runs. We have checked that our parameter constraints are not particularly affected by the choice between these two different methods.

\begin{table*}
\caption{\label{tabcuts} Scales used in various setups of our analysis, considering observations of cosmic shear, galaxy-galaxy lensing, and redshift-space multipole power spectra. The physical and angular scales are given at the respective bin mid-points. We also list the size of the data vectors of the different setups.}
\begin{tabular}{lcccccc}
\hline
Cosmological observations & scales & scales & scales & scales & scales & size\\
&$\xi_+ {\rm [arcmin]}$ & $\xi_- {\rm [arcmin]}$ & $\gamma_{\rm t} {\rm [arcmin]}$ & $P_0 [h~\rm Mpc^{-1}]$ & $P_2 [h~\rm Mpc^{-1}]$ & data vector \\
\hline
$\{\xi_+, \xi_-\}$      & $0.7$ -- $51$ & $6.0$ -- $210$ & --  & -- & -- & $130$\\
$\{\xi_+, \xi_-, \gamma_{\rm t}\}$      & $0.7$ -- $51$ & $6.0$ -- $210$ & $12$ -- $100$  & -- & -- & $194$\\
$\{\xi_+, \xi_-, \gamma_{\rm t}\}$-conserv      & $0.7$ -- $51$ & $6.0$ -- $210$ & $25$ -- $100$  & -- & -- &  $178$\\
$\{\xi_+, \xi_-, P_0, P_2\}$      & $0.7$ -- $51$ & $6.0$ -- $210$ & --  & $0.075$ -- $0.125$ & $0.075$ -- $0.125$ &  $146$\\
$\{\xi_+, \xi_-, P_0, P_2\}$-conserv      & $0.7$ -- $51$ & $6.0$ -- $210$ & --  & $0.075$ & $0.075$ & $138$\\
$\{\xi_+, \xi_-, \gamma_{\rm t}, P_0, P_2\}$      & $0.7$ -- $51$ & $6.0$ -- $210$ & $12$ -- $100$  & $0.075$ -- $0.125$ & $0.075$ -- $0.125$ &  $210$\\
$\{\xi_+, \xi_-, \gamma_{\rm t}, P_0, P_2\}$-conserv      & $0.7$ -- $51$ & $6.0$ -- $210$ & $25$ -- $100$  & $0.075$ & $0.075$ & $186$\\
$\{\xi_+, \xi_-, \gamma_{\rm t}, P_0, P_2\}$-large scales      & $25$ -- $51$ & $210$ & $51$ -- $100$  & $0.075$ & $0.075$ & $70$\\
\hline
\end{tabular}
\end{table*}

\subsection{Blinding}
Along the lines of the KiDS-450 analysis~\citep{Hildebrandt16}, we employed `blinding' of our data files to avoid confirmation bias (in the case of cosmic shear we were `double-blinded'). We generated three separate copies of the measurements and covariance (one true copy and two false copies), and randomly used `blind1' throughout the testing phase of our work. The multipole power spectra were not blinded, such that they would not change between the three copies of the data files. The simulated covariance matrices differed very slightly between blindings because the propagation of the m-correction error involved the measured shear correlation functions in each case.

Once all decisions had been made, we generated the core results of this paper with all three blindings, and then unblinded. We found that `blind2' contains the true copy of the measurements and covariance, and then proceeded to generate the remainder of our results without making any further decisions that could change the core results that were generated pre-unblinding. We refer to \citet{Hildebrandt16} for further details on the blinding scheme.

\section{Likelihood calculation, parameter priors, and model selection}
\label{cosmomcsec}

\subsection{Extended \cosmomc fitting pipeline for self-consistent cosmological analyses of WL and RSDs}

\subsubsection{Likelihood calculation}

In \citet{joudaki16}, as part of the cosmological analysis package \cosmomc\footnote{\texttt{http://cosmologist.info/cosmomc/}} \citep{Lewis:2002ah}, we released a new module in the Fortran 90 language for cosmological parameter analyses of tomographic weak gravitational lensing measurements, including key astrophysical systematics arising from intrinsic galaxy alignments, baryonic effects in the nonlinear matter power spectrum, and photometric redshift uncertainties. The original\footnote{\texttt{https://github.com/sjoudaki/cfhtlens_revisited}} and updated\footnote{\texttt{https://github.com/sjoudaki/kids450}} versions of this lensing module were subsequently used in the cosmological parameter analyses of CFHTLenS and KiDS data \citep{joudaki16,joudaki17,Hildebrandt16}.

As part of the current work, we are releasing an extended version of the \cosmomc module, so that in addition to tomographic cosmic shear ($\xi_+, \xi_-$), the module self-consistently allows for the analysis of tomographic galaxy-galaxy lensing ($\gamma_{\rm t}$) and multipole power spectrum ($P_0, P_2$) measurements from overlapping lensing and spectroscopic surveys, with new degrees of freedom that include the galaxy bias, pairwise velocity dispersion, and shot noise. As described in Section~\ref{covlab}, the module includes the full $N$-body simulated covariance between the observables. The code is internally parallelized and given a dual 8-core Intel Xeon processor (Sandy Bridge 2.6 GHz) computes the full likelihood of $\{\xi_+, \xi_-, \gamma_{\rm t}, P_0, P_2\}$ at a single cosmology in 0.13 seconds (and even less for a reduced vector), which is sufficiently fast for our MCMC purposes. This only refers to the likelihood module itself, and does not include e.g.~the time it takes \camb\footnote{\texttt{http://camb.info}} \citep{LCL} to compute the linear matter power spectrum. The full fitting pipeline is denoted \cosmolss to highlight that it incorporates the key probes of large-scale structure, and is made public at \sjaddress.

\subsubsection{Matter power spectrum}

In our cosmological analysis, the linear matter power spectrum obtained from \camb can be extended to nonlinear scales with either \halofit~\citep{Smith03, Takahashi12} or \hmcode\footnote{\texttt{https://github.com/alexander-mead/hmcode}} \citep{Mead15,Mead16}, where we advocate the latter in particular because of its ability to account for the impact of baryonic physics on the nonlinear matter power spectrum (e.g.~due to star formation, radiative cooling, and AGN feedback). \hmcode achieves this through calibration to the OverWhelmingly Large (OWL) hydrodynamical simulations (\citealt{Schaye10,Daalen11,semboloni11}; in addition to the Coyote dark matter simulations of \citealt{Coyote4} and references therein), where the baryonic feedback amplitude $B$ is varied as a free parameter in our analysis (in accordance with previous work, e.g.~\citealt{joudaki16, joudaki17, Hildebrandt16}). 

In \hmcode, the halo bloating parameter $\eta_{\hmcode}$ changes the halo density profile and is a function of the feedback amplitude, such that $\eta_{\hmcode} = 0.98 - 0.12B$. This expression for $\eta_{\hmcode}$, obtained from fitting to the OWL simulations, is a slight improvement to the original version~\citep{Mead15} and has a marginal impact on the cosmological parameter constraints. While it is possible to vary $\eta_{\hmcode}$ as a free parameter (i.e.~independently of $B$, see e.g.~\citealt{maccrann17}), and we allow for this capability in our public release of \cosmolss (as we did for the lensing-only pipelines), we only consider a one-parameter baryonic feedback model in the forthcoming analysis given the strong degeneracy between $\eta_{\hmcode}$ and $B$ in the fits to the OWL simulations.

In \citet{miratitan17}, a new 'cosmic emulator' is provided that is able to produce the matter power spectrum to $4\%$ accuracy ($k \leq 5~h~{\rm Mpc}^{-1}$, $z \leq 2$) within a space of eight cosmological parameters $(\Omega_{\mathrm m}h^2, \Omega_{\mathrm b}h^2, \sigma_8, h, n_{\mathrm s}, w_0, w_{\mathrm a}, \Omega_{\nu}h^2)$ by interpolating between high-accuracy simulation results. \citet{miratitan17} report that both \halofit and \hmcode disagree at the $\sim20\%$ level when compared to their new emulator. We have conducted independent tests that confirm this, but find that the worst errors arise exclusively for high neutrino masses. If the sum of neutrino masses is fixed at 0.05~{eV}, the agreement between the new emulator and \hmcode is $\sim5\%$, even with time-varying dark energy models. The discrepancy increases to $20\%$ approximately linearly when increasing the sum of neutrino masses from 0.05~{eV} to 0.95~{eV}.

\subsubsection{Astrophysical parameter space}
\label{astrospace}

The joint lensing/RSD fitting pipeline allows for a large number of degrees of freedom to be varied in MCMC analyses. For cosmic shear calculations limited to $\{\xi_+, \xi_-\}$, the code allows for the intrinsic alignment amplitude $A_{\rm IA}$, redshift dependence $\eta_{\rm IA}$, and luminosity dependence $\beta_{\rm IA}$ to be varied (along with $B$ and $\eta_\hmcode$ already mentioned). In our cosmological analysis, we always fix $\beta_{\rm IA} = 0$ as described in Section~\ref{cslab}, and only vary $\eta_{\rm IA}$ in the `extended systematics' scenario (Appendix~\ref{extsystlab}).

We incorporate the photometric redshift uncertainties in accordance with \citet{Hildebrandt16}, where we iterate over a large range of bootstrap realizations of the photometric redshifts obtained from the DIR method\footnote{While not considered in the analysis of KiDS data, the fitting pipeline also allows for additional degrees of freedom that shift the tomographic source distributions along the redshift axis (preserving their shapes, in accordance with the analysis of CFHTLenS; \citealt{joudaki16}).}. We do not include the sample variance associated with these photometric redshifts, which we estimate to be subdominant for KiDS cosmic shear (Appendix C3.1 in \citealt{Hildebrandt16}). We also do not introduce additional degrees of freedom to account for uncertainties in the multiplicative shear calibration, but propagate the uncertainties from \citet{fc16} into the covariance. We refer to Appendix A in \citet{joudaki17} on the impact of `unknown' additional uncertainty in either the shear and redshift calibration corrections on the cosmological parameter constraints.

Given galaxy-galaxy lensing measurements, the fitting pipeline allows for the galaxy bias $b$ of each lens sample to be varied freely. When multipole power spectrum measurements are considered, this extends to the shot noise $N_{\rm shot}$ and pairwise velocity dispersion $\sigma_{\rm v}$. Given our four galaxy samples, this implies 4 additional degrees of freedom when galaxy-galaxy lensing measurements are considered, and 12 additional degrees of freedom when multipole power spectrum measurements are considered. Hence, for the full data vector, even when ignoring any redshift or luminosity dependence of the intrinsic alignments, the fitting pipeline varies 14 astrophysical degrees of freedom in addition to any cosmological degrees of freedom. When considering fluctuations in the underlying cosmology, this increases to at least 19 degrees of freedom in $\Lambda$CDM (Section~\ref{Results}), and to at most 27 degrees of freedom in our binned modified gravity model (Section~\ref{Resultsextended})\footnote{When combined with Planck CMB temperature measurements, this increases to 20 and 28, respectively, given the introduction of the optical depth~$\tau$ as an additional degree of freedom. These numbers are even larger when counting the CMB nuisance parameters.}. We are able to complete MCMC runs allowing for such sizable parameter spaces owing in large part to the speed of the fitting pipeline.

\begin{table}
\begin{center}
\caption{Priors on the cosmological and astrophysical parameters varied in the MCMC runs. We always vary the `vanilla' cosmological parameters in the first third of the table simultaneously with the astrophysical parameters in the second third of the table (encapsulating intrinsic alignments, baryon feedback, galaxy bias, shot noise, velocity dispersion). In the table, $\theta_{\mathrm{s}}$ denotes the angular size of the sound horizon at the redshift of last scattering, and we emphasize that the Hubble constant is a derived parameter. Our prior on the baryonic feedback amplitude is wider than in \citet{joudaki17}, and our priors on the baryon density and Hubble constant are moreover wider than in \citet{Hildebrandt16}. While varied independently, we impose the same prior ranges on the galaxy bias, pairwise velocity dispersion, and shot noise for all four lens samples (2dFLOZ, 2dFHIZ, LOWZ, CMASS). We only vary the optical depth when including the CMB. The extended cosmological parameters in the last third of the table are varied simultaneously with the baseline parameters, and constrained in Section~\ref{Resultsextended}.
}
\begin{tabular}{lll}
\toprule
Parameter & Symbol & Prior\\
\midrule
Cold dark matter density & $\Omega_{\mathrm c}h^2$ & $[0.001, 0.99]$\\
Baryon density & $\Omega_{\mathrm b}h^2$ & $[0.013, 0.033]$\\
100 $\times$ approximation to $\theta_{\mathrm s}$ & $100 \theta_{\rm MC}$ & $[0.5, 10]$\\
Amplitude of scalar spectrum & $\ln{(10^{10} A_{\mathrm s})}$ & $[1.7, 5.0]$\\
Scalar spectral index & $n_{\mathrm s}$ & $[0.7, 1.3]$\\
Optical depth & $\tau$ & $[0.01, 0.8]$ \\
Dimensionless Hubble constant & $h$ & $[0.4, 1.0]$ \\
Pivot scale $[{\rm{Mpc}}^{-1}]$ & $k_{\rm pivot}$ & 0.05 \\
\midrule
IA amplitude & $A_{\rm IA}$ & $[-6, 6]$\\
{\it~~~--~extended case} &  & $[-20, 20]$\\
IA redshift dependence & $\eta_{\rm IA}$ & [0, 0]\\
{\it~~~--~extended case} &  & $[-20, 20]$\\
Feedback amplitude & $B$ & $[1, 4]$\\
{\it~~~--~extended case} &  & $[1, 10]$\\
Galaxy bias & $b_x$ & $[0, 4]$\\
{\it~~~--~extended case} &  & $[0, 10]$\\
Velocity dispersion $[h^{-1}{\rm Mpc}]$ & $\sigma_{{\rm v},x}$ & $[0, 10]$\\
{\it~~~--~extended case} &  & $[0, 100]$\\
Shot noise $[h^{-1}{\rm Mpc}]^3$ & $N_{{\rm shot},x}$ & $[0, 2300]$\\
{\it~~~--~extended case} &  & $[0, 3000]$\\
\midrule
MG bins (modifying grav. const.) & $Q_i$ & $[0, 10]$\\
MG bins (modifying deflect. light) & $\Sigma_j$ & $[0, 10]$\\
Sum of neutrino masses [eV] & $\sum m_\nu$ & $[0.06, 10]$\\
Constant dark energy EOS & $w$ & $[-3, 0]$\\
Present dark energy EOS & $w_0$ & $[-3, 0]$\\
Derivative of dark energy EOS & $w_a$ & $[-5, 5]$\\
Curvature & $\Omega_k$ & $[-0.15, 0.15]$\\
\bottomrule
\end{tabular}
\label{tabpri}
\end{center}
\end{table}

\subsection{Model selection and dataset concordance}
\label{modsec}

\subsubsection{Deviance information criterion}

To assess the relative statistical preference of cosmological models, we use the Deviance Information Criterion (DIC;~\citealt{spiegelhalter02}, also see \citealt{liddle07,trotta08,spiegelhalter14,joudaki16}), given by
\begin{equation}
{\rm{DIC}} \equiv {\chi^2_{\rm eff}(\hat{\theta})} + 2p_D. 
\label{dicdef}
\end{equation}
The first term of the DIC is the best-fit effective ${\chi^2_{\rm eff}(\hat{\theta})}  = -2 \ln {\mathcal{L}}_{{\rm max}}$, where $\hat{\theta}$ is the vector of varied parameters at the maximum likelihood ${\mathcal{L}}_{\rm max}$ of the data given the model. The goodness of fit of a model is then commonly quantified in terms of the reduced $\chi^2$, given by $\chi^2_{\rm red} = \chi^2_{\rm eff}/\nu$, where $\nu$ is the number of degrees of freedom. As a measure of the effective number of parameters, the second term of the DIC is the `Bayesian complexity', $p_D = \overline{\chi^2_{\rm eff}(\theta)} - {\chi^2_{\rm eff}(\hat{\theta})}$, where the bar denotes the mean taken over the posterior distribution. The DIC is therefore the sum of the goodness of fit of a model and its Bayesian complexity, and acts to penalize more complex models. For reference, a difference in $\chi^2_{\rm eff}$ of $\{5, 10\}$ between two models corresponds to a probability ratio of $\{1/12, 1/148\}$, and we therefore take a DIC difference of $\{5, 10\}$ to correspond to \{moderate, strong\} preference in favor of the model with the lower DIC. In our convention, a positive $\Delta {\rm DIC} = {\rm DIC_{extended}} - {\rm DIC}_{{\Lambda}{\rm CDM}}$ favors $\Lambda$CDM relative to an extended model.

\subsubsection{$\log \mathcal{I}$ concordance diagnostic}

In addition to model selection, we utilize the DIC to quantify the level of concordance between different datasets (in particular our lensing/RSD measurements compared to the Planck CMB temperature measurements). Our diagnostic is given by (\citealt{joudaki16}; Parkinson \& Joudaki in prep):
\begin{equation}
{\mathcal{I}}(D_1, D_2) \equiv \exp\{{-{\mathcal{G}}(D_1, D_2)/2}\}, 
\label{eqnlogi1}
\end{equation}
where $D_1$ and $D_2$ refer to two distinct datasets, and
\begin{equation}
{\mathcal{G}}(D_1, D_2) = {{{\rm{DIC}}(D_1 \cup D_2)} - {{\rm{DIC}}(D_1) - {{\rm{DIC}}(D_2)}}},
\label{eqnlogi2}
\end{equation}
such that ${{\rm{DIC}}(D_1 \cup D_2)}$ is the joint DIC of the two datasets. Here, the two datasets are taken to be concordant when $\log \mathcal{I}$ is positive and discordant when $\log \mathcal{I}$ is negative. The magnitude of $\log \mathcal{I}$ follows the Jeffreys' scale \citep{jeffreys,kr95}, such that $|\log \mathcal{I}| \gtrsim 1/2$ is considered `substantial', $|\log \mathcal{I}| \gtrsim 1$ is considered `strong', and $|\log \mathcal{I}| \gtrsim 2$ is considered `decisive' (where e.g. the last case corresponds to a probability ratio in excess of 100). 

The $\log \mathcal{I}$ diagnostic has been applied to cosmic shear and CMB temperature data, notably to assess the discordance between CFHTLenS and Planck \citep{joudaki16}, and between KiDS and Planck \citep{Hildebrandt16,joudaki17}. The main benefit of the diagnostic compared to analogous concordance tests (in particular tests grounded in the Bayesian evidence, see e.g.~\citealt{mrs06,raveri15}) is that it is trivially obtained from the same MCMC chains generated for parameter constraint purposes (while the $\log \mathcal{I}$ and evidence-based tests were shown to largely agree in \citealt{joudaki16}). 

\subsubsection{$T(S_8)$ concordance diagnostic}

As current cosmic shear measurements most strongly constrain $S_8$, we also quantify the tension $T$ in this parameter between different datasets (e.g.~\citealt{Hildebrandt16, joudaki17}):
\begin{equation}
T(S_8) = \left|\overline{S_8^{D_1}} - \overline{S_8^{D_2}}\right| / \sqrt{\sigma^2\left(S_8^{D_1}\right) + \sigma^2\left(S_8^{D_2}\right)} ,
\label{eqnts8}
\end{equation}
where the vertical bars denote absolute value, and $\sigma$ is the symmetric 68\% confidence interval about the parameter mean. Given a $\Lambda$CDM cosmological model, the tension in $S_8$ between KiDS and Planck was shown to be $2.3\sigma$ in \citet{Hildebrandt16}, and $2.1\sigma$ when employing the wider priors in \citet{joudaki17}. Both of these analyses moreover found $\log \mathcal{I}$ to be consistent with `substantial discordance' between the KiDS and Planck datasets, in qualitative agreement with the assessment of the $T(S_8)$ diagnostic. In Section~\ref{Results}, we examine to what extent these concordance tests change when further including galaxy-galaxy lensing and multipole power spectrum measurements in the cosmological analysis.

\subsection{Baseline configurations}
\label{basesec}

We now detail the baseline configurations used in all of our MCMC runs. Extensions to these configurations, in particular in the form of expanded parameter spaces, are discussed in each relevant subsection of Section~\ref{Resultsextended}. All varied parameters and their uniform priors are listed in Table~\ref{tabpri}.

\subsubsection{Cosmological degrees of freedom}

The baseline (or `vanilla') cosmological parameters varied in all MCMC runs are the cold dark matter density, $\Omega_{\mathrm c}h^2$, baryon density, $\Omega_{\mathrm b}h^2$, approximation to the angular size of the sound horizon, $\theta_{\rm MC}$, amplitude of the scalar spectrum, $\ln{(10^{10} A_{\mathrm s})}$, and scalar spectral index, $n_{\mathrm s}$. Here, `$\ln$' refers to the natural logarithm (while we take `$\log$' to refer to logarithms with base 10), and we define \{$A_{\mathrm s}, n_{\mathrm s}$\} at the pivot wavenumber $k_{\rm pivot} = 0.05~{\rm Mpc}^{-1}$. From these cosmological parameters, we derive the Hubble constant $H_0$ (or $h$ in its dimensionless form), and rms of the linear matter density field at present on $8~h^{-1} {\rm Mpc}$ scales, $\sigma_8$. When considering Planck CMB temperature measurements, we also always vary the optical depth to reionization, $\tau$.

The baseline cosmological model assumes flatness, a cosmological constant, no running of the spectral index, and three massless neutrinos (such that the effective number of neutrinos $N_{\rm eff} = 3.046$). We do not impose a minimum neutrino mass of 0.06 eV in our baseline cosmological model (motivated by neutrino oscillation experiments in the normal hierarchy, e.g. \citealt{pdg16} and references therein), as this has a negligible impact on our results. We do, however, constrain the sum of neutrino masses in Section~\ref{numasslab}. The primordial helium abundance, $Y_{\rm p}$, is determined in a manner consistent with Big Bang Nucleosynthesis where it is a function of the effective number of neutrinos and baryon density (e.g.~see equation~1 in \citealt{sj13}).

\subsubsection{Astrophysical degrees of freedom}

We vary the astrophysical parameters as described in Section~\ref{astrospace}. Our prior on the shot noise ($0 < N_{\rm shot} < 2300~h^{-3}{\rm Mpc}^3$) is motivated by the multipole power spectrum analysis of \citet{beutler14} for BOSS CMASS-DR11 (where the full sample is considered as compared to our sub-samples; also see \citealt{baldauf13}), which obtained $N_{\rm shot} = 1080 \pm 620~h^{-3}{\rm Mpc}^3$ when imposing $k_{\rm max} = 0.15~h~{\rm Mpc}^{-1}$ for their fitting range. Our prior can therefore be seen as approximately encompassing the $2\sigma$ upper bound of the \citet{beutler14} constraint. We also consider an extended scenario in which $0 < N_{\rm shot} < 3000~h^{-3}{\rm Mpc}^3$ that is approximately encompassing the $3\sigma$ upper bound. We do not allow for even wider priors on the shot noise to avoid biasing our cosmological parameter constraints given the inability of our data to constrain this degree of freedom, and the asymmetry created by the floor at $N_{\rm shot} = 0$. We discuss the impact of the shot noise on our cosmological parameter constraints in Section~\ref{xipmgtpllab}.

We explored the possibility of reducing the number of free parameters associated with the galaxy-galaxy lensing and multipole power spectrum measurements (e.g.~setting $b_{\rm 2dFLOZ} = b_{\rm LOWZ}$, $b_{\rm 2dFHIZ} = b_{\rm CMASS}$, $\sigma_{\rm v, \rm 2dFLOZ} = \sigma_{\rm v, LOWZ}$, $\sigma_{\rm v, \rm 2dFHIZ} = \sigma_{\rm v, CMASS}$, $N_{\rm 2dFLOZ} = N_{\rm LOWZ}$, or $N_{\rm 2dFHIZ} = N_{\rm CMASS}$), but found that even a subset of these simplifications would non-negligibly shift our cosmology parameter constraints. Instead, we take the conservative approach and vary all of these 12 parameters simultaneously in our MCMC runs (together with the cosmological and other astrophysical parameters).

\begin{table}
\begin{center}
\caption{Exploring changes in $\chi^2_{\rm eff}$ and DIC from extensions to the standard cosmological model. The vector $\{\xi_{\pm}, \gamma_{\rm t}, P_{0/2}\}$ is denoted `WL/RSD', and includes the full covariance between the observables (detailed in Section~\ref{covlab}). We consider fiducial and conservative cuts to the data (detailed in Table~\ref{tabcuts}). Given the difficulty to capture the impact of modified gravity on nonlinear scales, we moreover consider a `large scale' case (our most conservative case), where effectively only linear scales are included in the analysis (Table~\ref{tabcuts}). The reference $\Lambda$CDM model with fiducial treatment of the astrophysical systematics gives $\chi^2_{\rm eff} = 206.9$ and $\rm{DIC} = 236.1$ (for WL/RSD). Using conservative data cuts, $\{\chi^2_{\rm eff}, {\rm DIC}\} = \{184.3, 207.4\}$. We do not include results from the joint analysis of WL/RSD and Planck unless there is concordance between the datasets, as quantified by $\log \mathcal{I} > 0$ in Table~\ref{tabs8logi}. We also do not include the `Planck-only' results here, as they can be found in \citet{joudaki17}.
Negative values indicate preference in favor of the extended model compared to fiducial $\Lambda$CDM.
}
\begin{tabular}{p{4.3cm}>{\raggedleft}p{1.3cm}>{\raggedleft\arraybackslash}p{1.4cm}}
\toprule
Model & $\Delta\chi^2_{\rm eff}$ & $\Delta{\rm DIC}$\\
\midrule
$\Lambda$CDM (extended systematics) &  & \\
{~~~--~WL/RSD} & $-2.4$ & $1.3$\\
{~~~--~WL/RSD (conserv)} & $-2.2$ & $3.8$\\
Neutrino mass &  & \\
{~~~--~WL/RSD} & $0.49$ & $0.83$\\
{~~~--~WL/RSD (conserv)} & $0.79$ & $0.58$\\
Curvature &  & \\
{~~~--~WL/RSD} & $0.60$ & $-0.22$\\
{~~~--~WL/RSD (conserv)} & $-0.98$ & $0.85$\\
Dark energy (constant $w$) &  & \\
{~~~--~WL/RSD} & $1.4$ & $2.6$\\
{~~~--~WL/RSD (conserv)} & $0.48$ & $0.93$\\
Dark energy ($w_0-w_a$) &  & \\
{~~~--~WL/RSD} & $-0.35$ & $5.5$\\
{~~~--~WL/RSD (conserv)} & $-0.16$ & $2.0$\\
Modified gravity &  & \\
{~~~--~WL/RSD} & $-3.3$ & $0.78$\\
{~~~--~WL/RSD (conserv)} & $-1.7$ & $1.6$\\
{~~~--~WL/RSD (large scales)} & $0.22$ & $9.3$\\
{~~~--~WL/RSD (large scales)+Planck} & $-5.9$ & $2.1$\\
\bottomrule
\end{tabular}
\label{tabchidic}
\end{center}
\end{table}

\subsubsection{Convergence criterion}

In determining the convergence of our MCMC chains, we follow previous work (e.g.~\citealt{Hildebrandt16,joudaki17}) in using the \citet{Gelman92} $R$ statistic, where $R$ is the variance of chain means divided by the mean of chain variances. We enforce $(R - 1) < 2 \times 10^{-2}$, and stop the runs after further probing of the distribution tails.

\section{Combined probes in $\Lambda$CDM}
\label{Results}

We now present the cosmological and astrophysical parameter constraints from a systematic combination of the observables. We do not present results for galaxy-galaxy lensing and multipole power spectrum measurements on their own ($\gamma_{\rm t}$-only and $P_{0/2}$-only) given their weak cosmological parameter constraints, which arise from degeneracies with the large number of astrophysical degrees of freedom that are simultaneously varied in the analysis. However, we have confirmed that the different observables are concordant before combining them (via full MCMC runs). As a result, we keep the cosmic shear measurements as the base in our combinations. Unless specified, all two-sided bounds are at the 68\% confidence level (CL), and one-sided bounds at 95\% CL.

\begin{table}
\begin{center}
\caption{
Assessing the concordance between Planck-2015 (TT+lowP) and $\{\xi_{\pm}, \gamma_{\rm t}, P_{0/2}\}$, for KiDS overlapping with 2dFLenS and BOSS, as quantified by $\log \mathcal{I}$ and $T(S_8)$ defined in equations~(\ref{eqnlogi1}) and (\ref{eqnts8}), respectively.
}
\begin{tabular}{p{4.7cm}>{\raggedleft}p{1.1cm}>{\raggedleft\arraybackslash}p{1.2cm}}
\toprule
Model & $T(S_8)$ & $\log \mathcal{I}$\\
\midrule
$\Lambda$CDM &  & \\
{~~---~~fiducial systematics} & $2.6\sigma$ & $-3.1$\\
{~~---~~fiducial systematics (conserv)} & $3.0\sigma$ & $-1.3$\\
{~~---~~fiducial systematics (large scales)} & $3.0\sigma$ & $-1.9$\\
{~~---~~extended systematics} & $2.4\sigma$ & --\\
{~~---~~extended systematics (conserv)} & $2.9\sigma$ & --\\
Neutrino mass & $2.6\sigma$  & $-2.8$ \\
{~~---~~conservative} & $3.1\sigma$ & $-1.1$\\
Curvature & $3.8\sigma$ & $-4.8$ \\
{~~---~~conservative} & $4.0\sigma$ & $-2.5$\\
Dark energy (constant $w$) & $1.6\sigma$ & $-3.0$ \\
{~~---~~conservative} & $1.7\sigma$ & $-1.2$\\
Dark energy ($w_0-w_a$) & $1.7\sigma$ & $-2.1$ \\
{~~---~~conservative} & $1.9\sigma$ & $-0.28$\\
Modified gravity & $1.2\sigma$ & $-0.50$ \\
{~~---~~conservative} & $1.7\sigma$ & $-0.47$\\
{~~---~~large scales} & $0.037\sigma$ & $0.87$\\
\bottomrule
\end{tabular}
\label{tabs8logi}
\end{center}
\end{table}

\begin{figure*}
\hspace{-2.2em}
\vspace{-0.2em}
\resizebox{9.3cm}{!}
{\includegraphics{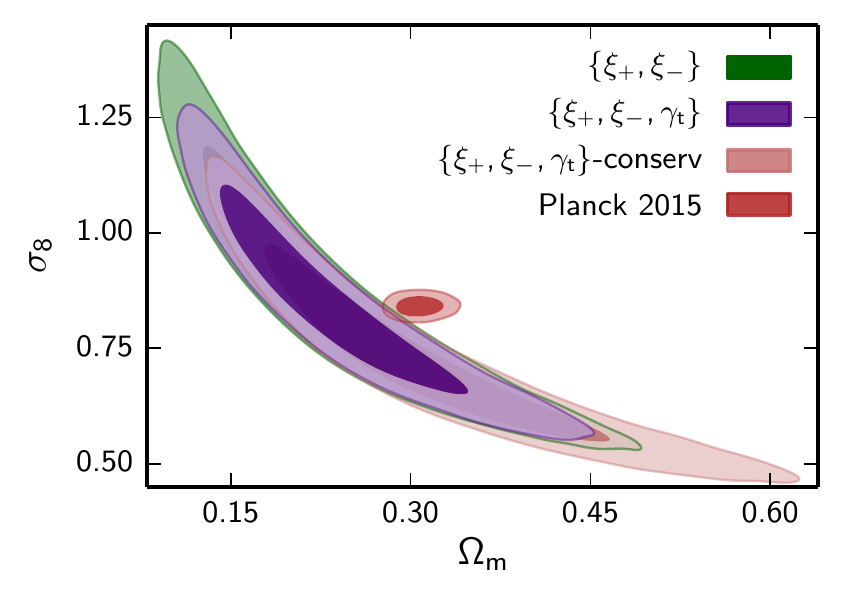}}
\resizebox{8.0cm}{!}
{\includegraphics{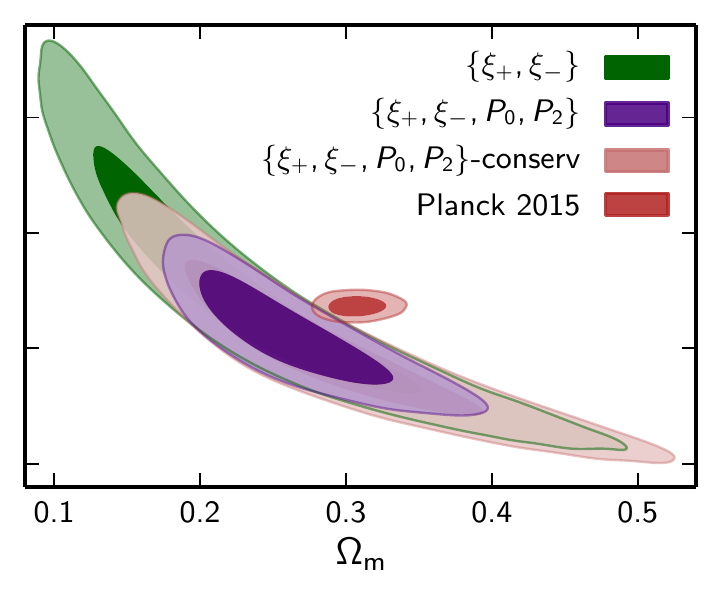}}
\vspace{-1.5em}
\caption{Left: Marginalized posterior contours in the $\sigma_8$ -- $\Omega_{\mathrm m}$ plane (inner 68\%~CL, outer 95\%~CL) from $\{\xi_+, \xi_-\}$ in green, $\{\xi_+, \xi_-, \gamma_{\rm t}\}$ in purple, and $\{\xi_+, \xi_-, \gamma_{\rm t}\}$ with conservative cuts to the data in pink. For comparison, we show the constraints from Planck 2015 CMB temperature measurements in red. Right: Same as left panel, but with $\{\xi_+, \xi_-, P_0, P_2\}$ instead of $\{\xi_+, \xi_-, \gamma_{\rm t}\}$.
}
\label{figxipmgtxipmpl}
\end{figure*}

\subsection{Cosmic shear $\{\xi_+, \xi_-\}$ - differences with past analyses}
\label{csalone}

To be self-consistent, our use of an $N$-body simulated covariance for all of the observables implies our cosmic shear constraints differ marginally from the reported constraints in \citet{Hildebrandt16} and \citet{joudaki17} given their use of an analytic covariance. Our cosmic shear constraints further differ from those in \citet{Hildebrandt16} and \citet{joudaki17} due to a wider prior on the baryonic feedback amplitude ($1<B<4$ as compared to $2<B<4$ previously) to better account for the `cosmo-OWL' simulations \citep{lebrun14}, which more strongly affect the matter power spectrum than the original OWL simulations. As shown in Table~\ref{tabpri}, our priors on the baryon density and Hubble constant follow \citet{joudaki17}, which are more conservative than the priors on these parameters in \citet{Hildebrandt16}. 

Despite these slightly different approaches between the different analyses, the cosmological constraints agree well. In particular, our cosmic shear constraint on $S_8 = \sigma_8 \sqrt{\Omega_{\mathrm m}/0.3} = 0.738^{+0.042}_{-0.046}$, which reflects a $T(S_8) = 2.2\sigma$ discordance with Planck-2015 (also see Table~\ref{tabbf}). For comparison, $S_8 = 0.745^{+0.038}_{-0.038}$ in \citet{Hildebrandt16} which reflects a $T(S_8) = 2.3\sigma$ discordance with Planck, and $S_8 = 0.752^{+0.040}_{-0.039}$ in \citet{joudaki17} which reflects a $T(S_8) = 2.1\sigma$ discordance.

However, there is one noticeable difference in our results relative to previous analyses in that the reduced $\chi^2$ is lower when employing the $N$-body simulated covariance compared to the analytic covariance (as noted in \citealt{Hildebrandt16}; arguably due to the slightly larger uncertainties predicted by the numerical covariance for the angular scales that carry the most information). As further discussed in Section~\ref{gfsec}, 
$\chi^2_{\rm red} = 1.17$ in our analysis compared to $\chi^2_{\rm red} = 1.33$ in \citet{Hildebrandt16} and \citet{joudaki17}. We note that our $\chi^2_{\rm red}$ is even closer to unity when using the \citet{hartlap07} correction to the inverse of the numerically estimated covariance instead of the \citet{sh15} correction to the likelihood (which breaks its Gaussianity). 

\begin{figure}
\resizebox{8.55cm}{!}
{\includegraphics{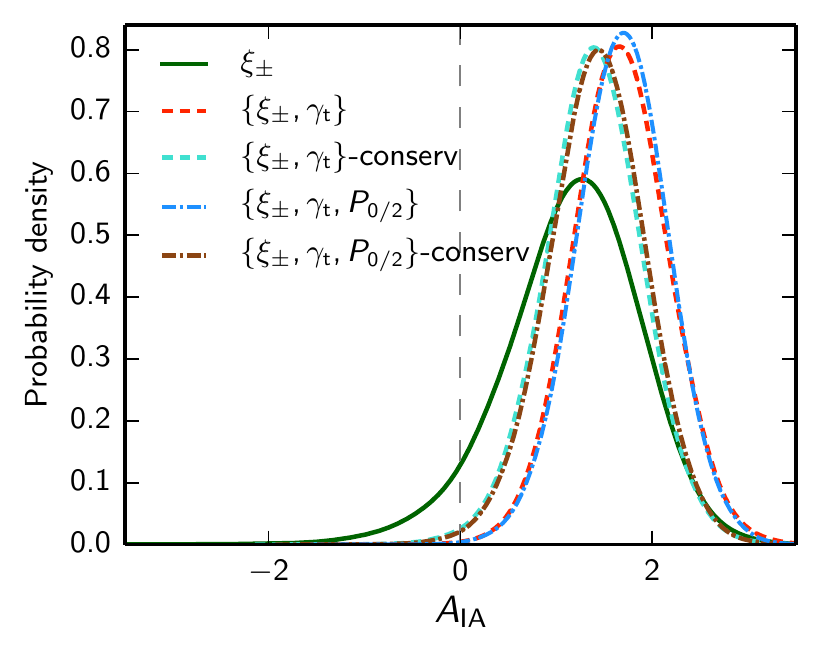}}
\vspace{-2.2em}
\caption{Marginalized posterior distributions for the intrinsic alignment amplitude from $\{\xi_+, \xi_-\}$ in solid green, $\{\xi_+, \xi_-, \gamma_{\rm t}\}$ in dashed red, $\{\xi_+, \xi_-, \gamma_{\rm t}\}$ with conservative cuts to the data in dashed cyan, $\{\xi_+, \xi_-, \gamma_{\rm t}, P_0, P_2\}$ in dot-dashed blue, and $\{\xi_+, \xi_-, \gamma_{\rm t}, P_0, P_2\}$ with conservative data cuts in dot-dashed brown. 
}
\label{figampia}
\end{figure}

\subsection{Cosmic shear and galaxy-galaxy lensing $\{\xi_+, \xi_-, \gamma_{\rm t}\}$}
\label{xipmgtlab}

\subsubsection{Cosmological constraints}

When combining cosmic shear and galaxy-galaxy lensing measurements, we find only a marginal improvement in the cosmological parameter constraints relative to cosmic shear alone (Fig~\ref{figxipmgtxipmpl}). This is true for both fiducial and conservative cuts to the $\gamma_{\rm t}$ measurements, which agree well with each other and with the constraints restricted to the $\xi_{\pm}$ measurements. We constrain $S_8 = 0.731^{+0.037}_{-0.042}$, which reflects a discordance with Planck at the level of $T(S_8) = 2.6\sigma$. Employing conservative $\gamma_{\rm t}$ cuts, $S_8 = 0.715^{+0.037}_{-0.042}$ and $T(S_8) = 2.9\sigma$. In other words, including the $\gamma_{\rm t}$ measurements improves the $S_8$ constraint by approximately $10\%$ compared to cosmic shear on its own, and increases the discordance with Planck by $0.4\sigma$ and $0.7\sigma$ for the fiducial and conservative $\gamma_{\rm t}$ scenarios, respectively.

An important reason for the marginal improvement in the parameter constraints is the strong degeneracy between the cosmological parameters and the galaxy bias (which modulates the amplitude of the $\gamma_{\rm t}$ measurements) of each of the four samples (i.e.~$b_{\rm 2dFLOZ}, b_{\rm 2dFHIZ}, b_{\rm LOWZ}, b_{\rm CMASS}$, discussed in Section~\ref{xipmgtpllab}). In Appendix~\ref{subgalbias}, we illustrate the significant improvement in the cosmological parameter constraints when fixing the galaxy bias of the different samples to their best-fit values.

\subsubsection{Astrophysical constraints}

Considering the $\xi_{\pm}$ and $\{\xi_{\pm}, \gamma_{\rm t}\}$ data vectors, we show marginalized posterior distributions for the IA amplitude in Fig.~\ref{figampia}. For cosmic shear alone, $A_{\rm IA} = 1.16^{+0.77}_{-0.60}$ (in agreement with $A_{\rm IA} = 1.15^{+0.71}_{-0.59}$ in \citealt{joudaki17}). Given the additional information from the `gI' piece of $\gamma_{\rm t}$, we find a 30\% improvement in the constraint on the IA amplitude, such that $A_{\rm IA} = 1.67^{+0.50}_{-0.49}$ for fiducial cuts to $\gamma_{\rm t}$ and $A_{\rm IA} = 1.39^{+0.50}_{-0.50}$ with conservative cuts (positive at $3.3\sigma$ and $2.7\sigma$, respectively). As in the `cosmic shear only' scenario, the baryonic feedback amplitude is unconstrained within its prior range.

\begin{figure*}
\hspace{-1.0em}
\resizebox{8.8cm}{!}{\includegraphics{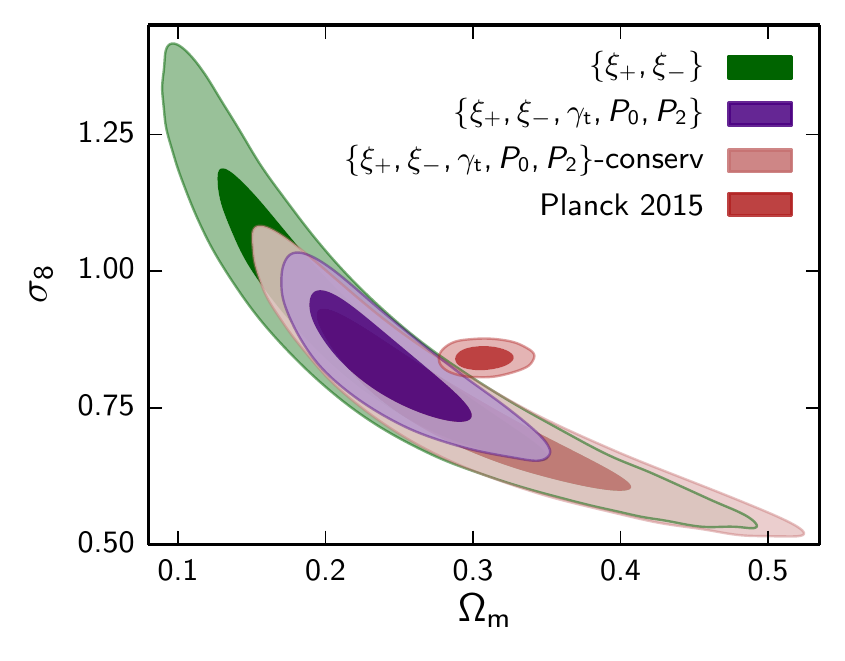}}
\resizebox{8.9cm}{!}{\includegraphics{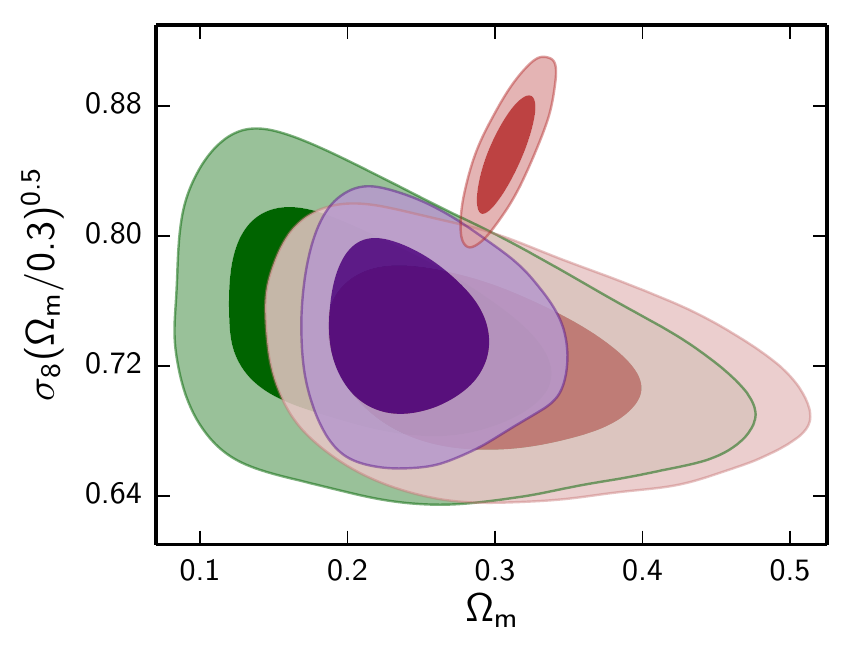}}
\vspace{-1.5em}
\caption{Marginalized posterior contours in the $\sigma_8$ -- $\Omega_{\mathrm m}$ plane (inner 68\%~CL, outer 95\%~CL) from observations of cosmic shear, galaxy-galaxy lensing, and redshift-space multipole power spectra for KiDS overlapping with 2dFLenS and BOSS. We show constraints from $\{\xi_+, \xi_-\}$ in green, $\{\xi_+, \xi_-, \gamma_{\rm t}, P_0, P_2\}$ in purple, and $\{\xi_+, \xi_-, \gamma_{\rm t}, P_0, P_2\}$ with conservative data cuts in pink. For comparison, we show the constraints from Planck 2015 in red.
}
\label{figxipmgtpl}
\end{figure*}

\subsection{Cosmic shear and redshift-space galaxy clustering $\{\xi_+, \xi_-, P_0, P_2\}$}
\label{xipmpllab}

\subsubsection{Cosmological constraints}

In combining cosmic shear and multipole power spectrum measurements, we find a noticeable improvement in the cosmological constraints along the lensing degeneracy direction (Fig~\ref{figxipmgtxipmpl}), in particular when employing fiducial cuts to the $\{P_0, P_2\}$ measurements (in spite of varying 12 additional nuisance parameters; by factors of 2.3 in $\sigma_8$ and 1.8 in $\Omega_{\rm m}$ relative to $\xi_{\pm}$). The cosmological constraints for the fiducial and conservative cases agree with each other and with cosmic shear alone (with a seeming preference for its high-$\Omega_{\rm m}$ tail). While we have chosen wide priors on the nuisance parameters, our constraints along the lensing degeneracy direction retain a dependence on the shot noise prior as further discussed in Sec.~\ref{snsec}.

Perpendicular to the lensing degeneracy direction, we measure $S_8 = 0.722^{+0.038}_{-0.037}$, which corresponds to a 15\% improvement in the constraint compared to cosmic shear alone, and reflects a discordance with Planck at the level of $T(S_8) = 2.9\sigma$. When considering conservative $P_{0/2}$ cuts, $S_8 = 0.717^{+0.039}_{-0.039}$ (roughly 10\% improvement) and $T(S_8) = 2.9\sigma$. As a result, regardless of whether the cosmic shear measurements are combined with galaxy-galaxy lensing or multipole power spectrum measurements, and regardless of the different cuts to the measurements, the discordance with Planck increases.

\subsubsection{Astrophysical constraints}

The multipole power spectra do not particularly improve the IA amplitude constraint compared to cosmic shear alone. However, they constrain the galaxy biases more strongly than galaxy-galaxy lensing (further discussed in Section~\ref{xipmgtpllab}). The baryonic feedback and shot noise parameters are unconstrained within their prior ranges. For fiducial cuts to the $P_{0/2}$ measurements, the 2dFLenS velocity dispersion parameters are bounded from above, such that $\{{\sigma_{\rm v, 2dFLOZ}, \sigma_{\rm v, 2dFHIZ}}\} < \{{5.6, 5.7}\}~h^{-1}{\rm Mpc}$. For BOSS, the bounds are two-sided: $\{{\sigma_{\rm v, LOWZ}, \sigma_{\rm v, CMASS}}\} = \{{3.4^{+1.4}_{-0.8}, 5.5^{+1.1}_{-0.8}}\}~h^{-1}{\rm Mpc}$. For conservative cuts, the velocity dispersion parameters are unconstrained within their prior ranges, with the exception of $\sigma_{\rm v, CMASS} < 7.6~h^{-1}{\rm Mpc}$. Our CMASS constraints agree with those given for the full survey in \citet{beutler14}. The constraints can be converted to units of ${\rm km~s^{-1}}$ by multiplying with the Hubble constant, and correspond to velocities of hundreds of ${\rm km~s^{-1}}$~as expected.

\subsection{Cosmic shear, galaxy-galaxy lensing, and redshift-space galaxy clustering $\{\xi_+, \xi_-, \gamma_{\rm t}, P_0, P_2\}$}
\label{xipmgtpllab}

\subsubsection{Cosmological constraints}

We show the key cosmological parameter constraints in the $\sigma_8$ -- $\Omega_{\rm m}$ plane in Fig.~\ref{figxipmgtpl}. Analogous to the $\{\xi_{\pm}, P_{0/2}\}$ data combination, the \{high-$\sigma_8$, low-$\Omega_{\rm m}$\} end of the underlying cosmic shear contour is seemingly disfavored (following an improvement on $\sigma_8$ by $\{60, 40\}\%$ and on $\Omega_{\rm m}$ by $\{50, 10\}\%$ for \{fiducial, conservative\} data cuts\footnote{The real impact is larger given the dependence of the `cosmic shear only' results along the lensing degeneracy direction on the cosmological priors~\citep{joudaki16}.}). Perpendicular to the lensing degeneracy direction, there is a minor narrowing of the contours, reflected in $S_8 = 0.742^{+0.035}_{-0.035}$ for fiducial data cuts, and $S_8 = 0.721^{+0.036}_{-0.036}$ with conservative cuts. The $\{\xi_\pm, \gamma_{\rm t}, P_{0/2}\}$ constraints  on $S_8$ are 8-9\% stronger than the respective constraints from $\{\xi_\pm, P_{0/2}\}$, 9-13\% stronger than the constraints from $\{\xi_\pm, \gamma_{\rm t}\}$, and 19-22\% stronger than the constraint from $\xi_\pm$. These improvements are relatively modest due in part to the currently incomplete overlap of KiDS with 2dFLenS and BOSS, the careful selection of scales for $\gamma_{\rm t}$ and $P_{0/2}$, and the large number of nuisance parameters that are simultaneously varied in the analysis (19 parameters for $\{{\xi_\pm, \gamma_{\rm t}, P_{0/2}}\}$ compared to 7 parameters for cosmic shear alone).

\begin{figure*}
\vspace{-1.55em}
\includegraphics[width=0.90\hsize]{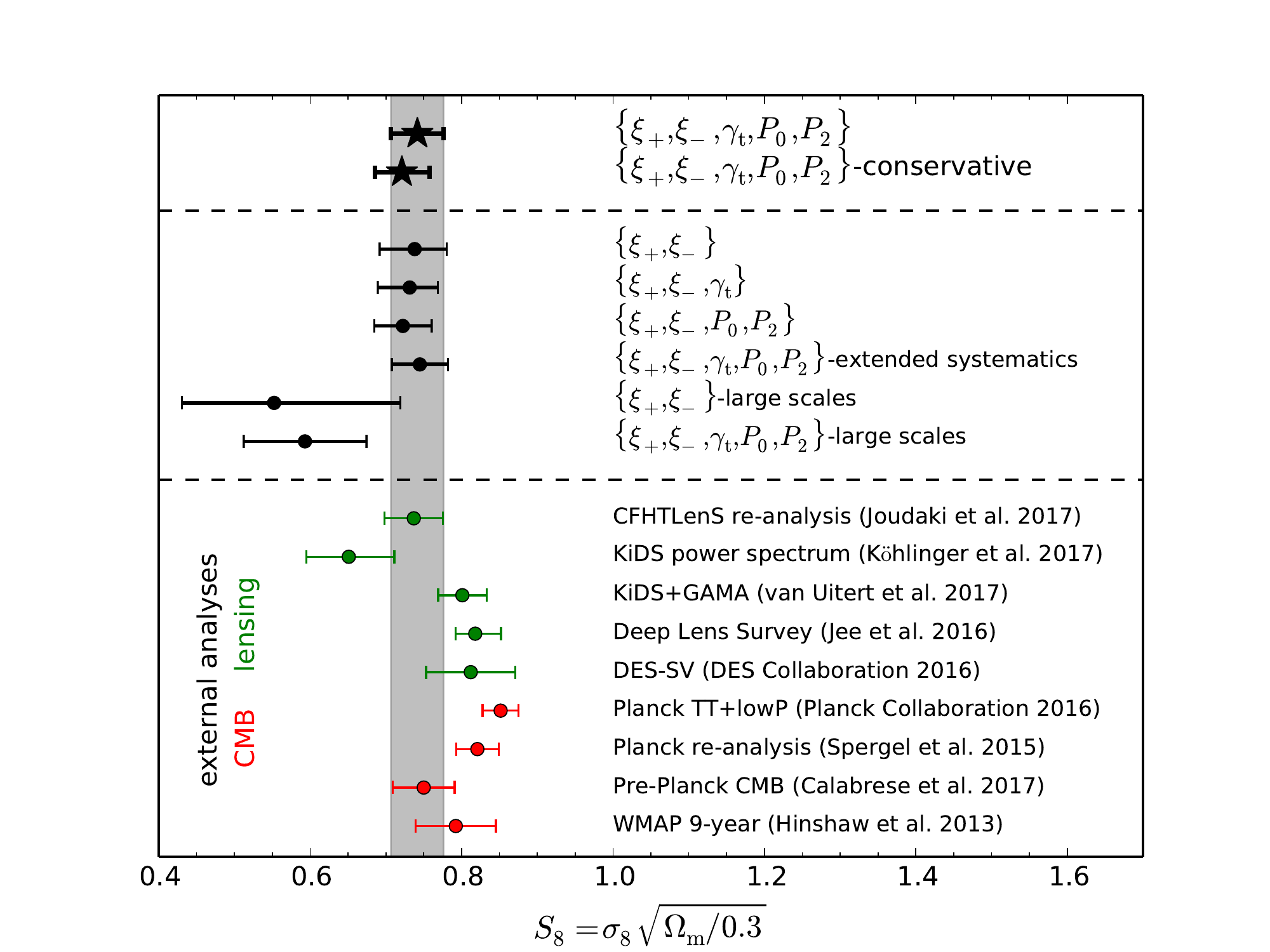}
\vspace{-0.2em}
\caption{Marginalized constraints on $S_8$ at 68\% CL from observations of cosmic shear, galaxy-galaxy lensing, and redshift-space multipole power spectra for KiDS overlapping with 2dFLenS and BOSS. We show the constraints from external CMB and lensing analyses for comparison. We have excluded the conservative sub-vector constraints on $S_8$ for clarity, but note that they are in agreement with the fiducial constraints. The large-scale $\xi_{\pm}$ constraint is from \citet{joudaki17}. Numerical constraints are provided in Table~\ref{tabbf}.
}
\label{figs8}
\end{figure*}

The fully combined fiducial and conservative $S_8$ constraints are in complete agreement relative to one another, and with the earlier sub-vector constraints (visualized in Fig~\ref{figs8}). However, the fully combined $S_8$ constraints are discordant with Planck at the level of $2.6\sigma$ and $3.0\sigma$ in the fiducial and conservative cases, respectively. In Appendices~\ref{tausec} and \ref{extsystlab}, we show that these discordances are largely unaffected by the new Planck HFI measurement of the reionization optical depth~\citep{plancktau} and by an extended treatment of the astrophysical systematics. We moreover evaluated the $\log \mathcal{I}$ diagnostic, which accounts for the discordance over the full parameter space. As shown in Table~\ref{tabs8logi}, $\log \mathcal{I} = -3.1$ for fiducial cuts to the data, which indicates `decisive' discordance with Planck, and $\log \mathcal{I} = -1.3$ with conservative cuts indicating `strong' discordance. Hence, despite the similar level of discordance with Planck as quantified by $S_8$, the discordance between the probes is larger in the fiducial scenario given the stronger constraints on the underlying parameter space (as can be seen in Fig~\ref{figxipmgtpl}).

\subsubsection{Shot noise prior dependence} 
\label{snsec}

The constraints are subject to an important caveat predominantly along the lensing degeneracy direction. As discussed in Section~\ref{basesec}, our fiducial shot noise prior $0 < N_{\rm shot} < 2300~h^{-3}{\rm Mpc}^3$ is motivated by the analysis of \citet{beutler14} for BOSS. While we expect $N_{\rm shot}$ on the order of 1000, our data is unable to constrain the shot noise on its own, and our results along the lensing degeneracy direction are sensitive to the choice of prior on this parameter (to lesser extent when employing conservative data cuts). Given the anti-correlation between $N_{\rm shot}$ and $\Omega_{\rm m}$, a lower bound on the shot noise prior shifts the constraints along the lensing degeneracy direction towards larger matter density (and smaller $\sigma_8$), while a higher upper bound shifts the constraints toward smaller matter density (and larger $\sigma_8$).

The prior dependence of the cosmological constraints along the lensing degeneracy direction was illustrated for cosmic shear alone in \citet{joudaki16}. We now further advise caution in the interpretation of cosmological constraints along the lensing degeneracy direction when including multipole power spectrum measurements. This includes the $\log \mathcal{I}$ diagnostic which is sensitive to the full parameter space. Meanwhile, the constraints on $S_8$ are more robust, as illustrated by the weak correlation of $S_8$ with the shot noise of each galaxy sample in Fig~\ref{figsubmore}. We expect the constraints along the lensing degeneracy direction to become increasingly insensitive to the shot noise prior as the overlap increases for forthcoming data releases of KiDS.

\subsubsection{Intrinsic alignment amplitude}

Our constraints on the IA amplitude improve only marginally compared to $\{\xi_{\pm}, \gamma_{\rm t}\}$ as expected (Fig.~\ref{figampia}). Given fiducial data cuts, $A_{\rm IA} = 1.69^{+0.48}_{-0.48}$
which reflects a $3.5\sigma$ preference for being positive (an additional $0.2\sigma$), and with conservative cuts $A_{\rm IA} = 1.42^{+0.50}_{-0.50}$ which reflects a $2.8\sigma$ preference (an additional $0.1\sigma$). As discussed in Section~\ref{Resultsextended} (and shown in \citealt{joudaki17} for cosmic shear alone), the IA amplitude constraints are largely robust to the underlying cosmological model. However, we note that the statistically significant deviation of the IA amplitude from zero could partly be a reflection of unaccounted systematics (e.g.~in the photometric redshifts, see Appendix A in \citealt{joudaki17}). In Appendix~B, we further present the extended constraints on the IA redshift dependence $\eta_{\rm IA}$, finding that it is consistent with zero following a 60\% improvement in the bound relative to cosmic shear alone.

\subsubsection{Galaxy bias: 2dFLenS and BOSS}

We show our constraints on the galaxy bias from the analysis of the $\{\xi_{\pm}, \gamma_{\rm t}\}$ and $\{\xi_{\pm}, \gamma_{\rm t}, P_{0/2}\}$ data combinations in Fig.~\ref{figbias}, noting an agreement between their respective constraints (albeit with minor tendencies of tension in the 2dFHIZ bias). We find that the galaxy biases of the different samples all peak around $b \sim 2$ as expected (see e.g.~\citealt{beutler14, gm16, blake16}), and in agreement between the fiducial and conservative data analyses. The constraints from $\{\xi_{\pm}, P_{0/2}\}$ are moreover similar to those from $\{\xi_{\pm}, \gamma_{\rm t}, P_{0/2}\}$. Next, we therefore only quote the galaxy bias constraints from the fully joint analysis.

Beginning with 2dFLenS, we constrain $b_{\rm 2dFLOZ} = 1.75^{+0.17}_{-0.27}$ for fiducial data cuts, and $b_{\rm 2dFLOZ} = 1.89^{+0.27}_{-0.37}$ with conservative cuts. Beyond the factor of 1.5 difference in constraining power between the two cases, these constraints are factors of 2.5 and 1.9 stronger than the respective constraints from $\{\xi_{\pm}, \gamma_{\rm t}\}$. We moreover constrain $b_{\rm 2dFHIZ} = \{1.91^{+0.16}_{-0.30}, 2.02^{+0.25}_{-0.36}\}$ in the \{fiducial, conservative\} analyses. These constraints are comparable to those for 2dFLOZ, and stronger by factors of 2.6 and 2.3 relative to the respective constraints from $\{\xi_{\pm}, \gamma_{\rm t}\}$.

Turning to BOSS, we constrain $b_{\rm LOWZ} = 2.03^{+0.17}_{-0.31}$ for fiducial data cuts, and $b_{\rm LOWZ} = 2.15^{+0.29}_{-0.41}$ with conservative cuts, corresponding to a factor of 1.5 difference in their relative constraining powers, and factors of 2.4 and 1.7 improvement relative to the respective constraints from $\{\xi_{\pm}, \gamma_{\rm t}\}$. We moreover constrain $b_{\rm CMASS} = \{1.85^{+0.16}_{-0.29}, 1.95^{+0.25}_{-0.34}\}$ in the \{fiducial, conservative\} analyses. These constraints are stronger by factors of 2.3 and 2.1 relative to the respective constraints from $\{\xi_{\pm}, \gamma_{\rm t}\}$, and comparable to those for LOWZ.

\subsubsection{Additional nuisance parameters: baryonic feedback, velocity dispersion}

The baryonic feedback amplitude is unconstrained within its prior range in our conservative scenario (same as in the $\{\xi_{\pm}, \gamma_{\rm t}\}$ and $\{\xi_{\pm}, \gamma_{\rm t}, P_{0/2}\}$ cases), but constrained from above when employing fiducial cuts to the data, such that $B < 3.3$ (95\% CL). This constraint on the feedback amplitude can be compared to the `dark matter only' scenario at $B \simeq 3.1$ \citep{Mead15}. The marginalized posterior peaks at $B = 1.6$ (in the conservative scenario at $B = 2.0$), suggesting a preference for strong baryonic feedback (for comparison, the `DBLIM' and `AGN' scenarios in the OWL simulations give $B$ of 2.4 and 2.0, respectively). Meanwhile, the constraints on the pairwise velocity dispersion closely resemble the constraints given for $\{\xi_{\pm}, P_{0/2}\}$. In Fig.~\ref{figsubmore}, we show the marginalized posterior distributions of all of our astrophysical parameters, and their relative correlations.

\begin{figure*}
\resizebox{7.9cm}{!}
{\includegraphics{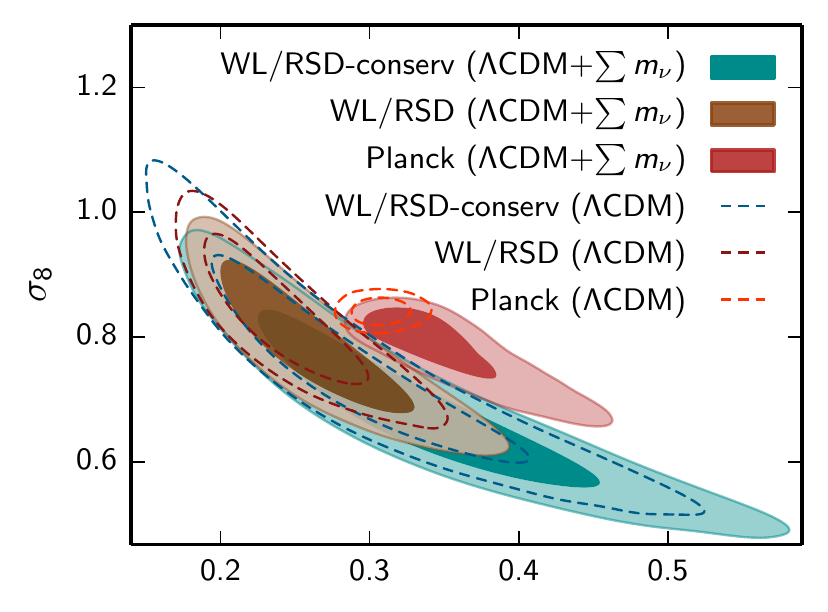}}
\resizebox{7.9cm}{!}
{\includegraphics{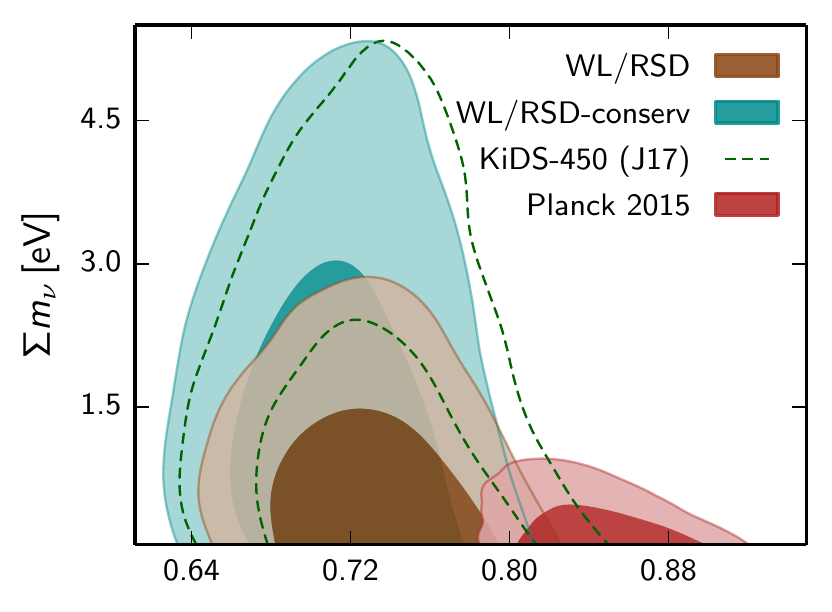}}
\resizebox{7.8cm}{!}
{\includegraphics{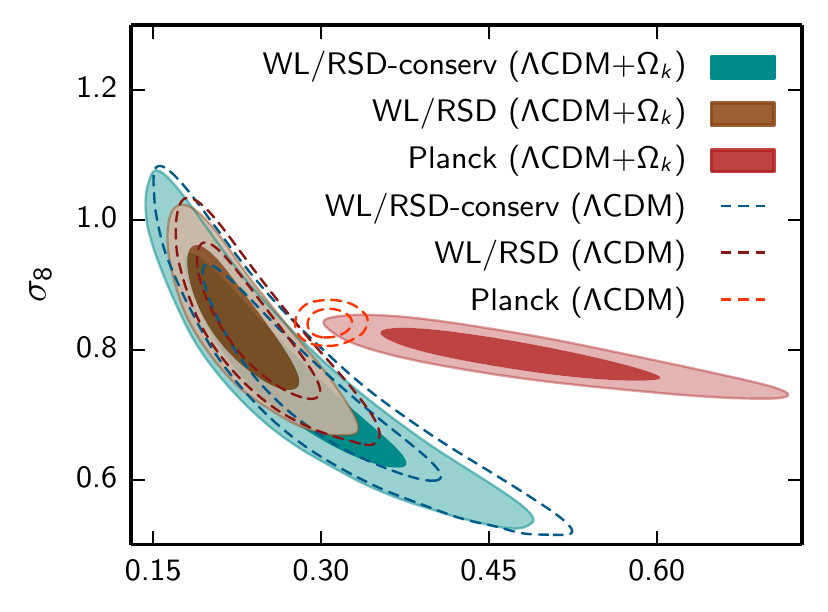}}
\resizebox{8.1cm}{!}
{\includegraphics{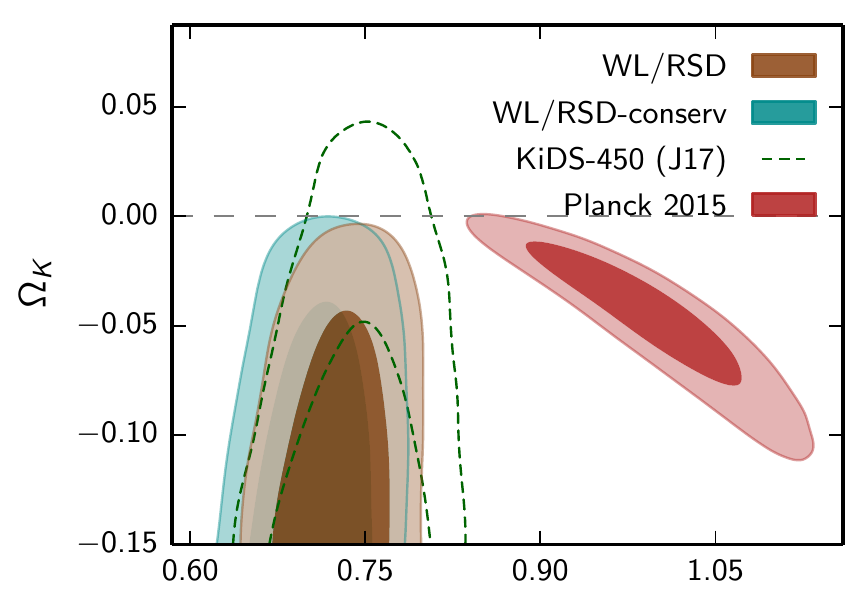}}
\resizebox{8.0cm}{!}{\includegraphics{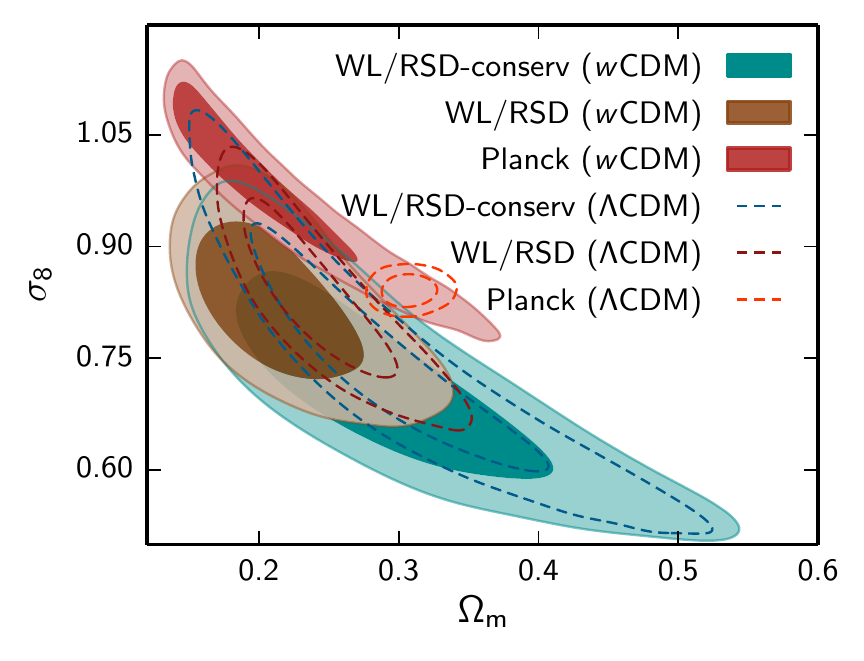}}
\resizebox{8.0cm}{!}{\includegraphics{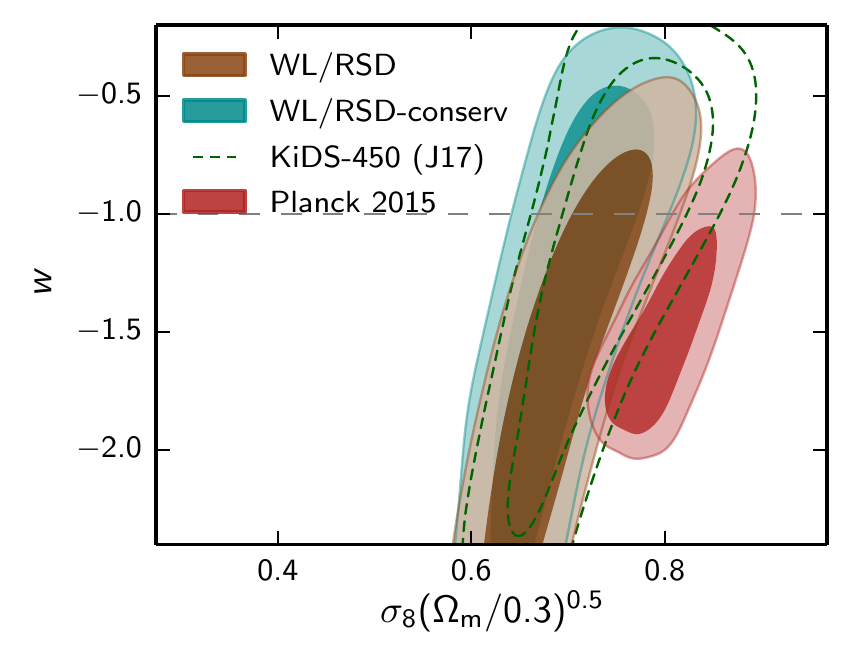}}
\vspace{-1.3em}
\caption{Left: Marginalized posterior contours in the $\sigma_8$ -- $\Omega_{\mathrm m}$ plane (inner 68\%~CL, outer 95\%~CL) in a universe with massive neutrinos, curvature, and evolving dark energy (constant $w$). We show the constraints from $\{\xi_{\pm}, \gamma_{\rm t}, P_{0/2}\}$ in brown, with conservative cuts to the data in cyan, and Planck CMB temperature in red. For comparison, we also show dashed contours assuming fiducial $\Lambda$CDM. Right: Marginalized posterior contours in the plane with $S_8$ for $\sum m_{\nu}$, $\Omega_k$, and $w$, respectively. For comparison, we also show the contour when restricted to $\xi_{\pm}$ in dashed green (from \citealt{joudaki17}, with narrower prior on the baryon feedback and use of analytic covariance).
}
\label{figxipmgtplnumassetc}
\end{figure*}

\subsection{Combined probes: effectively linear scales} 
\label{largescaleslab}

Given the sensitivity of weak gravitational lensing to highly nonlinear scales in the matter power spectrum, we also considered a `large scales' case with effectively linear cuts to the measurements. As detailed in Table~\ref{tabcuts}, we only retain the measurements at $\theta = \{24.9, 50.7\}$~arcmin in $\xi^{ij}_+(\theta)$ and $\theta = 210$~arcmin in $\xi^{ij}_{-}(\theta)$. We further restrict the measurements to $\theta = \{50.7, 103\}$~arcmin in $\gamma_{\rm t}^{j}(\theta)$ and $k = 0.075~h~{\rm Mpc}^{-1}$ in $\{P_0, P_2\}$. 

For cosmic shear alone, these cuts widen the uncertainty and lower the constraint on $S_8 = 0.55^{+0.17}_{-0.12}$ \citep{joudaki17}, increasing the agreement with the KiDS power spectrum analysis in \citet{kohlinger17}. In our fully combined analysis, we obtain a similarly shifted constraint but with nearly a factor of two smaller uncertainty, as $S_8 = 0.59^{+0.08}_{-0.08}$. While the large-scale constraint on $S_8$ seems to suggest a potential discordance with the fiducial and conservative constraints, in Section~\ref{modgrav} we show that the different cases are in agreement in the $\sigma_8$ -- $\Omega_{\rm m}$ plane (with somewhat different tilts in the contours that affect the $S_8$ constraints). The large-scale $S_8$ constraint maintains the discordance with Planck at the level of $T(S_8) = 3.0\sigma$, supported by $\log \mathcal{I} = -1.9$ over the full parameter space. 

As we discard a wealth of small-scale information, the constraint on the IA amplitude degrades by more than a factor of two compared to the fiducial and conservative cases, such that $A_{\rm IA} = 1.39^{+1.10}_{-0.98}$. This large-scale constraint on the IA amplitude is however a factor of two stronger than the corresponding constraint from cosmic shear alone. As the galaxy biases are most strongly determined by the multipole power spectra, the constraints are comparable in the large-scale and conservative cases given their identical cuts to $P_{0/2}$.

\subsection{Goodness of fit of the different combinations of probes}
\label{gfsec}

In the $\{\xi_{\pm}, \gamma_{\rm t}\}$ analysis, $\chi^2_{\rm red} = 1.09$ for both fiducial and conservative cuts to $\gamma_{\rm t}$ (shown in Table~\ref{tabbf}). This reflects a slight improvement in the goodness of fit compared to cosmic shear alone, where $\chi^2_{\rm red} = 1.17$. Given $\{\xi_{\pm}, P_{0/2}\}$, $\chi^2_{\rm red} = 1.17$ for fiducial cuts to $\{P_0, P_2\}$ and $\chi^2_{\rm red} = 1.22$ with conservative cuts. As a result, we do not obtain the same substantial improvement in the goodness of fit as when galaxy-galaxy lensing is included.

In the fully combined analysis, $\chi^2_{\rm red} = 1.09$ for fiducial data cuts, and $\chi^2_{\rm red} = 1.11$ with conservative cuts. Again, the inclusion of the multipole power spectra does not particularly modify $\chi^2_{\rm red}$. An extended treatment of the astrophysical systematics (Appendix~\ref{extsystlab}) improves the fit by $\Delta\chi^2_{\rm eff} \approx -2$ (compared to the fiducial systematics model, for both fiducial and conservative data cuts), but is not favored in a model selection sense as $\Delta{\rm DIC}$ is positive (see Table~\ref{tabchidic}). In the `large-scale' scenario, $\chi^2_{\rm red} = 1.31$ indicating a noticeably degraded fit to the data when effectively only linear scales are considered.

\section{Extended cosmological models}
\label{Resultsextended}

We now examine the constraints in extended cosmological models from the full $\{\xi_{\pm}, \gamma_{\rm t}, P_{0/2}\}$ data combination. The extended cosmologies include neutrino mass, curvature, evolving dark energy, and modified gravity. Beyond constraining the additional cosmological degrees of freedom, we assess the level of concordance with Planck, and determine the viability of the models relative to $\Lambda$CDM. The astrophysical constraints are generally robust to the underlying cosmology (for the model extensions considered), and we will therefore not quote them. The main exception to this is the IA amplitude, which we discuss when noticeably affected.

\subsection{Neutrino mass}
\label{numasslab}

\subsubsection{Background}

Massive neutrinos suppress the clustering of matter below the neutrino free-streaming scale (and thereby also suppress the weak lensing correlation functions on small angular scales, illustrated in e.g.~\citealt{joudaki17}). 
To model the impact of neutrino mass in the nonlinear matter power spectrum, we use \hmcode which is calibrated to the massive neutrino simulations of \citet*{massara14}. The ratio of power spectra (with and without neutrino mass) is calibrated to the few percent level for $k \leq 10~h~{\rm{Mpc}}^{-1}$, $z \leq 1$, and sum of neutrino masses $\sum m_{\nu} \leq 0.60~{\rm eV}$. The overall performance of \hmcode degrades for higher neutrino masses when compared to \cite{miratitan17}, but is adequate at the level of constraining power of our datasets. While \hmcode's accuracy is a minor improvement compared to other methods such as \halofit \citep{Smith03, Takahashi12, bird12}, its main benefit is being able to simultaneously account for the impact of baryons in the nonlinear matter power spectrum.

\subsubsection{Constraints on $S_8$ and discordance with Planck}

We show the impact of neutrino mass on the marginalized parameter constraints in the $\sigma_8$ -- $\Omega_{\rm m}$ plane in Fig.~\ref{figxipmgtplnumassetc}. Similar to the `cosmic shear only' analysis in \citet{joudaki17}, the $\{\xi_{\pm}, \gamma_{\rm t}, P_{0/2}\}$ and Planck contours move along the lensing degeneracy direction toward smaller values of $\sigma_8$ and larger matter density, while effectively preserving the constraints on $S_8$ (both at $0.3\sigma$ to $0.4\sigma$ level). The discordance between the combined probes and Planck is $T(S_8) = 2.6\sigma$ for fiducial cuts to the data and $T(S_8) = 3.1\sigma$ with conservative cuts, in agreement with the discordances in $\Lambda$CDM to within $0.1\sigma$. The discordance over the full parameter space is also comparable to that in $\Lambda$CDM (as quantified by $\log \mathcal{I}$ in Table~\ref{tabs8logi}).

\subsubsection{Constraints on neutrino mass and model selection}

In Fig.~\ref{figxipmgtplnumassetc}, we moreover show marginalized constraints in the $\sum m_{\nu}$ -- $S_8$ plane, where the sum of neutrino masses $\sum m_{\nu} < 2.2$~eV (95\% CL) for fiducial data cuts, and $\sum m_{\nu} < 4.0$~eV with conservative cuts. As expected, the constraint on $\sum m_{\nu}$ is stronger in the fiducial case as we retain measurements on smaller scales where the impact of neutrino mass is larger. Despite the fully combined analysis, the neutrino mass constraint in the conservative scenario is comparable to that from cosmic shear alone (where $\sum m_{\nu} < 4.0$~eV in \citealt{joudaki17}, albeit with narrower prior on the baryonic feedback as discussed in Section~\ref{csalone}).

Our neutrino mass constraints are not competitive with those obtained from other probes (e.g.~Planck, and its combination with BAOs, \citealt{planck15}), but we expect them to increasingly improve with future data releases of KiDS (and the resulting increase in spectroscopic overlap). As shown in Table~\ref{tabchidic}, the extended cosmology is not favored compared to $\Lambda$CDM, such that $\Delta {\rm DIC} \lesssim 1$ for both data cuts. We do not quote results from the joint analysis with Planck given the discordance between the datasets (aside from using the chains of the joint MCMC to obtain $\log \mathcal{I}$).

\subsection{Curvature}
\label{curvsec}

An increase in $\Omega_k$ decreases the weak lensing signal fairly uniformly across the angular scales probed by KiDS (as shown in \citealt{joudaki17}), while the multipole power spectra are integrals over the matter power spectrum (equation~\ref{eqn:p2}), which experiences a boost that becomes asymptotically less significant on smaller scales (and where the relative impact of the Alcock-Paczynski scaling factors grows). In Fig~\ref{figxipmgtplnumassetc}, we find that our constraints in the $\sigma_8$ -- $\Omega_{\rm m}$ plane are not significantly affected by $\Omega_k$ (in contrast to the significant widening of the contour given KiDS cosmic shear alone; \citealt{joudaki17}), with a small shift perpendicular to the lensing degeneracy direction away from Planck for both data cuts. Given the degeneracy between the curvature, matter density, and Hubble constant in the angular diameter distance to the last-scattering surface, the Planck CMB temperature constraints in this plane are highly degraded, with nearly horizontal elongation of the Planck contour along increasing $\Omega_{\rm m}$.

While our $S_8$ constraints are only affected between $0.3\sigma$ to $0.4\sigma$ for the two different data cuts, the Planck constraint on $S_8$ degrades by a factor of $2.5$ and the posterior mean increases by~$2.3\sigma$ (relative to the uncertainty in a flat universe). As a result, the tension with Planck increases to $T(S_8) = 3.8\sigma$ using fiducial data cuts, and $T(S_8) = 4.0\sigma$ with conservative cuts. Considering the full parameter space, $\log \mathcal{I} < -2$ in both cases indicating `decisive' discordance with Planck. These results can be contrasted with the discordance between Planck and KiDS cosmic shear alone, where $T(S_8) = 3.5\sigma$ and $\log \mathcal{I} = -1.7$ \citep{joudaki17}.

In Fig~\ref{figxipmgtplnumassetc}, we moreover show the constraints on $\Omega_k$ in the plane with $S_8$. Improving on the constraint from KiDS cosmic shear alone, $\Omega_k < -0.026$ (95\%~CL) for fiducial data cuts, and $\Omega_k < -0.022$ with conservative cuts. In comparison, Planck constrains $\Omega_k < -0.0048$, such that all datasets point towards positive curvature. From a model selection standpoint, however, our analysis shows no preference away from flatness (as $\Delta {\rm DIC} \simeq 0$ for both data cuts) while Planck weakly favors a curved universe ($\Delta {\rm DIC} = -4.3$; \citealt{joudaki17}). Given the discordance of our measurements with Planck, we do not quote their joint constraints.

\subsection{Dark energy (constant $w$)}
\label{constwlab}

\subsubsection{Background}

An increase in the dark energy equation of state provides a scale-dependent suppression of the matter power spectrum (relative to a cosmological constant), which in the case of the weak lensing correlation functions is counteracted by an increase in the lensing kernel (discussed in~\citealt{joudaki17}). We model the impact of evolving dark energy on the nonlinear matter power spectrum with \hmcode, which is calibrated to both constant and time-varying equations of state \citep{Mead16}.

In \citet{joudaki17}, we showed that evolving dark energy is able to alleviate the discordance between KiDS cosmic shear and Planck CMB temperature measurements, and between the direct Hubble constant measurement of \citet{riess16} and that inferred by Planck. In the joint analysis of KiDS and Planck, the constant equation of state $(w)$ model was found to be weakly favored compared to $\Lambda$CDM in a model selection sense, while it was moderately favored in the case of a time-varying equation of state ($w_0 - w_a$ parameterization). We examine to what extent these results change with our combined probes.

\subsubsection{Constraints on $S_8$ and discordance with Planck}

In the $\sigma_8$ -- $\Omega_{\rm m}$ plane (Fig.~\ref{figxipmgtplnumassetc}), we find a substantial widening of our marginalized posterior contours perpendicular to the lensing degeneracy direction (translated into roughly 50\% increase in the $S_8$ uncertainty relative to $\Lambda$CDM). Given the degeneracy between the equation of state and matter density in the angular diameter distance to the last-scattering surface, the Planck contour widens towards lower values of $\Omega_{\rm m}$ (and higher $\sigma_8$). As a result, the Planck and KiDS contours overlap, and the discordance between the datasets is alleviated (shown in \citealt{joudaki17}). However, the combined probes constrain the lensing degeneracy direction more strongly than KiDS on its own, such that the \{low $\Omega_{\rm m}$, high $\sigma_8$\} tail is no longer favored, and the overlap with Planck is no longer present. We note that this picture comes with a caveat given the sensitivity of our results along the lensing degeneracy direction to the shot noise prior (as discussed in Section~\ref{xipmgtpllab}).

For fiducial data cuts $T(S_8) = 1.6\sigma$, and for conservative cuts $T(S_8) = 1.7\sigma$. These results follow a roughly 15\% tightening and $0.8\sigma$ decrease in $S_8$ compared to KiDS cosmic shear alone, where $T(S_8) \lesssim 1\sigma$ \citep{joudaki17}. Considering the full parameter space, the $\log \mathcal{I}$ statistic is even less forgiving, indicating at least strong discordance between the combined probes and Planck, in contrast to substantial-to-strong concordance between KiDS and Planck alone.

\begin{figure*}
\begin{center}
\resizebox{8.6cm}{!}{\includegraphics{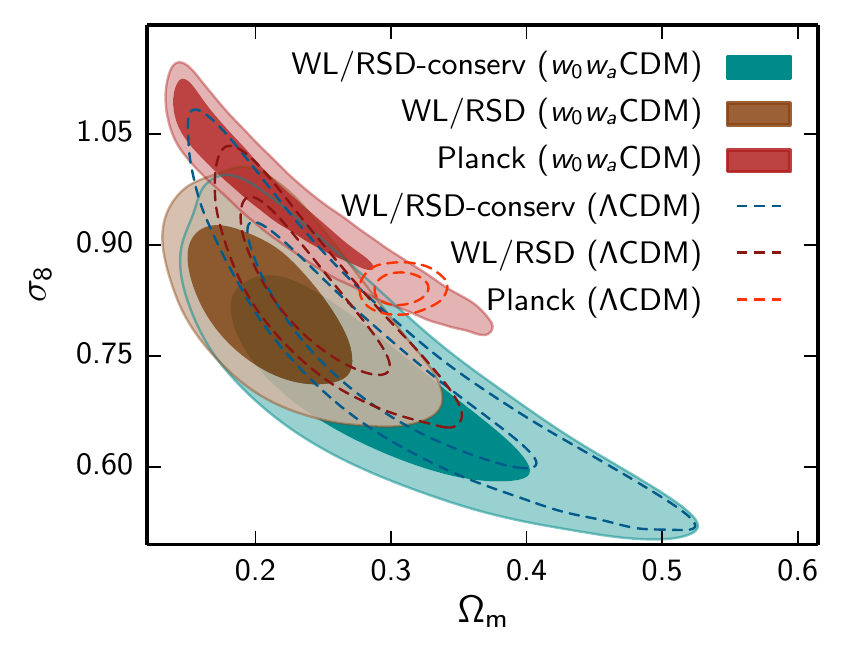}}
\resizebox{8.8cm}{!}{\includegraphics{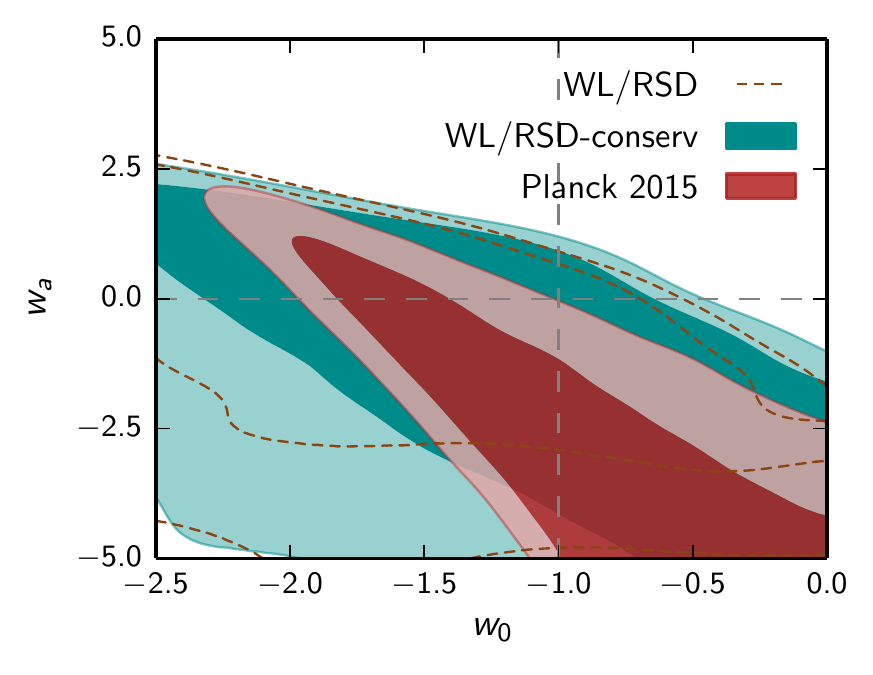}}
\end{center}
\vspace{-2.0em}
\caption{Left: Marginalized posterior contours in the $\sigma_8$ -- $\Omega_{\mathrm m}$ plane (inner 68\%~CL, outer 95\%~CL) in a universe with a time-dependent dark energy equation of state. We show the constraints from $\{\xi_{\pm}, \gamma_{\rm t}, P_{0/2}\}$ in brown, with conservative cuts to the data in cyan, and Planck CMB temperature in red. For comparison, we also show dashed contours assuming fiducial $\Lambda$CDM. Right: Marginalized posterior contours in the $w_0$ -- $w_a$ plane. The dashed horizontal and vertical lines intersect at the $\Lambda$CDM prediction.
}
\label{figxipmgtplw0wa}
\end{figure*}

\subsubsection{Constraints on dark energy, intrinsic alignment amplitude, and model selection}

We show the dark energy equation of state constraints in the plane with $S_8$ in Fig.~\ref{figxipmgtplnumassetc}. In contrast to KiDS alone, where $w < -0.24$ (95\%~CL), the combined probes favor a component that accelerates the universe at 99.93\%~confidence level in the fiducial analysis, and at 98\%~confidence in the conservative analysis (such that $w < -0.73$ at 95\% CL in the fiducial analysis, and $w < -0.37$ in the conservative analysis). 
In contrast to the curvature and neutrino mass cosmologies, the constraint on the IA amplitude is non-negligibly degraded in this extended cosmology (given the non-negligible degradation perpendicular to the lensing degeneracy direction), such that a positive amplitude is favored at $2.7\sigma$ and $2.3\sigma$ in the fiducial and conservative cases, respectively (compared to $3.5\sigma$ and $2.8\sigma$ in $\Lambda$CDM).

From a model selection standpoint, the combined probes do not favor the extended cosmology, as $\Delta {\rm DIC} \gtrsim 0$. While the same holds for KiDS alone \citep{joudaki17}, the concordance between KiDS and Planck allowed for their joint analysis, which in turn weakly favors the extended cosmology. Here, we do not perform a joint analysis of $\{\xi_{\pm}, \gamma_{\rm t}, P_{0/2}\}$ and Planck given their relative discordance (other than to compute the $\log \mathcal{I}$ statistic).

\subsection{Dark energy ($w_0-w_a$)}
\label{w0walab}

\subsubsection{Background}

Following a deviation from the cosmological constant scenario, there is no strong theoretical motivation to keep the dark energy equation of state constant (e.g.~\citealt{cds98,zws99}). We therefore also examine an evolving dark energy model with a time-varying equation of state in the form of the `$w_0 - w_a$ parameterization' \citep{cp01,linder03}. Executing a Taylor expansion of the equation of state to first order in the scale factor, $a$, we obtain 
\begin{equation}
w(a) = w_0 + (1-a) w_a, 
\end{equation}
where $w_0$ is the present equation of state, and $w_a = - {{\mathrm d}w}/{{\mathrm d}{a}}|_{a=1}$ (also expressed as $w_a = -2 {{\mathrm d}w}/{{\mathrm d}\ln{a}}|_{a=1/2}$; \citealt{linder03}). 

While a positive $w_a$ increases $w(a)$ with time, such that its impact on the observables is qualitatively similar to that described for a constant $w > -1$, the two $\{w_0, w_a\}$ degrees of freedom allow for a greater range of phenomenological scenarios to be realized. This dark energy model was considered in the analysis of KiDS cosmic shear in \citet{joudaki17}, where similar to the constant~$w$ scenario it was found to alleviate the discordance between KiDS and Planck, and between \citet{riess16} and Planck. In combining KiDS and Planck with a uniform \citet{riess16} prior on the Hubble constant, this model was further found moderately favored compared to $\Lambda$CDM (as evidenced by $\Delta {\rm DIC} \lesssim -6$). However, considering theoretical stability conditions, the favored dark energy region cannot be accommodated by minimally-coupled single-field quintessence, but would seemingly require multiple scalar fields \citep{peirone17}. We examine to what extent these results are impacted by the galaxy-galaxy lensing and multipole power spectrum measurements.

\begin{figure*}
\begin{center}
\resizebox{8.8cm}{!}{\includegraphics{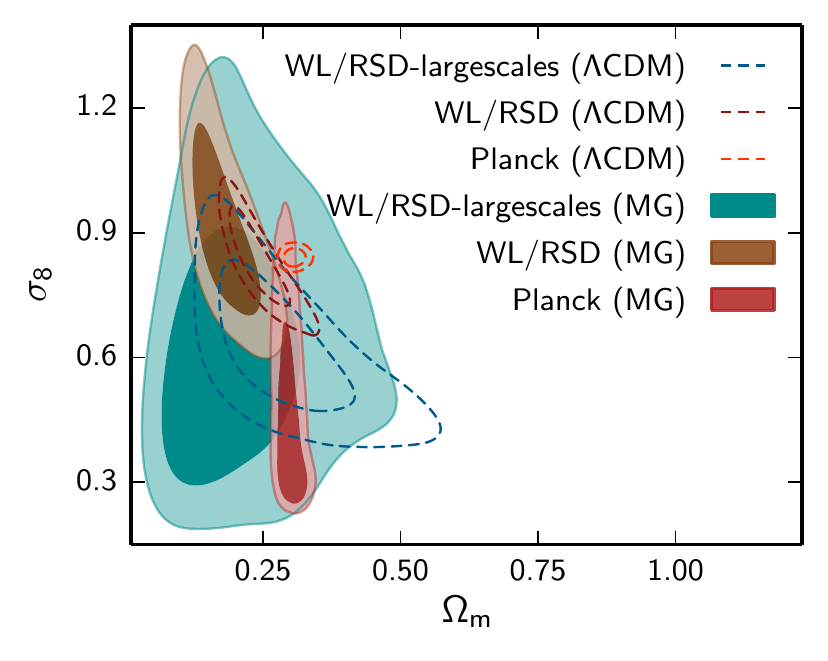}}
\resizebox{8.8cm}{!}{\includegraphics{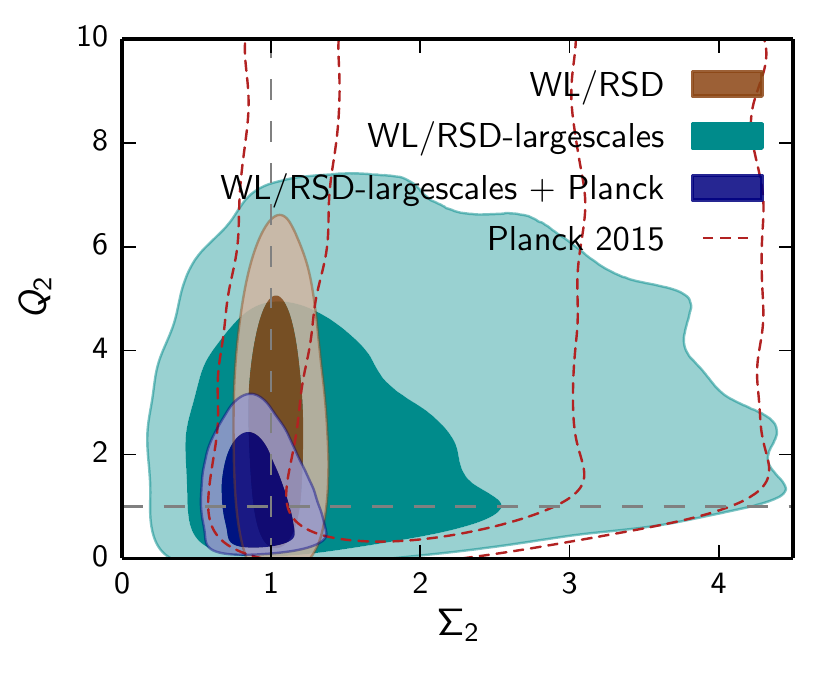}}
\end{center}
\vspace{-2.0em}
\caption{Left: Marginalized posterior contours in the $\sigma_8$ -- $\Omega_{\mathrm m}$ plane (inner 68\%~CL, outer 95\%~CL) in a universe with modified gravity. We show the constraints from $\{\xi_{\pm}, \gamma_{\rm t}, P_{0/2}\}$ in brown, with `large scale' cuts to the data in cyan (our most conservative case), and Planck CMB temperature in red. For comparison, we also show dashed contours assuming fiducial $\Lambda$CDM.
Right: Marginalized posterior contours in the $Q_2$ -- $\Sigma_2$ plane (where the indices represent a particular combination of bins, such that $z < 1$ and $k > 0.05~h~{\rm{Mpc}}^{-1}$). In addition to the cases described, we include $\{\xi_{\pm}, \gamma_{\rm t}, P_{0/2}\}$ with large-scale cuts jointly analyzed with Planck in blue. The dashed horizontal and vertical lines intersect at the GR prediction ($Q = \Sigma = 1$).
}
\label{figxipmgtplmg}
\end{figure*}

\subsubsection{Constraints on $S_8$ and discordance with Planck}

Similar to the constant $w$ cosmology, our marginalized posterior contours exhibit a discordance with Planck in the $\sigma_8$ -- $\Omega_{\rm m}$ plane (Fig.~\ref{figxipmgtplw0wa}). The $S_8$ constraints are approximately 60\% weaker than in $\Lambda$CDM, and 10-20\% stronger than from cosmic shear alone. The discordances are encapsulated through $T(S_8) = 1.7\sigma$ for fiducial data cuts, and $T(S_8) = 1.9\sigma$ with conservative cuts (in contrast to $T(S_8) = 0.9\sigma$ between KiDS and Planck alone). Accounting for the full parameter space with the $\log \mathcal{I}$ statistic, the conservative scenario is only weakly discordant with Planck, while the fiducial scenario is decisively discordant (as the stronger constraints allow for potentially larger discordances). In other words, the substantial concordance between KiDS and Planck alone in \citet{joudaki17} is broken by the improved constraints along the lensing degeneracy direction by the multipole power spectra (again assuming our shot noise prior is approximately correct; Section~\ref{xipmgtpllab}). 

\subsubsection{Constraints on dark energy, intrinsic alignment amplitude, and model selection}

We show the marginalized constraints on $\{w_0, w_a\}$ in Fig.~\ref{figxipmgtplw0wa}. The constraints are weak for both data cuts, with results in agreement with a cosmological constant. While the constraints favor the `large $w_0$, small $w_a$' corner (even more so for the KiDS-only constraints; \citealt{joudaki17}), the fiducial scenario shows an indication to move out of the corner. Similar to the constant $w$ cosmology, the constraint on the IA amplitude is degraded, with a positive amplitude favored at $\{2.6\sigma, 2.3\sigma\}$ for \{fiducial, conservative\} data cuts. From a model selection standpoint, the extended cosmology is not favored relative to $\Lambda$CDM, as $\Delta {\rm DIC}$ is positive (at a level of $5.5$ and $2.0$ for the fiducial and conservative cases, respectively). 
In \citet{joudaki17}, while KiDS alone did not favor the $w_0 - w_a$ cosmology ($\Delta {\rm DIC} \simeq 1.0$), the combination of KiDS and Planck moderately favored the extended cosmology ($\Delta {\rm DIC} = -6.8$, reduced to $-6.4$ with a \citealt{riess16} prior on $H_0$). Here, we do not combine $\{\xi_{\pm}, \gamma_{\rm t}, P_{0/2}\}$ with Planck given their relative discordance (established by $\log \mathcal{I} < 0$).

\begin{figure*}
\hspace{-0.05cm}
\includegraphics[width=0.98\hsize]
{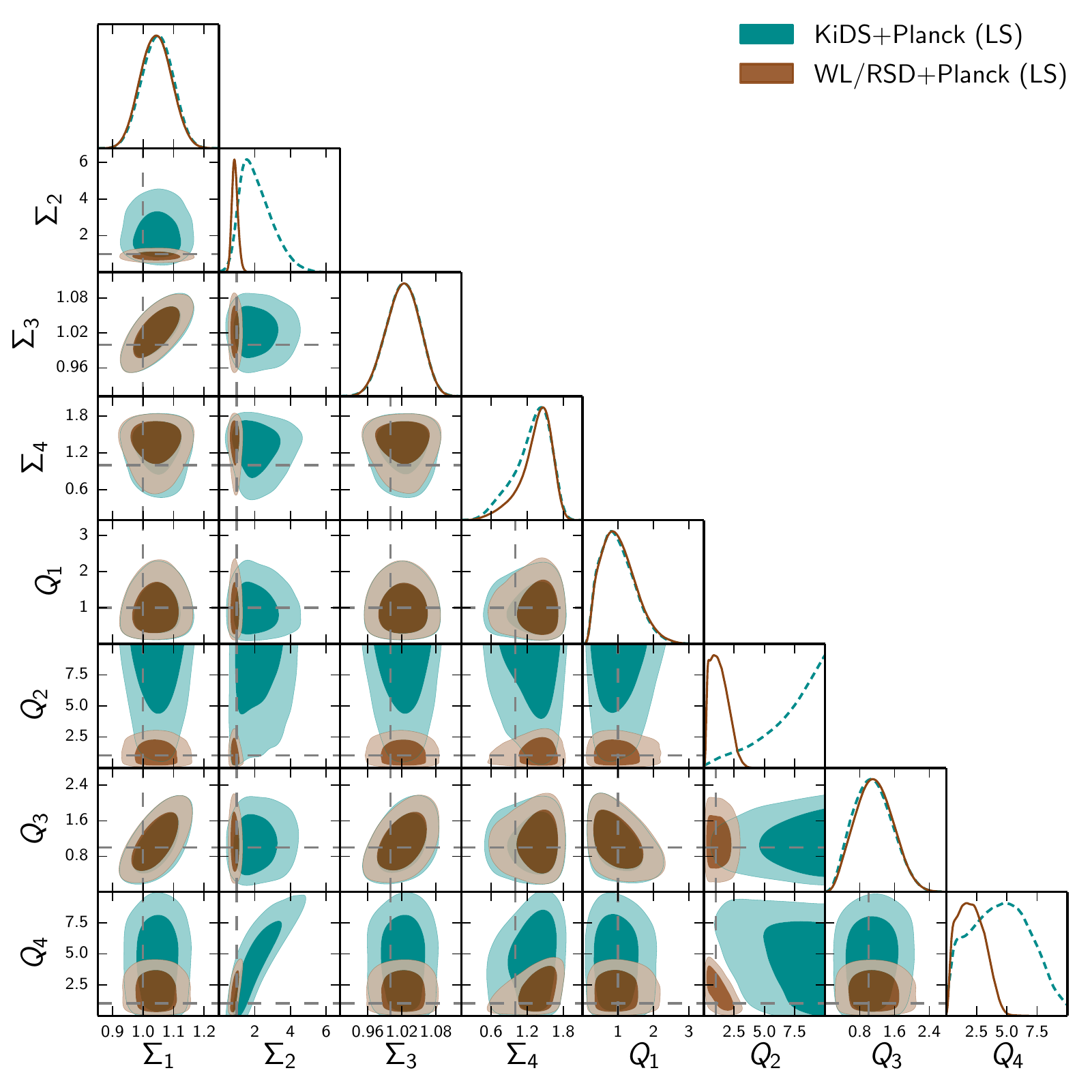}
\vspace{-1.3em}
\caption{\label{figmgsub} 
Marginalized posterior distributions of the modified gravity parameters $\{Q_i, \Sigma_j\}$ and their correlation 
for $\xi_{\pm}$ with large-scale cuts to the data together with Planck in cyan, and $\{\xi_{\pm}, \gamma_{\rm t}, P_{0/2}\}$ with large scale cuts to the data together with Planck in brown (simultaneously varying 5 vanilla cosmological parameters, 14 astrophysical WL/RSD parameters, optical depth to reionization, and additional CMB parameters). The indices of the MG parameters indicate transitions at $k = 0.05~h~{\rm{Mpc}}^{-1}$ and $z = 1$, such that `1' refers to $\{{\rm low}~z, {\rm low}~k\}$, `2' refers to $\{{\rm low}~z, {\rm high}~k\}$, `3' refers to $\{{\rm high}~z, {\rm low}~k\}$, and `4' refers to $\{{\rm high}~z, {\rm high}~k\}$. The GR prediction is given by the intersection of the horizontal and vertical dashed lines ($Q = \Sigma = 1$). 
}
\end{figure*}

\subsection{Modified gravity}
\label{modgrav}

\subsubsection{Background}
For purposes of universality, we do not consider specific models of modified gravity, but instead explore 
model-independent modifications of the metric potentials, $\phi$ and $\psi$, describing spatial and temporal perturbations to the metric in the conformal Newtonian gauge, respectively. While distinct modified gravity models affect the metric potentials differently (e.g.~\citealt{ps16}), the zeroth-order approach is to search for {\it any} deviations from GR. Considering the first-order perturbed Einstein equations (e.g.~\citealt{mb95}), we modify the Poisson equation,
\begin{equation}
-k^2 \phi = 4 \pi G a^2 \sum_i \rho_i \Delta_i Q(k,a),
\end{equation}
where $G$ is Newton's gravitational constant, $\rho_i$ is the density of species $i$, with a fractional overdensity $\delta_i$, and $Q(k,a)$ provides non-standard modifications of the Poisson equation in scale and time, such that $Q \equiv 1$ in GR (e.g.~\citealt{jz08,bt10,dipd15}). We can therefore consider the product of $G$ and $Q(k,a)$ to encapsulate an `effective gravitational constant' that is both scale and time dependent: $G_{\rm eff}(k,a) = G \times Q(k,a)$. As in the analysis of \citet{joudaki17}, we moreover modify the sum of the metric potentials probed by weak gravitational lensing with $\Sigma(k,a)$, such that
\begin{equation}
\begin{split}
-k^2 (\psi + \phi) = &~8 \pi G a^2 \sum_i \rho_i \Delta_i \Sigma(k,a) \\
+&~12 \pi G a^2 \sum_i \rho_i \sigma_i (1+w_i) Q(k,a),
\end{split}
\end{equation}
where $\sigma_i$ is the anisotropic shear stress, and $w_i$ is the equation of state. We thereby allow for the two metric potentials to differ even in the absence of anisotropic stress, whereas $\Sigma \equiv 1$ in GR. The two parameters $\{Q, \Sigma\}$ take on specific functional forms in distinct modified gravity scenarios. In our analysis, we bin the two parameters in $\{k, z\}$, such that we constrain a total of eight modified gravity parameters, as described in the forthcoming subsection.

To capture the modifications to General Relativity with our combined lensing and RSD probes, we have integrated our updated version of \cosmomc (described in Section~\ref{cosmomcsec}) with the \isitgr package \citep{dossett11,dossett12}. The combination of weak gravitational lensing and redshift space distortions is particularly complementary as the former mainly probes the sum of the metric potentials $\psi+\phi$ modifying the relativistic deflection of light, while the latter probes the potential $\psi$ modifying the growth of large-scale structure. This complementarity has for instance been encapsulated in the gravitational slip statistic $E_G$ (\citealt{zhang07}, also see \citealt{leonard15}), measured in e.g.~\citet{reyes10,blake16eg, pullen16,alam17}.

The complementarity between weak lensing and redshift-space distortion measurements has also been used for \cfhtlens, WiggleZ, and 6dFGS in \citet{simpson13}, where no evidence for deviations from GR were found.
For model-independent constraints on deviations from General Relativity with other data combinations, see e.g.~\citet{daniel10,zhao10,johnson15,planckmg15,dVMS16,mueller16}. For the `KiDS-only' constraints, see \citet{joudaki17}. In the current work, we move beyond previous analyses in presenting self-consistent constraints on modified gravity from overlapping lensing and spectroscopic surveys, including the full covariance between the observables.

\subsubsection{Parameterization}

We bin $\{Q, \Sigma\}$ in $k$ and $z$, with divisions at $k = 0.05~h~{\rm{Mpc}}^{-1}$ and $z = 1$ for consistency with the KiDS-only analysis in \citet{joudaki17}. Our specific divisions allow for further complementarity with the CMB, but we recommend the exploration of other choices. We thereby consider four parameters in $Q_{\{1, 2, 3, 4\}}$ and four parameters in $\Sigma_{\{1,2,3,4\}}$, such that `$1$' refers to the $\{{\rm low}~z, {\rm low}~k\}$ bin, `$2$' refers to the $\{{\rm low}~z, {\rm high}~k\}$ bin, `$3$' refers to the $\{{\rm high}~z, {\rm low}~k\}$ bin, and `$4$' refers to the $\{{\rm high}~z, {\rm high}~k\}$ bin. In the MCMC runs, we simultaneously vary these 8 modified gravity parameters along with the 5 vanilla cosmological parameters and 14 astrophysical parameters (listed in Table~\ref{tabpri}), equaling a total of 27 free parameters (additional parameters such as the optical depth are varied when including the CMB).

In the modified gravity runs, we keep a $\Lambda$CDM background expansion. In computing the weak lensing observables, we further modify the likelihood code to directly integrate over the sum of the metric potentials instead of the matter power spectrum. While the lensing observables are useful in constraining $\Sigma$, also known as $G_{\rm light}$, the multipole power spectrum measurements are useful in constraining $2\Sigma - Q$, also known as $G_{\rm matter}$ (e.g.~\citealt{dl13}). While there is merit to the $\{G_{\rm matter}, G_{\rm light}\}$ parameterization, we continue with the $\{Q, \Sigma\}$ convention to be consistent with the analysis in \citet{joudaki17}. Ultimately, given equivalent priors, the cosmological inferences from a full variation of these parameters are equivalent.

\subsubsection{Avoiding nonlinearities: large-scale cuts}

Instead of the standard approach of fiducial and conservative data cuts, we consider fiducial and `large-scale' cuts in Fig.~\ref{figxipmgtplmg} (but see Tables~\ref{tabchidic}~and~\ref{tabs8logi} for conservative results). As shown in Table~\ref{tabcuts}, the large-scale cut removes nonlinear scales from the analysis, such that it effectively corresponds to a linear cut. Concretely, we keep only two angular bins in $\xi_+$ centred at $\theta = \{24.9, 50.7\}$ arcmin, one angular bin in $\xi_-$ centred at $\theta = 210$ arcmin, two angular bins in $\gamma_{\rm t}$ centred at $\theta = \{50.7, 103\}$ arcmin, and one physical bin in $\{P_0, P_2\}$ centred at $k = 0.075~h~{\rm Mpc}^{-1}$. We consider this cut because there is no adequate modeling of the nonlinear corrections to the matter power spectrum (and no screening mechanism) for the model-independent modified gravity scenario considered.

\subsubsection{Constraints on $S_8$ and discordance with Planck}

We find a substantial widening of the marginalized posterior contours in the $\sigma_8$ -- $\Omega_{\rm m}$ plane in Fig.~\ref{figxipmgtplmg}, translated into factors of 2.3 and 2.6 degradation in the $S_8$ constraint compared to $\Lambda$CDM for the fiducial and large-scale cases, respectively. Concretely, $S_8 = 0.69^{+0.07}_{-0.09}$ for fiducial data cuts, and $S_8 = 0.45^{+0.13}_{-0.26}$  with large-scale cuts. The contours have significantly narrowed compared to the KiDS-only analysis (e.g. see Figure~13 in \citealt{joudaki17}); the constraint on $S_8$ alone has improved by a factor of three between the respective fiducial cases. The fiducial $S_8$ constraint is more than a factor of two stronger than the large-scale and Planck constraints (which are of broadly comparable level).

In $\Lambda$CDM, the fiducial and large-scale contours are concordant relative to one another, and in discordance with Planck. In the extended cosmology, the fiducial contour moves towards \{larger $\sigma_8$, smaller $\Omega_{\rm m}$\} and is largely discordant with Planck, which is driven towards small values of $\sigma_8$ for a given matter density (due to the weaker growth allowed, as compared to that inferred by the CMB-measured $A_{\rm s}$ in $\Lambda$CDM). While the $S_8$ constraints are not in tension, as $T(S_8) = 1.2\sigma$, the discordance over the full parameter space is unequivocal, as $\log \mathcal{I} = -0.50$, indicating weak-to-substantial discordance between the datasets. Meanwhile, the large-scale contour is in concordance with Planck, largely as a result of its diminished constraining power. This is manifested over the full parameter space, as $\log \mathcal{I} = 0.87$, corresponding to substantial concordance between the datasets. While not shown, the Planck discordance for conservative data cuts in comparable to that of the fiducial scenario (such that $T(S_8) = 1.7\sigma$ and $\log \mathcal{I} = -0.47$). 

\subsubsection{Constraints on modified gravity}

We show marginalized constraints in the $Q_2$ -- $\Sigma_2$ plane in Fig.~\ref{figxipmgtplmg}, corresponding to the particular combination of modified gravity bins where $k > 0.05~h~{\rm Mpc}^{-1}$ and $z < 1$. Given the complementarity of the weak lensing and RSD measurements, there is significant improvement in the constraints compared to KiDS alone (cf.~Figure 13 in \citealt{joudaki17}). For fiducial data cuts, $Q_2 = 2.8^{+1.1}_{-2.0}$ and $\Sigma_2 = 1.04^{+0.11}_{-0.14}$, while KiDS alone is unable to constrain $Q_2$ within its prior range and $\Sigma_2 = 1.23^{+0.34}_{-0.70}$~\citep{joudaki17}. These constraints are consistent with the GR prediction $(Q = 1, \Sigma = 1)$, which also holds for the six other modified gravity parameters. In the large-scale scenario, $Q_2 = 2.53^{+0.70}_{-2.17}$ and $\Sigma_2 = 1.57^{+0.26}_{-1.02}$, while the equivalent KiDS-only constraints are $Q_2 > 1.4$ (pushing against the upper bound at $Q_2 = 10$) and $\Sigma_2 < 8.5$ \citep{joudaki17}. The large-scale $\{\xi_\pm, \gamma_{\rm t}, P_{0/2}\}$ constraints on the six other modified gravity parameters are all consistent with GR.

Given the concordance between the large-scale measurements and Planck, we further show their combined constraints in Fig.~\ref{figxipmgtplmg}. While Planck alone provides weaker constraints than the large-scale lensing/RSD probes in the $Q_2$ -- $\Sigma_2$ plane, their combination is complementary and improves the constraint on $Q_2 = 1.28^{+0.41}_{-1.00}$ and $\Sigma_2 = 0.90^{+0.14}_{-0.18}$. These large-scale `lensing/RSD + Planck' constraints can be compared to the large-scale `KiDS+Planck' constraints in \citet{joudaki17}, where $Q_2 > 2.2$ (restricted by the upper bound) and $\Sigma_2 = 2.13^{+0.58}_{-1.10}$. The substantial constraint improvement for `lensing/RSD + Planck' compared to `KiDS+Planck' illustrates the importance of the $\{\gamma_{\rm t}, P_{0/2}\}$ measurements included in our analysis. In the full submatrix of `lensing/RSD + Planck' constraints in Fig~\ref{figmgsub}, we find no deviations from GR for any of the modified gravity degrees of freedom (even in the plane of $Q_2$--$Q_4$ where `KiDS+Planck' favors a discrepancy). The complementarity between the measurements will improve as the overlap between KiDS and 2dFLenS/BOSS increases. 

\subsubsection{IA amplitude and galaxy bias: benefits of combined analysis with Planck}

As in $\Lambda$CDM, the strong preference for a positive IA amplitude reduces to approximately $1\sigma$ 
in the large-scale scenario. However, in combining the large-scale measurements and Planck (given their concordance in the extended cosmology), the constraint improves by more than a factor of two, and a positive IA amplitude is again favored at non-trivial significance. Concretely, $A_{\rm IA} = 1.63^{+0.63}_{-0.63}$, which
corresponds to a $2.5\sigma$ preference. As the combined analysis with Planck narrows down the cosmological parameter space, it also improves the galaxy bias constraints by approximately a factor of 3.5, further showcasing the potential gains that could be obtained when the datasets are concordant.

\subsubsection{Model selection} 

In evaluating the statistical preference of the extended model relative to $\Lambda$CDM, the fiducial and conservative cases marginally improve the goodness of fit (by $\Delta\chi^2_{\rm eff}$ of -3.3 and -1.7, respectively) but are not favored by the data in a model selection sense (as $\Delta {\rm DIC} \gtrsim 0$). The  large-scale scenario, which is concordant with Planck, is strongly disfavored by the data (as $\Delta {\rm DIC} = 9.3$). In the joint analysis of the large-scale measurements and Planck, the goodness of fit improves substantially relative to $\Lambda$CDM (by $\Delta\chi^2_{\rm eff} = -5.9$), but the extended model is again not favored in a model selection sense (as $\Delta {\rm DIC} = 2.1$). These results are consistent with those obtained for KiDS alone (and when combined with Planck), where $\Delta {\rm DIC} > 0$ \citep{joudaki17}. As our large modified gravity parameter space is penalized by the DIC, we advocate further investigations of modified gravity considering more specific models and refined parameter spaces. 

\section{Conclusions}
\label{conclusionslab}

We have presented a combined analysis of cosmic shear tomography, galaxy-galaxy lensing tomography, and redshift-space galaxy clustering from the Kilo Degree Survey (KiDS-450) overlapping with the 2-degree Field Lensing Survey (2dFLenS) and Baryon Oscillation Spectroscopic Survey (BOSS). Both 2dFLenS and BOSS are restricted to the overlapping areas with KiDS, and divided into `low-redshift' and `high-redshift' galaxy samples, covering $0.15 < z < 0.43$ (2dFLOZ, LOWZ) and $0.43 < z < 0.70$ (2dFHIZ, CMASS). The observables extracted from these surveys are the tomographic two-point shear correlation functions, $\xi_{\pm}^{ij}(\theta)$, tomographic galaxy-galaxy lensing angular cross-correlation, $\gamma_{\rm t}^j(\theta)$, and redshift-space multipole power spectra encapsulated by the monopole $P_0(k)$ and quadrupole $P_2(k)$. We computed the full covariance between the observables, scales, and samples using a large suite of 930 $N$-body simulations.

As part of this work, we have developed and made public a coherent pipeline in \cosmomc for obtaining the cosmological parameter constraints from the observables. This fitting pipeline builds upon the `cosmic shear only' pipeline used in the KiDS-450 analysis, and allows for a large space of astrophysical parameters to be varied in the MCMC, including intrinsic galaxy alignments (amplitude, redshift, and luminosity dependence), baryonic feedback in the nonlinear matter power spectrum (amplitude and bloating using \hmcode), photometric redshift uncertainties (via bootstrap realizations, and via nuisance parameters that shift the tomographic distributions), galaxy bias, pairwise velocity dispersion, and shot noise (the latter three for each galaxy sample). 

In the first part of our cosmological analysis, we examined the parameter constraints in $\Lambda$CDM from a systematic combination of the observables. Given $\xi_{\pm}$ as the basis, these combinations are $\{\xi_{\pm}, \gamma_{\rm t}\}$, $\{\xi_{\pm}, P_{0/2}\}$, and $\{\xi_{\pm}, \gamma_{\rm t}, P_{0/2}\}$. In verifying the internal consistency of the measurements, we constrained $S_8 = 0.742^{+0.035}_{-0.035}$ in the fully combined fiducial data analysis, and $S_8 = 0.721^{+0.036}_{-0.036}$ in the conservative analysis. These $S_8$ constraints correspond to 19-22\% improvements relative to cosmic shear alone, and are discordant with Planck at $2.6\sigma$ and $3.0\sigma$ in the fiducial and conservative analyses, respectively. Owing to the multipole power spectra, the most striking improvement in our constraints are along the lensing degeneracy direction, where $\{\sigma_8, \Omega_{\rm m}\}$ improve by up to $\{60\%, 50\%\}$, and seemingly favor the \{low-$\sigma_8$, high-$\Omega_{\rm m}$\} tail of the underlying cosmic shear contour. Using the $\log \mathcal{I}$ diagnostic to account for the full parameter space, the discordance is \{`decisive', `strong'\} for the \{fiducial, conservative\} cases. 

By virtue of the galaxy-galaxy lensing measurements, the IA amplitude constraint improves by 30\% relative cosmic shear alone, and is positive at $3.5\sigma$ and $2.8\sigma$ in the fiducial and conservative analyses, respectively. We note that the positive preference away from zero may partly be a result of unknown systematics, e.g.~in the photometric redshift distributions \citep{joudaki17}. We constrained the baryonic feedback amplitude to $B < 3.3$ (95\%~CL) in the fiducial data analysis, with a posterior mean at $B = 1.6$, indicating a preference away from the `dark matter only' scenario at $B \simeq 3.1$. The linear biases of the four galaxy samples all peak around $b\sim2$ (for both data analyses and sub-combinations). The cosmological constraints are robust to widened astrophysical priors and IA redshift dependence ($\eta_{\rm IA}$, which improves by 60\% compared to cosmic shear alone, and agrees with zero). Restricting all measurements to effectively linear scales, the discordance with Planck is maintained at $3.0\sigma$.

In the second part of our analysis, we examined the fully combined constraints on extended cosmological models, including neutrino mass, curvature, evolving dark energy (both constant and time-varying equations of state), and modified gravity. We find that none of the extended cosmologies are simultaneously favored in a model selection sense and able to resolve the dataset discordance with Planck. Unlike KiDS cosmic shear alone \citep{joudaki17},
the inability to resolve the discordance holds for models of evolving dark energy due to the improved constraints along the lensing degeneracy direction (assuming the shot noise prior is not too restrictive). In the fiducial data analysis, we constrain the sum of neutrino masses $\sum m_{\nu} < 2.2~{\rm eV}$, the curvature density parameter $\Omega_k < -0.026$, and the dark energy equation of state $w < -0.73$ (all at 95\% CL). We moreover constrain $\{w_0, w_a\}$ to be consistent with a cosmological constant, and do not find a model selection preference for the extended cosmology (analogous to KiDS alone, different from KiDS+Planck;~\citealt{joudaki17}).

As a result of the complementarity between weak lensing and redshift space distortions, modified gravity is the extended cosmology where our constraints improve the most, as their combination allows for constraints on modified gravity degrees of freedom that are not simultaneously bounded with either probe alone. The 8 modified gravity parameters encapsulating $\{k, z\}$ binned modifications to the Poisson equation and relativistic deflection of light are all consistent with the GR prediction. The constraint on $S_8$ improves by up to a factor of 3 compared to KiDS alone. Executing effectively linear cuts to the data, the combined probes are concordant with Planck, both as encapsulated by $S_8$ and over the full parameter space. However, the extended cosmology is not favored in comparison with GR, as $\Delta {\rm DIC} > 0$.

Another KiDS study of a similar nature has run in parallel and has been simultaneously released by \citet{vu17}. In that work, cosmic shear measurements from KiDS-450 are combined with galaxy-galaxy lensing and angular clustering from GAMA \citep{liske15}. These two studies utilize different lens samples, clustering statistics, covariance methodologies, fitted scales, nuisance parameters and priors, and yet both find that the addition of the lens samples produces a $\sim 20\%$ improvement in constraints on the $S_8$ parameter, and a significant tightening of the fit of the intrinsic alignment amplitude. Although the overall results of these investigations are statistically consistent, we note that the addition of the GAMA data reduces the discordance between the lensing and Planck $\Lambda$CDM parameter fits to the point where it is no longer significant, whereas the addition of the 2dFLenS/BOSS data maintains the statistical tensions. 

To elaborate further, KiDS/2dFLenS/BOSS and KiDS/GAMA results agree at the level of $1.2\sigma$ in $S_8$, and even though the two constraints are correlated, their difference is not significant and falls within a systematic error budget reflecting different analysis choices. In~\citet{vu17}, the cosmic shear constraints are in complete agreement with our KiDS-only and KiDS/2dFLenS/BOSS results, and it is when the clustering information from GAMA is included that a mild discrepancy appears. This is evident also from the fact that the cosmic shear and `galaxy clustering + galaxy-galaxy lensing' results in~\citet{vu17} differ by $1.6\sigma$ relative to one another. As the KiDS dataset continues to grow, future investigations will continue to explore the impact of different scales, statistics, datasets and models on the concordance of lensing, clustering, and CMB datasets.

We make the fitting pipeline and data that were used in this analysis public at \sjaddress. The KiDS data are moreover publicly available at \kidsaddress, and~the~2dFLenS data will become public at \twodflensaddress.

\section*{Acknowledgements}
We much appreciate useful discussions with Erminia Calabrese, Jason Dossett, Pedro Ferreira, Ren\'{e}e Hlo\v{z}ek, Manoj Kaplinghat, Antony Lewis, and Michael Wang. We thank Robin Humble and Jarrod Hurley for HPC support. We acknowledge the use of \camb and \cosmomc packages (\citealt*{LCL}; \citealt{Lewis:2002ah}). This work was supported by resources awarded under Astronomy Australia Ltd's merit allocation scheme on the gSTAR national facility at Swinburne University of Technology and on the National Computational Infrastructure. gSTAR is funded by Swinburne and the Australian Government's Education Investment Fund. This work was supported by computational resources provided by the Australian Government through the Pawsey Supercomputing Centre under the National Computational Merit Allocation Scheme. This work was supported by the Flagship Allocation Scheme of the NCI National Facility at the ANU. Computations for the $N$-body simulations were performed in part on the Orcinus supercomputer at the WestGrid HPC consortium (www.westgrid.ca), in part on the GPC supercomputer at the SciNet HPC Consortium.
SciNet is funded by: the Canada Foundation for Innovation under the auspices of Compute Canada; the Government of Ontario; Ontario Research Fund - Research Excellence; and the University of Toronto. 

Parts of this research were conducted by the Australian Research Council Centre of Excellence for All-sky Astrophysics (CAASTRO), through project number CE110001020. SJ acknowledges support from the Beecroft Trust and ERC 693024. CB and DP acknowledge the support of the Australian Research Council through the awards of Future Fellowships. MA and CH acknowledge support from the European Research Council under grant number 647112. JHD acknowledge support from the European Commission under a Marie-Sk{\l}odwoska-Curie European Fellowship (EU project 656869). HHi is supported by an Emmy Noether grant (No. Hi 1495/2-1) of the Deutsche Forschungsgemeinschaft. HHo acknowledges support from the European Research Council under FP7 grant number 279396. DK and PS are supported by the Deutsche Forschungsgemeinschaft in the framework of the TR33 `The Dark Universe'. KK acknowledges support by the Alexander von Humboldt Foundation. AM acknowledges support from a CITA National Fellowship.
LM is supported by STFC grant ST/N000919/1. MV acknowledges support from the European Research Council under FP7 grant number 279396 and the Netherlands Organisation for Scientific Research (NWO) through grants 614.001.103. CW was supported by Australian Research Council Laureate Grant FL0992131.

KiDS is based on data from observations made with ESO Telescopes at the La Silla Paranal Observatory under programme IDs 177.A-3016, 177.A-3017 and 177.A-3018. 2dFLenS is based on data acquired through the Australian Astronomical Observatory, under program A/2014B/008. It would not have been possible without the dedicated work of the staff of the AAO in the development and support of the 2dF-AAOmega system, and the running of the AAT.


\bibliographystyle{mnras}
\bibliography{lensingrsd}

\begin{thebibliography}{}
\makeatletter
\relax
\def\mn@urlcharsother{\let\do\@makeother \do\$\do\&\do\#\do\^\do\_\do\%\do\~}
\def\mn@doi{\begingroup\mn@urlcharsother \@ifnextchar [ {\mn@doi@}
  {\mn@doi@[]}}
\def\mn@doi@[#1]#2{\def\@tempa{#1}\ifx\@tempa\@empty \href
  {http://dx.doi.org/#2} {doi:#2}\else \href {http://dx.doi.org/#2} {#1}\fi
  \endgroup}
\def\mn@eprint#1#2{\mn@eprint@#1:#2::\@nil}
\def\mn@eprint@arXiv#1{\href {http://arxiv.org/abs/#1} {{\tt arXiv:#1}}}
\def\mn@eprint@dblp#1{\href {http://dblp.uni-trier.de/rec/bibtex/#1.xml}
  {dblp:#1}}
\def\mn@eprint@#1:#2:#3:#4\@nil{\def\@tempa {#1}\def\@tempb {#2}\def\@tempc
  {#3}\ifx \@tempc \@empty \let \@tempc \@tempb \let \@tempb \@tempa \fi \ifx
  \@tempb \@empty \def\@tempb {arXiv}\fi \@ifundefined
  {mn@eprint@\@tempb}{\@tempb:\@tempc}{\expandafter \expandafter \csname
  mn@eprint@\@tempb\endcsname \expandafter{\@tempc}}}

\bibitem[\protect\citeauthoryear{Abbott et~al.}{Abbott et~al.}{2016}]{dessv}
Abbott T.,  et~al., 2016, \mn@doi [Phys. Rev.] {10.1103/PhysRevD.94.022001},
  D94, 022001

\bibitem[\protect\citeauthoryear{{Ade} et~al.,}{{Ade} et~al.}{2014}]{planck13}
{Ade} P.~A.~R.,  et~al., 2014, \mn@doi [\aap] {10.1051/0004-6361/201321591},
  \href {http://adsabs.harvard.edu/abs/2014A%26A...571A..16P} {571, A16}

\bibitem[\protect\citeauthoryear{{Ade} et~al.,}{{Ade} et~al.}{2016a}]{planck15}
{Ade} P.~A.~R.,  et~al., 2016a, \mn@doi [\aap] {10.1051/0004-6361/201525830},
  \href {http://adsabs.harvard.edu/abs/2016A%26A...594A..13P} {594, A13}

\bibitem[\protect\citeauthoryear{{Ade} et~al.,}{{Ade}
  et~al.}{2016b}]{planckmg15}
{Ade} P.~A.~R.,  et~al., 2016b, \mn@doi [\aap] {10.1051/0004-6361/201525814},
  \href {http://adsabs.harvard.edu/abs/2016A%26A...594A..14P} {594, A14}

\bibitem[\protect\citeauthoryear{{Aghanim} et~al.,}{{Aghanim}
  et~al.}{2016a}]{planck15like}
{Aghanim} N.,  et~al., 2016a, \mn@doi [\aap] {10.1051/0004-6361/201526926},
  \href {http://adsabs.harvard.edu/abs/2016A%26A...594A..11P} {594, A11}

\bibitem[\protect\citeauthoryear{Aghanim et~al.}{Aghanim
  et~al.}{2016b}]{plancktau}
Aghanim N.,  et~al., 2016b, \mn@doi [Astron. Astrophys.]
  {10.1051/0004-6361/201628890}, 596, A107

\bibitem[\protect\citeauthoryear{{Alam}, {Miyatake}, {More}, {Ho}  \&
  {Mandelbaum}}{{Alam} et~al.}{2017}]{alam17}
{Alam} S.,  {Miyatake} H.,  {More} S.,  {Ho} S.,   {Mandelbaum} R.,  2017,
  \mn@doi [\mnras] {10.1093/mnras/stw3056}, \href
  {http://adsabs.harvard.edu/abs/2017MNRAS.465.4853A} {465, 4853}

\bibitem[\protect\citeauthoryear{{Alcock} \& {Paczynski}}{{Alcock} \&
  {Paczynski}}{1979}]{ap79}
{Alcock} C.,  {Paczynski} B.,  1979, \mn@doi [Nature] {10.1038/281358a0}, \href
  {http://adsabs.harvard.edu/abs/1979Natur.281..358A} {281, 358}

\bibitem[\protect\citeauthoryear{{Amon} et~al.,}{{Amon} et~al.}{2017}]{amon17}
{Amon} A.,  et~al., 2017, preprint, \href
  {http://adsabs.harvard.edu/abs/2017arXiv170704105A} {} (\mn@eprint {arXiv}
  {1707.04105})

\bibitem[\protect\citeauthoryear{{Anderson} et~al.,}{{Anderson}
  et~al.}{2014}]{anderson14}
{Anderson} L.,  et~al., 2014, \mn@doi [\mnras] {10.1093/mnras/stu523}, \href
  {http://adsabs.harvard.edu/abs/2014MNRAS.441...24A} {441, 24}

\bibitem[\protect\citeauthoryear{{Baker}, {Ferreira}  \& {Skordis}}{{Baker}
  et~al.}{2013}]{baker13}
{Baker} T.,  {Ferreira} P.~G.,   {Skordis} C.,  2013, \mn@doi [\prd]
  {10.1103/PhysRevD.87.024015}, \href
  {http://adsabs.harvard.edu/abs/2013PhRvD..87b4015B} {87, 024015}

\bibitem[\protect\citeauthoryear{{Baldauf}, {Seljak}, {Smith}, {Hamaus}  \&
  {Desjacques}}{{Baldauf} et~al.}{2013}]{baldauf13}
{Baldauf} T.,  {Seljak} U.,  {Smith} R.~E.,  {Hamaus} N.,   {Desjacques} V.,
  2013, \mn@doi [\prd] {10.1103/PhysRevD.88.083507}, \href
  {http://adsabs.harvard.edu/abs/2013PhRvD..88h3507B} {88, 083507}

\bibitem[\protect\citeauthoryear{{Ballinger}, {Peacock}  \&
  {Heavens}}{{Ballinger} et~al.}{1996}]{ballinger96}
{Ballinger} W.~E.,  {Peacock} J.~A.,   {Heavens} A.~F.,  1996, \mn@doi [\mnras]
  {10.1093/mnras/282.3.877}, \href
  {http://adsabs.harvard.edu/abs/1996MNRAS.282..877B} {282, 877}

\bibitem[\protect\citeauthoryear{{Bean} \& {Tangmatitham}}{{Bean} \&
  {Tangmatitham}}{2010}]{bt10}
{Bean} R.,  {Tangmatitham} M.,  2010, \mn@doi [\prd]
  {10.1103/PhysRevD.81.083534}, \href
  {http://adsabs.harvard.edu/abs/2010PhRvD..81h3534B} {81, 083534}

\bibitem[\protect\citeauthoryear{{Begeman}, {Belikov}, {Boxhoorn}  \&
  {Valentijn}}{{Begeman} et~al.}{2013}]{begeman13}
{Begeman} K.,  {Belikov} A.~N.,  {Boxhoorn} D.~R.,   {Valentijn} E.~A.,  2013,
  \mn@doi [Experimental Astronomy] {10.1007/s10686-012-9311-4}, \href
  {http://adsabs.harvard.edu/abs/2013ExA....35....1B} {35, 1}

\bibitem[\protect\citeauthoryear{{Ben{\'{\i}}tez}}{{Ben{\'{\i}}tez}}{2000}]{benitez2000}
{Ben{\'{\i}}tez} N.,  2000, \mn@doi [\apj] {10.1086/308947}, \href
  {http://adsabs.harvard.edu/abs/2000ApJ...536..571B} {536, 571}

\bibitem[\protect\citeauthoryear{{Bertschinger} \& {Zukin}}{{Bertschinger} \&
  {Zukin}}{2008}]{bz08}
{Bertschinger} E.,  {Zukin} P.,  2008, \mn@doi [\prd]
  {10.1103/PhysRevD.78.024015}, \href
  {http://adsabs.harvard.edu/abs/2008PhRvD..78b4015B} {78, 024015}

\bibitem[\protect\citeauthoryear{{Beutler} et~al.,}{{Beutler}
  et~al.}{2011}]{beutler11}
{Beutler} F.,  et~al., 2011, \mn@doi [\mnras]
  {10.1111/j.1365-2966.2011.19250.x}, \href
  {http://adsabs.harvard.edu/abs/2011MNRAS.416.3017B} {416, 3017}

\bibitem[\protect\citeauthoryear{{Beutler} et~al.,}{{Beutler}
  et~al.}{2012}]{beutler12}
{Beutler} F.,  et~al., 2012, \mn@doi [\mnras]
  {10.1111/j.1365-2966.2012.21136.x}, \href
  {http://adsabs.harvard.edu/abs/2012MNRAS.423.3430B} {423, 3430}

\bibitem[\protect\citeauthoryear{{Beutler} et~al.,}{{Beutler}
  et~al.}{2014}]{beutler14}
{Beutler} F.,  et~al., 2014, \mn@doi [\mnras] {10.1093/mnras/stu1051}, \href
  {http://adsabs.harvard.edu/abs/2014MNRAS.443.1065B} {443, 1065}

\bibitem[\protect\citeauthoryear{{Bianchi}, {Gil-Mar{\'{\i}}n}, {Ruggeri}  \&
  {Percival}}{{Bianchi} et~al.}{2015}]{bianchi15}
{Bianchi} D.,  {Gil-Mar{\'{\i}}n} H.,  {Ruggeri} R.,   {Percival} W.~J.,  2015,
  \mn@doi [\mnras] {10.1093/mnrasl/slv090}, \href
  {http://adsabs.harvard.edu/abs/2015MNRAS.453L..11B} {453, L11}

\bibitem[\protect\citeauthoryear{{Bird}, {Viel}  \& {Haehnelt}}{{Bird}
  et~al.}{2012}]{bird12}
{Bird} S.,  {Viel} M.,   {Haehnelt} M.~G.,  2012, \mn@doi [\mnras]
  {10.1111/j.1365-2966.2011.20222.x}, \href
  {http://adsabs.harvard.edu/abs/2012MNRAS.420.2551B} {420, 2551}

\bibitem[\protect\citeauthoryear{{Blake} et~al.,}{{Blake}
  et~al.}{2011}]{blake11}
{Blake} C.,  et~al., 2011, \mn@doi [\mnras] {10.1111/j.1365-2966.2011.19592.x},
  \href {http://adsabs.harvard.edu/abs/2011MNRAS.418.1707B} {418, 1707}

\bibitem[\protect\citeauthoryear{{Blake} et~al.,}{{Blake}
  et~al.}{2012}]{blake12}
{Blake} C.,  et~al., 2012, \mn@doi [\mnras] {10.1111/j.1365-2966.2012.21473.x},
  \href {http://adsabs.harvard.edu/abs/2012MNRAS.425..405B} {425, 405}

\bibitem[\protect\citeauthoryear{{Blake} et~al.,}{{Blake}
  et~al.}{2016a}]{blake16eg}
{Blake} C.,  et~al., 2016a, \mn@doi [\mnras] {10.1093/mnras/stv2875}, \href
  {http://adsabs.harvard.edu/abs/2016MNRAS.456.2806B} {456, 2806}

\bibitem[\protect\citeauthoryear{{Blake} et~al.,}{{Blake}
  et~al.}{2016b}]{blake16}
{Blake} C.,  et~al., 2016b, \mn@doi [\mnras] {10.1093/mnras/stw1990}, \href
  {http://adsabs.harvard.edu/abs/2016MNRAS.462.4240B} {462, 4240}

\bibitem[\protect\citeauthoryear{{Bridle} \& {King}}{{Bridle} \&
  {King}}{2007}]{BK07}
{Bridle} S.,  {King} L.,  2007, \mn@doi [New Journal of Physics]
  {10.1088/1367-2630/9/12/444}, \href
  {http://adsabs.harvard.edu/abs/2007NJPh....9..444B} {9, 444}

\bibitem[\protect\citeauthoryear{{Cai} \& {Bernstein}}{{Cai} \&
  {Bernstein}}{2012}]{cb12}
{Cai} Y.-C.,  {Bernstein} G.,  2012, \mn@doi [\mnras]
  {10.1111/j.1365-2966.2012.20676.x}, \href
  {http://adsabs.harvard.edu/abs/2012MNRAS.422.1045C} {422, 1045}

\bibitem[\protect\citeauthoryear{{Calabrese} et~al.,}{{Calabrese}
  et~al.}{2013}]{calabrese13}
{Calabrese} E.,  et~al., 2013, \mn@doi [\prd] {10.1103/PhysRevD.87.103012},
  \href {http://adsabs.harvard.edu/abs/2013PhRvD..87j3012C} {87, 103012}

\bibitem[\protect\citeauthoryear{{Calabrese} et~al.,}{{Calabrese}
  et~al.}{2017}]{calabrese17}
{Calabrese} E.,  et~al., 2017, \mn@doi [\prd] {10.1103/PhysRevD.95.063525},
  \href {http://adsabs.harvard.edu/abs/2017PhRvD..95f3525C} {95, 063525}

\bibitem[\protect\citeauthoryear{{Caldwell}, {Dave}  \&
  {Steinhardt}}{{Caldwell} et~al.}{1998}]{cds98}
{Caldwell} R.~R.,  {Dave} R.,   {Steinhardt} P.~J.,  1998, \mn@doi [Physical
  Review Letters] {10.1103/PhysRevLett.80.1582}, \href
  {http://adsabs.harvard.edu/abs/1998PhRvL..80.1582C} {80, 1582}

\bibitem[\protect\citeauthoryear{{Chevallier} \& {Polarski}}{{Chevallier} \&
  {Polarski}}{2001}]{cp01}
{Chevallier} M.,  {Polarski} D.,  2001, \mn@doi [International Journal of
  Modern Physics D] {10.1142/S0218271801000822}, \href
  {http://adsabs.harvard.edu/abs/2001IJMPD..10..213C} {10, 213}

\bibitem[\protect\citeauthoryear{{Daniel} \& {Linder}}{{Daniel} \&
  {Linder}}{2013}]{dl13}
{Daniel} S.~F.,  {Linder} E.~V.,  2013, \mn@doi [\jcap]
  {10.1088/1475-7516/2013/02/007}, \href
  {http://adsabs.harvard.edu/abs/2013JCAP...02..007D} {2, 007}

\bibitem[\protect\citeauthoryear{Daniel, Linder, Smith, Caldwell, Cooray,
  Leauthaud  \& Lombriser}{Daniel et~al.}{2010}]{daniel10}
Daniel S.~F.,  Linder E.~V.,  Smith T.~L.,  Caldwell R.~R.,  Cooray A.,
  Leauthaud A.,   Lombriser L.,  2010, \mn@doi [Phys. Rev.]
  {10.1103/PhysRevD.81.123508}, D81, 123508

\bibitem[\protect\citeauthoryear{Dawson et~al.}{Dawson et~al.}{2013}]{dawson13}
Dawson K.~S.,  et~al., 2013, \mn@doi [Astron. J.] {10.1088/0004-6256/145/1/10},
  145, 10

\bibitem[\protect\citeauthoryear{{Di Valentino}, {Melchiorri}  \& {Silk}}{{Di
  Valentino} et~al.}{2016}]{dVMS16}
{Di Valentino} E.,  {Melchiorri} A.,   {Silk} J.,  2016, \mn@doi [\prd]
  {10.1103/PhysRevD.93.023513}, \href
  {http://adsabs.harvard.edu/abs/2016PhRvD..93b3513D} {93, 023513}

\bibitem[\protect\citeauthoryear{{Dossett} \& {Ishak}}{{Dossett} \&
  {Ishak}}{2012}]{dossett12}
{Dossett} J.~N.,  {Ishak} M.,  2012, \mn@doi [\prd]
  {10.1103/PhysRevD.86.103008}, \href
  {http://adsabs.harvard.edu/abs/2012PhRvD..86j3008D} {86, 103008}

\bibitem[\protect\citeauthoryear{{Dossett}, {Ishak}  \&
  {Moldenhauer}}{{Dossett} et~al.}{2011}]{dossett11}
{Dossett} J.~N.,  {Ishak} M.,   {Moldenhauer} J.,  2011, \mn@doi [\prd]
  {10.1103/PhysRevD.84.123001}, \href
  {http://adsabs.harvard.edu/abs/2011PhRvD..84l3001D} {84, 123001}

\bibitem[\protect\citeauthoryear{{Dossett}, {Ishak}, {Parkinson}  \&
  {Davis}}{{Dossett} et~al.}{2015}]{dipd15}
{Dossett} J.~N.,  {Ishak} M.,  {Parkinson} D.,   {Davis} T.~M.,  2015, \mn@doi
  [\prd] {10.1103/PhysRevD.92.023003}, \href
  {http://adsabs.harvard.edu/abs/2015PhRvD..92b3003D} {92, 023003}

\bibitem[\protect\citeauthoryear{{Dvornik} et~al.,}{{Dvornik}
  et~al.}{2017}]{dvornik17}
{Dvornik} A.,  et~al., 2017, \mn@doi [\mnras] {10.1093/mnras/stx705}, \href
  {http://adsabs.harvard.edu/abs/2017MNRAS.468.3251D} {468, 3251}

\bibitem[\protect\citeauthoryear{Eisenstein et~al.}{Eisenstein
  et~al.}{2011}]{eisenstein11}
Eisenstein D.~J.,  et~al., 2011, \mn@doi [Astron. J.]
  {10.1088/0004-6256/142/3/72}, 142, 72

\bibitem[\protect\citeauthoryear{{Erben} et~al.,}{{Erben}
  et~al.}{2013}]{erben13}
{Erben} T.,  et~al., 2013, \mn@doi [\mnras] {10.1093/mnras/stt928}, \href
  {http://adsabs.harvard.edu/abs/2013MNRAS.433.2545E} {433, 2545}

\bibitem[\protect\citeauthoryear{{Eriksen} \& {Gazta{\~n}aga}}{{Eriksen} \&
  {Gazta{\~n}aga}}{2015}]{eg15}
{Eriksen} M.,  {Gazta{\~n}aga} E.,  2015, \mn@doi [\mnras]
  {10.1093/mnras/stv1093}, \href
  {http://adsabs.harvard.edu/abs/2015MNRAS.451.1553E} {451, 1553}

\bibitem[\protect\citeauthoryear{{Fenech Conti}, {Herbonnet}, {Hoekstra},
  {Merten}, {Miller}  \& {Viola}}{{Fenech Conti} et~al.}{2017}]{fc16}
{Fenech Conti} I.,  {Herbonnet} R.,  {Hoekstra} H.,  {Merten} J.,  {Miller} L.,
    {Viola} M.,  2017, \mn@doi [\mnras] {10.1093/mnras/stx200}, \href
  {http://adsabs.harvard.edu/abs/2017MNRAS.467.1627F} {467, 1627}

\bibitem[\protect\citeauthoryear{{Font-Ribera}, {McDonald}, {Mostek}, {Reid},
  {Seo}  \& {Slosar}}{{Font-Ribera} et~al.}{2014}]{fr14}
{Font-Ribera} A.,  {McDonald} P.,  {Mostek} N.,  {Reid} B.~A.,  {Seo} H.-J.,
  {Slosar} A.,  2014, \mn@doi [\jcap] {10.1088/1475-7516/2014/05/023}, \href
  {http://adsabs.harvard.edu/abs/2014JCAP...05..023F} {5, 023}

\bibitem[\protect\citeauthoryear{Gelman \& Rubin}{Gelman \&
  Rubin}{1992}]{Gelman92}
Gelman A.,  Rubin D.,  1992, Statistical Science, 7, 457

\bibitem[\protect\citeauthoryear{{Gil-Mar{\'{\i}}n} et~al.,}{{Gil-Mar{\'{\i}}n}
  et~al.}{2016}]{gm16}
{Gil-Mar{\'{\i}}n} H.,  et~al., 2016, \mn@doi [\mnras] {10.1093/mnras/stw1096},
  \href {http://adsabs.harvard.edu/abs/2016MNRAS.460.4188G} {460, 4188}

\bibitem[\protect\citeauthoryear{{Guzik}, {Jain}  \& {Takada}}{{Guzik}
  et~al.}{2010}]{guzik10}
{Guzik} J.,  {Jain} B.,   {Takada} M.,  2010, \mn@doi [\prd]
  {10.1103/PhysRevD.81.023503}, \href
  {http://adsabs.harvard.edu/abs/2010PhRvD..81b3503G} {81, 023503}

\bibitem[\protect\citeauthoryear{{Harnois-D{\'e}raps} \& {van
  Waerbeke}}{{Harnois-D{\'e}raps} \& {van Waerbeke}}{2015}]{hdw15}
{Harnois-D{\'e}raps} J.,  {van Waerbeke} L.,  2015, \mn@doi [\mnras]
  {10.1093/mnras/stv794}, \href
  {http://adsabs.harvard.edu/abs/2015MNRAS.450.2857H} {450, 2857}

\bibitem[\protect\citeauthoryear{{Harnois-D{\'e}raps}, {Pen}, {Iliev}, {Merz},
  {Emberson}  \& {Desjacques}}{{Harnois-D{\'e}raps} et~al.}{2013}]{hd13}
{Harnois-D{\'e}raps} J.,  {Pen} U.-L.,  {Iliev} I.~T.,  {Merz} H.,  {Emberson}
  J.~D.,   {Desjacques} V.,  2013, \mn@doi [\mnras] {10.1093/mnras/stt1591},
  \href {http://adsabs.harvard.edu/abs/2013MNRAS.436..540H} {436, 540}

\bibitem[\protect\citeauthoryear{{Hartlap}, {Simon}  \& {Schneider}}{{Hartlap}
  et~al.}{2007}]{hartlap07}
{Hartlap} J.,  {Simon} P.,   {Schneider} P.,  2007, \mn@doi [\aap]
  {10.1051/0004-6361:20066170}, \href
  {http://adsabs.harvard.edu/abs/2007A%26A...464..399H} {464, 399}

\bibitem[\protect\citeauthoryear{{Heitmann}, {Lawrence}, {Kwan}, {Habib}  \&
  {Higdon}}{{Heitmann} et~al.}{2014}]{Coyote4}
{Heitmann} K.,  {Lawrence} E.,  {Kwan} J.,  {Habib} S.,   {Higdon} D.,  2014,
  \mn@doi [\apj] {10.1088/0004-637X/780/1/111}, \href
  {http://adsabs.harvard.edu/abs/2014ApJ...780..111H} {780, 111}

\bibitem[\protect\citeauthoryear{{Heymans} et~al.,}{{Heymans}
  et~al.}{2012}]{heymans12}
{Heymans} C.,  et~al., 2012, \mn@doi [\mnras]
  {10.1111/j.1365-2966.2012.21952.x}, \href
  {http://adsabs.harvard.edu/abs/2012MNRAS.427..146H} {427, 146}

\bibitem[\protect\citeauthoryear{{Heymans} et~al.,}{{Heymans}
  et~al.}{2013}]{Heymans13}
{Heymans} C.,  et~al., 2013, \mn@doi [\mnras] {10.1093/mnras/stt601}, \href
  {http://adsabs.harvard.edu/abs/2013MNRAS.432.2433H} {432, 2433}

\bibitem[\protect\citeauthoryear{{Hikage}, {Mandelbaum}, {Takada}  \&
  {Spergel}}{{Hikage} et~al.}{2013}]{hmts13}
{Hikage} C.,  {Mandelbaum} R.,  {Takada} M.,   {Spergel} D.~N.,  2013, \mn@doi
  [\mnras] {10.1093/mnras/stt1446}, \href
  {http://adsabs.harvard.edu/abs/2013MNRAS.435.2345H} {435, 2345}

\bibitem[\protect\citeauthoryear{{Hildebrandt} et~al.,}{{Hildebrandt}
  et~al.}{2012}]{hildebrandt12}
{Hildebrandt} H.,  et~al., 2012, \mn@doi [\mnras]
  {10.1111/j.1365-2966.2012.20468.x}, \href
  {http://adsabs.harvard.edu/abs/2012MNRAS.421.2355H} {421, 2355}

\bibitem[\protect\citeauthoryear{{Hildebrandt} et~al.,}{{Hildebrandt}
  et~al.}{2017}]{Hildebrandt16}
{Hildebrandt} H.,  et~al., 2017, \mn@doi [\mnras] {10.1093/mnras/stw2805},
  \href {http://adsabs.harvard.edu/abs/2017MNRAS.465.1454H} {465, 1454}

\bibitem[\protect\citeauthoryear{Hinshaw et~al.}{Hinshaw et~al.}{2013}]{wmap9}
Hinshaw G.,  et~al., 2013, \mn@doi [Astrophys. J. Suppl.]
  {10.1088/0067-0049/208/2/19}, 208, 19

\bibitem[\protect\citeauthoryear{{Hirata} \& {Seljak}}{{Hirata} \&
  {Seljak}}{2004}]{HS04}
{Hirata} C.~M.,  {Seljak} U.,  2004, \mn@doi [\prd]
  {10.1103/PhysRevD.70.063526}, \href
  {http://adsabs.harvard.edu/abs/2004PhRvD..70f3526H} {70, 063526}

\bibitem[\protect\citeauthoryear{{Hu} \& {Jain}}{{Hu} \& {Jain}}{2004}]{hj04}
{Hu} W.,  {Jain} B.,  2004, \mn@doi [\prd] {10.1103/PhysRevD.70.043009}, \href
  {http://adsabs.harvard.edu/abs/2004PhRvD..70d3009H} {70, 043009}

\bibitem[\protect\citeauthoryear{{Jain} \& {Zhang}}{{Jain} \&
  {Zhang}}{2008}]{jz08}
{Jain} B.,  {Zhang} P.,  2008, \mn@doi [\prd] {10.1103/PhysRevD.78.063503},
  \href {http://adsabs.harvard.edu/abs/2008PhRvD..78f3503J} {78, 063503}

\bibitem[\protect\citeauthoryear{{Jee}, {Tyson}, {Hilbert}, {Schneider},
  {Schmidt}  \& {Wittman}}{{Jee} et~al.}{2016}]{jee2016}
{Jee} M.~J.,  {Tyson} J.~A.,  {Hilbert} S.,  {Schneider} M.~D.,  {Schmidt} S.,
   {Wittman} D.,  2016, \mn@doi [\apj] {10.3847/0004-637X/824/2/77}, \href
  {http://adsabs.harvard.edu/abs/2016ApJ...824...77J} {824, 77}

\bibitem[\protect\citeauthoryear{{Jeffreys}}{{Jeffreys}}{1961}]{jeffreys}
{Jeffreys} H.,  1961, Theory of probability, 3rd edn , Oxford Classics series
  (reprinted 1998), Oxford University Press, Oxford, UK

\bibitem[\protect\citeauthoryear{{Joachimi} \& {Bridle}}{{Joachimi} \&
  {Bridle}}{2010}]{JB10}
{Joachimi} B.,  {Bridle} S.~L.,  2010, \mn@doi [\aap]
  {10.1051/0004-6361/200913657}, \href
  {http://adsabs.harvard.edu/abs/2010A%26A...523A...1J} {523, A1}

\bibitem[\protect\citeauthoryear{{Joachimi}, {Mandelbaum}, {Abdalla}  \&
  {Bridle}}{{Joachimi} et~al.}{2011}]{Joachimi11}
{Joachimi} B.,  {Mandelbaum} R.,  {Abdalla} F.~B.,   {Bridle} S.~L.,  2011,
  \mn@doi [\aap] {10.1051/0004-6361/201015621}, \href
  {http://adsabs.harvard.edu/abs/2011A%26A...527A..26J} {527, A26}

\bibitem[\protect\citeauthoryear{{Joachimi} et~al.,}{{Joachimi}
  et~al.}{2015}]{joachimi15}
{Joachimi} B.,  et~al., 2015, \mn@doi [\ssr] {10.1007/s11214-015-0177-4}, \href
  {http://adsabs.harvard.edu/abs/2015SSRv..tmp...65J} {}

\bibitem[\protect\citeauthoryear{Johnson, Blake, Dossett, Koda, Parkinson  \&
  Joudaki}{Johnson et~al.}{2016}]{johnson15}
Johnson A.,  Blake C.,  Dossett J.,  Koda J.,  Parkinson D.,   Joudaki S.,
  2016, \mn@doi [Mon. Not. Roy. Astron. Soc.] {10.1093/mnras/stw447}, 458, 2725

\bibitem[\protect\citeauthoryear{{Joudaki}}{{Joudaki}}{2013}]{sj13}
{Joudaki} S.,  2013, \mn@doi [\prd] {10.1103/PhysRevD.87.083523}, \href
  {http://adsabs.harvard.edu/abs/2013PhRvD..87h3523J} {87, 083523}

\bibitem[\protect\citeauthoryear{{Joudaki} \& {Kaplinghat}}{{Joudaki} \&
  {Kaplinghat}}{2012}]{JK12}
{Joudaki} S.,  {Kaplinghat} M.,  2012, \mn@doi [\prd]
  {10.1103/PhysRevD.86.023526}, \href
  {http://adsabs.harvard.edu/abs/2012PhRvD..86b3526J} {86, 023526}

\bibitem[\protect\citeauthoryear{{Joudaki} et~al.,}{{Joudaki}
  et~al.}{2017a}]{joudaki16}
{Joudaki} S.,  et~al., 2017a, \mn@doi [\mnras] {10.1093/mnras/stw2665}, \href
  {http://adsabs.harvard.edu/abs/2017MNRAS.465.2033J} {465, 2033}

\bibitem[\protect\citeauthoryear{{Joudaki} et~al.,}{{Joudaki}
  et~al.}{2017b}]{joudaki17}
{Joudaki} S.,  et~al., 2017b, \mn@doi [\mnras] {10.1093/mnras/stx998}, \href
  {http://adsabs.harvard.edu/abs/2017MNRAS.471.1259J} {471, 1259}

\bibitem[\protect\citeauthoryear{{Kaiser}}{{Kaiser}}{1987}]{kaiser87}
{Kaiser} N.,  1987, \mn@doi [\mnras] {10.1093/mnras/227.1.1}, \href
  {http://adsabs.harvard.edu/abs/1987MNRAS.227....1K} {227, 1}

\bibitem[\protect\citeauthoryear{{Kass} \& {Raftery}}{{Kass} \&
  {Raftery}}{1995}]{kr95}
{Kass} R.~E.,  {Raftery} A.~E.,  1995, J. Am. Stat. Ass., 90, 773

\bibitem[\protect\citeauthoryear{{Kaufman}}{{Kaufman}}{1967}]{kaufman67}
{Kaufman} G.~M.,  1967, Some Bayesian Moment Formulae, Report No. 6710, Centre
  for Operations Research and Econometrics, Catholic University of Louvain,
  Heverlee, Belgium

\bibitem[\protect\citeauthoryear{{Kilbinger}, {Bonnett}  \&
  {Coupon}}{{Kilbinger} et~al.}{2014}]{2014ascl.soft02026K}
{Kilbinger} M.,  {Bonnett} C.,   {Coupon} J.,  2014, {athena: Tree code for
  second-order correlation functions, Astrophysics Source Code Library
  (ascl:1402.026)} (\mn@eprint {ascl} {1402.026})

\bibitem[\protect\citeauthoryear{{Kilbinger} et~al.,}{{Kilbinger}
  et~al.}{2017}]{Kilbinger17}
{Kilbinger} M.,  et~al., 2017, \mn@doi [\mnras] {10.1093/mnras/stx2082}, \href
  {http://adsabs.harvard.edu/abs/2017MNRAS.472.2126K} {472, 2126}

\bibitem[\protect\citeauthoryear{{K{\"o}hlinger}, {Viola}, {Valkenburg},
  {Joachimi}, {Hoekstra}  \& {Kuijken}}{{K{\"o}hlinger}
  et~al.}{2016}]{kohlinger15}
{K{\"o}hlinger} F.,  {Viola} M.,  {Valkenburg} W.,  {Joachimi} B.,  {Hoekstra}
  H.,   {Kuijken} K.,  2016, \mn@doi [\mnras] {10.1093/mnras/stv2762}, \href
  {http://adsabs.harvard.edu/abs/2016MNRAS.456.1508K} {456, 1508}

\bibitem[\protect\citeauthoryear{{K{\"o}hlinger} et~al.,}{{K{\"o}hlinger}
  et~al.}{2017}]{kohlinger17}
{K{\"o}hlinger} F.,  et~al., 2017, \mn@doi [\mnras] {10.1093/mnras/stx1820},
  \href {http://adsabs.harvard.edu/abs/2017MNRAS.471.4412K} {471, 4412}

\bibitem[\protect\citeauthoryear{{Kuijken} et~al.,}{{Kuijken}
  et~al.}{2015}]{kuijken15}
{Kuijken} K.,  et~al., 2015, \mn@doi [\mnras] {10.1093/mnras/stv2140}, \href
  {http://adsabs.harvard.edu/abs/2015MNRAS.454.3500K} {454, 3500}

\bibitem[\protect\citeauthoryear{{Lawrence} et~al.,}{{Lawrence}
  et~al.}{2017}]{miratitan17}
{Lawrence} E.,  et~al., 2017, \mn@doi [\apj] {10.3847/1538-4357/aa86a9}, \href
  {http://adsabs.harvard.edu/abs/2017ApJ...847...50L} {847, 50}

\bibitem[\protect\citeauthoryear{{Le Brun}, {McCarthy}, {Schaye}  \&
  {Ponman}}{{Le Brun} et~al.}{2014}]{lebrun14}
{Le Brun} A.~M.~C.,  {McCarthy} I.~G.,  {Schaye} J.,   {Ponman} T.~J.,  2014,
  \mn@doi [\mnras] {10.1093/mnras/stu608}, \href
  {http://adsabs.harvard.edu/abs/2014MNRAS.441.1270L} {441, 1270}

\bibitem[\protect\citeauthoryear{{Leonard}, {Ferreira}  \& {Heymans}}{{Leonard}
  et~al.}{2015}]{leonard15}
{Leonard} C.~D.,  {Ferreira} P.~G.,   {Heymans} C.,  2015, \mn@doi [\jcap]
  {10.1088/1475-7516/2015/12/051}, \href
  {http://adsabs.harvard.edu/abs/2015JCAP...12..051L} {12, 051}

\bibitem[\protect\citeauthoryear{{Lewis} \& {Bridle}}{{Lewis} \&
  {Bridle}}{2002}]{Lewis:2002ah}
{Lewis} A.,  {Bridle} S.,  2002, \mn@doi [\prd] {10.1103/PhysRevD.66.103511},
  \href {http://adsabs.harvard.edu/abs/2002PhRvD..66j3511L} {66, 103511}

\bibitem[\protect\citeauthoryear{{Lewis}, {Challinor}  \& {Lasenby}}{{Lewis}
  et~al.}{2000}]{LCL}
{Lewis} A.,  {Challinor} A.,   {Lasenby} A.,  2000, \mn@doi [\apj]
  {10.1086/309179}, \href {http://adsabs.harvard.edu/abs/2000ApJ...538..473L}
  {538, 473}

\bibitem[\protect\citeauthoryear{{Liddle}}{{Liddle}}{2007}]{liddle07}
{Liddle} A.~R.,  2007, \mn@doi [\mnras] {10.1111/j.1745-3933.2007.00306.x},
  \href {http://adsabs.harvard.edu/abs/2007MNRAS.377L..74L} {377, L74}

\bibitem[\protect\citeauthoryear{{Limber}}{{Limber}}{1954}]{Limber}
{Limber} D.~N.,  1954, \mn@doi [\apj] {10.1086/145870}, \href
  {http://adsabs.harvard.edu/abs/1954ApJ...119..655L} {119, 655}

\bibitem[\protect\citeauthoryear{{Linder}}{{Linder}}{2003}]{linder03}
{Linder} E.~V.,  2003, \mn@doi [Physical Review Letters]
  {10.1103/PhysRevLett.90.091301}, \href
  {http://adsabs.harvard.edu/abs/2003PhRvL..90i1301L} {90, 091301}

\bibitem[\protect\citeauthoryear{{Liske} et~al.,}{{Liske}
  et~al.}{2015}]{liske15}
{Liske} J.,  et~al., 2015, \mn@doi [\mnras] {10.1093/mnras/stv1436}, \href
  {http://adsabs.harvard.edu/abs/2015MNRAS.452.2087L} {452, 2087}

\bibitem[\protect\citeauthoryear{{Loverde} \& {Afshordi}}{{Loverde} \&
  {Afshordi}}{2008}]{LA08}
{Loverde} M.,  {Afshordi} N.,  2008, \mn@doi [\prd]
  {10.1103/PhysRevD.78.123506}, \href
  {http://adsabs.harvard.edu/abs/2008PhRvD..78l3506L} {78, 123506}

\bibitem[\protect\citeauthoryear{{Ma} \& {Bertschinger}}{{Ma} \&
  {Bertschinger}}{1995}]{mb95}
{Ma} C.-P.,  {Bertschinger} E.,  1995, \mn@doi [\apj] {10.1086/176550}, \href
  {http://adsabs.harvard.edu/abs/1995ApJ...455....7M} {455, 7}

\bibitem[\protect\citeauthoryear{{MacCrann}, {Zuntz}, {Bridle}, {Jain}  \&
  {Becker}}{{MacCrann} et~al.}{2015}]{maccrann15}
{MacCrann} N.,  {Zuntz} J.,  {Bridle} S.,  {Jain} B.,   {Becker} M.~R.,  2015,
  \mn@doi [\mnras] {10.1093/mnras/stv1154}, \href
  {http://adsabs.harvard.edu/abs/2015MNRAS.451.2877M} {451, 2877}

\bibitem[\protect\citeauthoryear{MacCrann et~al.}{MacCrann
  et~al.}{2016}]{maccrann17}
MacCrann N.,  et~al., 2016, \mn@doi [Mon. Not. Roy. Astron. Soc.]
  {10.1093/mnras/stw2849}

\bibitem[\protect\citeauthoryear{{Marshall}, {Rajguru}  \& {Slosar}}{{Marshall}
  et~al.}{2006}]{mrs06}
{Marshall} P.,  {Rajguru} N.,   {Slosar} A.,  2006, \mn@doi [\prd]
  {10.1103/PhysRevD.73.067302}, \href
  {http://adsabs.harvard.edu/abs/2006PhRvD..73f7302M} {73, 067302}

\bibitem[\protect\citeauthoryear{{Massara}, {Villaescusa-Navarro}  \&
  {Viel}}{{Massara} et~al.}{2014}]{massara14}
{Massara} E.,  {Villaescusa-Navarro} F.,   {Viel} M.,  2014, \mn@doi [\jcap]
  {10.1088/1475-7516/2014/12/053}, \href
  {http://adsabs.harvard.edu/abs/2014JCAP...12..053M} {12, 053}

\bibitem[\protect\citeauthoryear{{McQuinn} \& {White}}{{McQuinn} \&
  {White}}{2013}]{mw13}
{McQuinn} M.,  {White} M.,  2013, \mn@doi [\mnras] {10.1093/mnras/stt914},
  \href {http://adsabs.harvard.edu/abs/2013MNRAS.433.2857M} {433, 2857}

\bibitem[\protect\citeauthoryear{{Mead}, {Peacock}, {Heymans}, {Joudaki}  \&
  {Heavens}}{{Mead} et~al.}{2015}]{Mead15}
{Mead} A.~J.,  {Peacock} J.~A.,  {Heymans} C.,  {Joudaki} S.,   {Heavens}
  A.~F.,  2015, \mn@doi [\mnras] {10.1093/mnras/stv2036}, \href
  {http://adsabs.harvard.edu/abs/2015MNRAS.454.1958M} {454, 1958}

\bibitem[\protect\citeauthoryear{{Mead}, {Heymans}, {Lombriser}, {Peacock},
  {Steele}  \& {Winther}}{{Mead} et~al.}{2016}]{Mead16}
{Mead} A.~J.,  {Heymans} C.,  {Lombriser} L.,  {Peacock} J.~A.,  {Steele}
  O.~I.,   {Winther} H.~A.,  2016, \mn@doi [\mnras] {10.1093/mnras/stw681},
  \href {http://adsabs.harvard.edu/abs/2016MNRAS.459.1468M} {459, 1468}

\bibitem[\protect\citeauthoryear{{Miller} et~al.,}{{Miller}
  et~al.}{2013}]{miller13}
{Miller} L.,  et~al., 2013, \mn@doi [\mnras] {10.1093/mnras/sts454}, \href
  {http://adsabs.harvard.edu/abs/2013MNRAS.429.2858M} {429, 2858}

\bibitem[\protect\citeauthoryear{{More}, {Miyatake}, {Mandelbaum}, {Takada},
  {Spergel}, {Brownstein}  \& {Schneider}}{{More} et~al.}{2015}]{more15}
{More} S.,  {Miyatake} H.,  {Mandelbaum} R.,  {Takada} M.,  {Spergel} D.~N.,
  {Brownstein} J.~R.,   {Schneider} D.~P.,  2015, \mn@doi [\apj]
  {10.1088/0004-637X/806/1/2}, \href
  {http://adsabs.harvard.edu/abs/2015ApJ...806....2M} {806, 2}

\bibitem[\protect\citeauthoryear{{Mueller}, {Percival}, {Linder}, {Alam},
  {Zhao}, {S{\'a}nchez}  \& {Beutler}}{{Mueller} et~al.}{2016}]{mueller16}
{Mueller} E.-M.,  {Percival} W.,  {Linder} E.,  {Alam} S.,  {Zhao} G.-B.,
  {S{\'a}nchez} A.~G.,   {Beutler} F.,  2016, preprint, \href
  {http://adsabs.harvard.edu/abs/2016arXiv161200812M} {} (\mn@eprint {arXiv}
  {1612.00812})

\bibitem[\protect\citeauthoryear{{Parejko} et~al.,}{{Parejko}
  et~al.}{2013}]{parejko13}
{Parejko} J.~K.,  et~al., 2013, \mn@doi [\mnras] {10.1093/mnras/sts314}, \href
  {http://adsabs.harvard.edu/abs/2013MNRAS.429...98P} {429, 98}

\bibitem[\protect\citeauthoryear{Patrignani et~al.}{Patrignani
  et~al.}{2016}]{pdg16}
Patrignani C.,  et~al., 2016, \mn@doi [Chin. Phys.]
  {10.1088/1674-1137/40/10/100001}, C40, 100001

\bibitem[\protect\citeauthoryear{{Peirone}, {Martinelli}, {Raveri}  \&
  {Silvestri}}{{Peirone} et~al.}{2017}]{peirone17}
{Peirone} S.,  {Martinelli} M.,  {Raveri} M.,   {Silvestri} A.,  2017, \mn@doi
  [\prd] {10.1103/PhysRevD.96.063524}, \href
  {http://adsabs.harvard.edu/abs/2017PhRvD..96f3524P} {96, 063524}

\bibitem[\protect\citeauthoryear{{Pogosian} \& {Silvestri}}{{Pogosian} \&
  {Silvestri}}{2016}]{ps16}
{Pogosian} L.,  {Silvestri} A.,  2016, \mn@doi [\prd]
  {10.1103/PhysRevD.94.104014}, \href
  {http://adsabs.harvard.edu/abs/2016PhRvD..94j4014P} {94, 104014}

\bibitem[\protect\citeauthoryear{{Pullen}, {Alam}, {He}  \& {Ho}}{{Pullen}
  et~al.}{2016}]{pullen16}
{Pullen} A.~R.,  {Alam} S.,  {He} S.,   {Ho} S.,  2016, \mn@doi [\mnras]
  {10.1093/mnras/stw1249}, \href
  {http://adsabs.harvard.edu/abs/2016MNRAS.460.4098P} {460, 4098}

\bibitem[\protect\citeauthoryear{Raveri}{Raveri}{2016}]{raveri15}
Raveri M.,  2016, \mn@doi [Phys. Rev.] {10.1103/PhysRevD.93.043522}, D93,
  043522

\bibitem[\protect\citeauthoryear{Reyes, Mandelbaum, Seljak, Baldauf, Gunn,
  Lombriser  \& Smith}{Reyes et~al.}{2010}]{reyes10}
Reyes R.,  Mandelbaum R.,  Seljak U.,  Baldauf T.,  Gunn J.~E.,  Lombriser L.,
   Smith R.~E.,  2010, \mn@doi [Nature] {10.1038/nature08857}, 464, 256

\bibitem[\protect\citeauthoryear{{Riess} et~al.,}{{Riess}
  et~al.}{2016}]{riess16}
{Riess} A.~G.,  et~al., 2016, \mn@doi [\apj] {10.3847/0004-637X/826/1/56},
  \href {http://adsabs.harvard.edu/abs/2016ApJ...826...56R} {826, 56}

\bibitem[\protect\citeauthoryear{{Schaye} et~al.,}{{Schaye}
  et~al.}{2010}]{Schaye10}
{Schaye} J.,  et~al., 2010, \mn@doi [\mnras]
  {10.1111/j.1365-2966.2009.16029.x}, \href
  {http://adsabs.harvard.edu/abs/2010MNRAS.402.1536S} {402, 1536}

\bibitem[\protect\citeauthoryear{{Sellentin} \& {Heavens}}{{Sellentin} \&
  {Heavens}}{2016}]{sh15}
{Sellentin} E.,  {Heavens} A.~F.,  2016, \mn@doi [\mnras]
  {10.1093/mnrasl/slv190}, \href
  {http://adsabs.harvard.edu/abs/2016MNRAS.456L.132S} {456, L132}

\bibitem[\protect\citeauthoryear{{Semboloni}, {Hoekstra}, {Schaye}, {van
  Daalen}  \& {McCarthy}}{{Semboloni} et~al.}{2011}]{semboloni11}
{Semboloni} E.,  {Hoekstra} H.,  {Schaye} J.,  {van Daalen} M.~P.,   {McCarthy}
  I.~G.,  2011, \mn@doi [\mnras] {10.1111/j.1365-2966.2011.19385.x}, \href
  {http://adsabs.harvard.edu/abs/2011MNRAS.417.2020S} {417, 2020}

\bibitem[\protect\citeauthoryear{{Simpson} et~al.,}{{Simpson}
  et~al.}{2013}]{simpson13}
{Simpson} F.,  et~al., 2013, \mn@doi [\mnras] {10.1093/mnras/sts493}, \href
  {http://adsabs.harvard.edu/abs/2013MNRAS.429.2249S} {429, 2249}

\bibitem[\protect\citeauthoryear{{Singh}, {Mandelbaum}  \& {More}}{{Singh}
  et~al.}{2015}]{singh15}
{Singh} S.,  {Mandelbaum} R.,   {More} S.,  2015, \mn@doi [\mnras]
  {10.1093/mnras/stv778}, \href
  {http://adsabs.harvard.edu/abs/2015MNRAS.450.2195S} {450, 2195}

\bibitem[\protect\citeauthoryear{{Singh}, {Mandelbaum}, {Seljak}, {Slosar}  \&
  {Vazquez Gonzalez}}{{Singh} et~al.}{2017}]{singh16}
{Singh} S.,  {Mandelbaum} R.,  {Seljak} U.,  {Slosar} A.,   {Vazquez Gonzalez}
  J.,  2017, \mn@doi [\mnras] {10.1093/mnras/stx1828}, \href
  {http://adsabs.harvard.edu/abs/2017MNRAS.471.3827S} {471, 3827}

\bibitem[\protect\citeauthoryear{{Smith} et~al.,}{{Smith}
  et~al.}{2003}]{Smith03}
{Smith} R.~E.,  et~al., 2003, \mn@doi [\mnras]
  {10.1046/j.1365-8711.2003.06503.x}, \href
  {http://adsabs.harvard.edu/abs/2003MNRAS.341.1311S} {341, 1311}

\bibitem[\protect\citeauthoryear{{Song}, {Zhao}, {Bacon}, {Koyama}, {Nichol}
  \& {Pogosian}}{{Song} et~al.}{2011}]{song11}
{Song} Y.-S.,  {Zhao} G.-B.,  {Bacon} D.,  {Koyama} K.,  {Nichol} R.~C.,
  {Pogosian} L.,  2011, \mn@doi [\prd] {10.1103/PhysRevD.84.083523}, \href
  {http://adsabs.harvard.edu/abs/2011PhRvD..84h3523S} {84, 083523}

\bibitem[\protect\citeauthoryear{Spergel, Flauger  \& Hložek}{Spergel
  et~al.}{2015}]{spergel15}
Spergel D.~N.,  Flauger R.,   Hložek R.,  2015, \mn@doi [Phys. Rev.]
  {10.1103/PhysRevD.91.023518}, D91, 023518

\bibitem[\protect\citeauthoryear{{Spiegelhalter}, {Best}  \&
  {Carlin}}{{Spiegelhalter} et~al.}{2002}]{spiegelhalter02}
{Spiegelhalter} D.,  {Best} N.~G.,   {Carlin} B.~P.,  2002,
  J.~Royal.~Stat.~Soc.~B, 64, 583

\bibitem[\protect\citeauthoryear{{Spiegelhalter}, {Best}, {Carlin}  \& van~der
  Linde}{{Spiegelhalter} et~al.}{2014}]{spiegelhalter14}
{Spiegelhalter} D.,  {Best} N.~G.,  {Carlin} B.~P.,   van~der Linde A.,  2014,
  J.~Royal.~Stat.~Soc.~B, 76, 485

\bibitem[\protect\citeauthoryear{{Takahashi}, {Sato}, {Nishimichi}, {Taruya}
  \& {Oguri}}{{Takahashi} et~al.}{2012}]{Takahashi12}
{Takahashi} R.,  {Sato} M.,  {Nishimichi} T.,  {Taruya} A.,   {Oguri} M.,
  2012, \mn@doi [\apj] {10.1088/0004-637X/761/2/152}, \href
  {http://adsabs.harvard.edu/abs/2012ApJ...761..152T} {761, 152}

\bibitem[\protect\citeauthoryear{{Taylor} \& {Hamilton}}{{Taylor} \&
  {Hamilton}}{1996}]{th96}
{Taylor} A.~N.,  {Hamilton} A.~J.~S.,  1996, \mn@doi [\mnras]
  {10.1093/mnras/282.3.767}, \href
  {http://adsabs.harvard.edu/abs/1996MNRAS.282..767T} {282, 767}

\bibitem[\protect\citeauthoryear{{Trotta}}{{Trotta}}{2008}]{trotta08}
{Trotta} R.,  2008, \mn@doi [Contemporary Physics] {10.1080/00107510802066753},
  \href {http://adsabs.harvard.edu/abs/2008ConPh..49...71T} {49, 71}

\bibitem[\protect\citeauthoryear{{Wolf} et~al.,}{{Wolf} et~al.}{2017}]{wolf17}
{Wolf} C.,  et~al., 2017, \mn@doi [\mnras] {10.1093/mnras/stw3151}, \href
  {http://adsabs.harvard.edu/abs/2017MNRAS.466.1582W} {466, 1582}

\bibitem[\protect\citeauthoryear{{Zhang}, {Liguori}, {Bean}  \&
  {Dodelson}}{{Zhang} et~al.}{2007}]{zhang07}
{Zhang} P.,  {Liguori} M.,  {Bean} R.,   {Dodelson} S.,  2007, \mn@doi
  [Physical Review Letters] {10.1103/PhysRevLett.99.141302}, \href
  {http://adsabs.harvard.edu/abs/2007PhRvL..99n1302Z} {99, 141302}

\bibitem[\protect\citeauthoryear{{Zhao}, {Giannantonio}, {Pogosian},
  {Silvestri}, {Bacon}, {Koyama}, {Nichol}  \& {Song}}{{Zhao}
  et~al.}{2010}]{zhao10}
{Zhao} G.-B.,  {Giannantonio} T.,  {Pogosian} L.,  {Silvestri} A.,  {Bacon}
  D.~J.,  {Koyama} K.,  {Nichol} R.~C.,   {Song} Y.-S.,  2010, \mn@doi [\prd]
  {10.1103/PhysRevD.81.103510}, \href
  {http://adsabs.harvard.edu/abs/2010PhRvD..81j3510Z} {81, 103510}

\bibitem[\protect\citeauthoryear{{Zlatev}, {Wang}  \& {Steinhardt}}{{Zlatev}
  et~al.}{1999}]{zws99}
{Zlatev} I.,  {Wang} L.,   {Steinhardt} P.~J.,  1999, \mn@doi [Physical Review
  Letters] {10.1103/PhysRevLett.82.896}, \href
  {http://adsabs.harvard.edu/abs/1999PhRvL..82..896Z} {82, 896}

\bibitem[\protect\citeauthoryear{{de Jong} et~al.,}{{de Jong}
  et~al.}{2015}]{dejong15}
{de Jong} J.~T.~A.,  et~al., 2015, \mn@doi [\aap]
  {10.1051/0004-6361/201526601}, \href
  {http://adsabs.harvard.edu/abs/2015A%26A...582A..62D} {582, A62}

\bibitem[\protect\citeauthoryear{{de Jong} et~al.,}{{de Jong}
  et~al.}{2017}]{dejong17}
{de Jong} J.~T.~A.,  et~al., 2017, \mn@doi [\aap]
  {10.1051/0004-6361/201730747}, \href
  {http://adsabs.harvard.edu/abs/2017A%26A...604A.134D} {604, A134}

\bibitem[\protect\citeauthoryear{{de Putter}, {Dor{\'e}}  \& {Takada}}{{de
  Putter} et~al.}{2013}]{deputter13}
{de Putter} R.,  {Dor{\'e}} O.,   {Takada} M.,  2013, preprint, \href
  {http://adsabs.harvard.edu/abs/2013arXiv1308.6070D} {} (\mn@eprint {arXiv}
  {1308.6070})

\bibitem[\protect\citeauthoryear{{de Putter}, {Dor{\'e}}  \& {Das}}{{de Putter}
  et~al.}{2014}]{deputter14}
{de Putter} R.,  {Dor{\'e}} O.,   {Das} S.,  2014, \mn@doi [\apj]
  {10.1088/0004-637X/780/2/185}, \href
  {http://adsabs.harvard.edu/abs/2014ApJ...780..185D} {780, 185}

\bibitem[\protect\citeauthoryear{{van Daalen}, {Schaye}, {Booth}  \& {Dalla
  Vecchia}}{{van Daalen} et~al.}{2011}]{Daalen11}
{van Daalen} M.~P.,  {Schaye} J.,  {Booth} C.~M.,   {Dalla Vecchia} C.,  2011,
  \mn@doi [\mnras] {10.1111/j.1365-2966.2011.18981.x}, \href
  {http://adsabs.harvard.edu/abs/2011MNRAS.415.3649V} {415, 3649}

\bibitem[\protect\citeauthoryear{{van Uitert} et~al.,}{{van Uitert}
  et~al.}{2017}]{vu17}
{van Uitert} E.,  et~al., 2017, preprint, \href
  {http://adsabs.harvard.edu/abs/2017arXiv170605004V} {} (\mn@eprint {arXiv}
  {1706.05004})

\makeatother
\end{thebibliography}

\appendix

\section{Impact of Planck with $\tau$ prior}
\label{tausec}

In addition to our comparisons with Planck CMB temperature and low-multipole polarization data (TT+lowP) of \citet{planck15}, which we denote `Planck 2015', we also considered the impact of the more recent polarization data of the Planck High Frequency Instrument (HFI) from \citet{plancktau}. The new HFI data particularly improves the constraint on the reionization optical depth, such that $\tau = 0.055 \pm 0.009$ (from the \simlow likelihood), which can be compared to our TT+lowP constraint of $\tau = 0.076 \pm 0.019$.

\begin{figure}
\hspace{-1.2em}
\resizebox{8.7cm}{!}
{\includegraphics{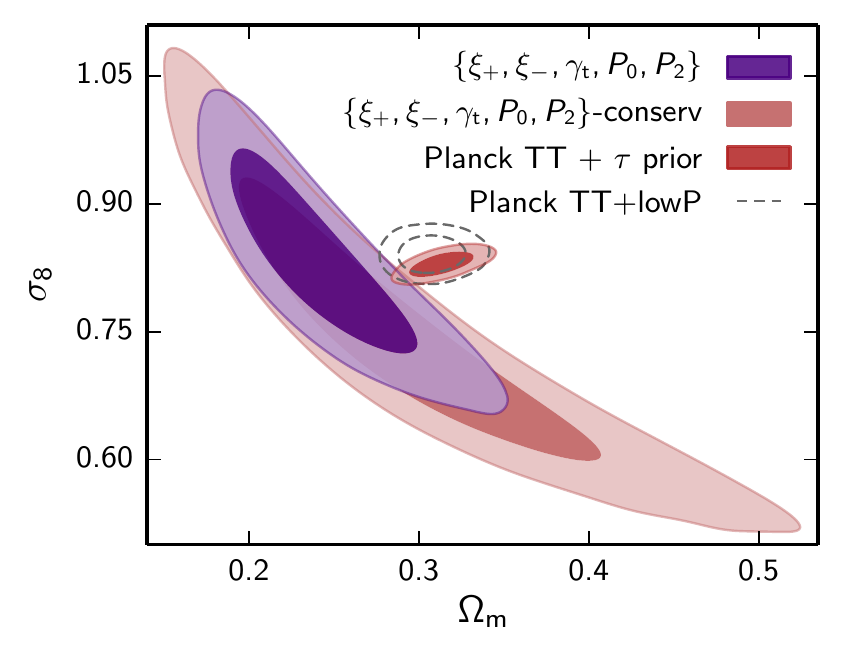}}
\vspace{-2.6em}
\caption{Marginalized posterior contours in the $\sigma_8$ -- $\Omega_{\mathrm m}$ plane (inner 68\%~CL, outer 95\%~CL) from $\{\xi_+, \xi_-, \gamma_{\rm t}, P_0, P_2\}$ in purple, and $\{\xi_+, \xi_-, \gamma_{\rm t}, P_0, P_2\}$ with conservative cuts to the data in pink. We moreover show the CMB constraints from Planck `TT + $\tau$ prior' in red, and Planck `TT+lowP' (also denoted `Planck 2015') in dashed grey.
}
\label{figtau}
\end{figure}

In Fig.~\ref{figtau}, we therefore assess the impact of this new $\tau$ prior (without the `lowP' meausurements) 
on cosmological parameter constraints from Planck in the $\sigma_8$ -- $\Omega_{\rm m}$ plane. While a Gaussian prior on the optical depth does not affect our $\{\xi_{\pm}, \gamma_{\rm t}, P_{0/2}\}$ constraints (as the observables are not functions of the optical depth), it does cause a narrowing of the Planck contour in this plane, preferring smaller $\sigma_8$ and larger $\Omega_{\rm m}$. As the Planck contour transforms along the lensing degeneracy direction, the discordances between the observables considered in this work and Planck remain. Concretely, the Planck `TT+lowP' constraint on $S_8 = 0.852^{+0.024}_{-0.024}$, while the Planck `TT + $\tau$ prior' constraint on $S_8 = 0.848^{+0.024}_{-0.024}$. As a result, the $S_8$ discordances are only marginally affected by the \simlow prior on the optical depth (a decrease by $0.1\sigma$; in agreement with \citealt{plancktau}), while the smaller area of overlap between the contours in Fig.~\ref{figtau} indicate that the discordances have increased over the full parameter space. 

\section{Galaxy bias}
\label{subgalbias}

\subsection{Fixing the galaxy bias in $\gamma_{\rm t}$}

\begin{figure}
\vspace{-0.0cm}
\hspace{0cm}
\resizebox{8.59cm}{!}{\includegraphics{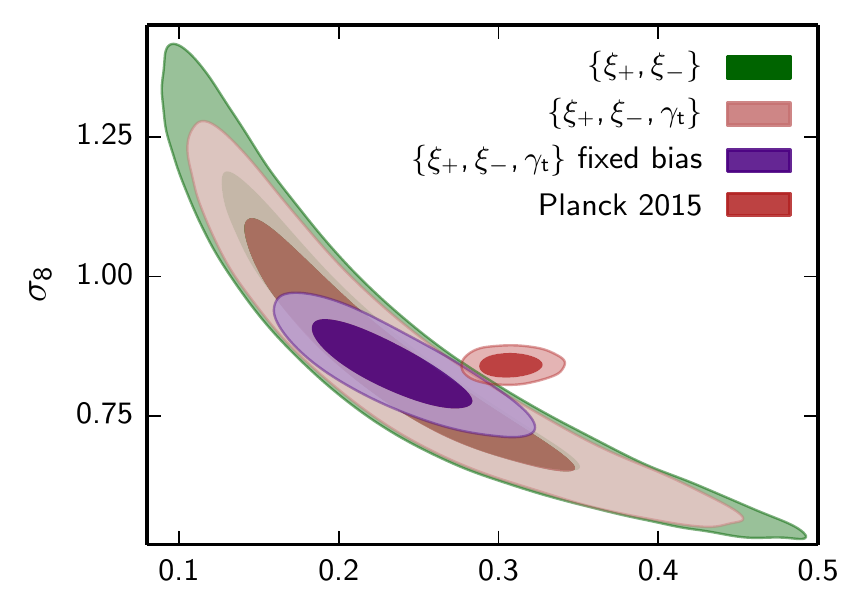}}
\resizebox{8.46cm}{!}{\includegraphics{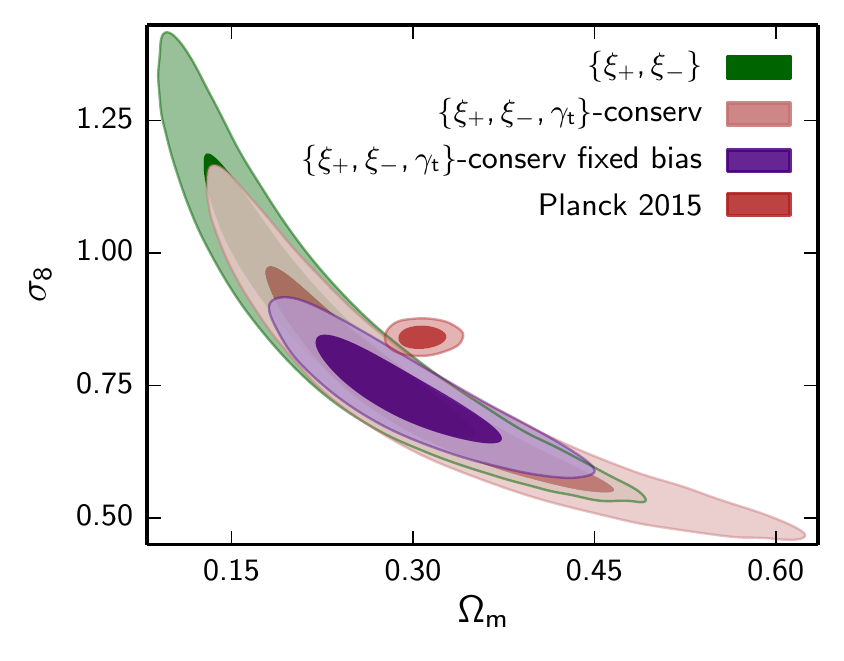}}
\vspace{-2.5em}
\caption{Marginalized posterior contours in the $\sigma_8$ -- $\Omega_{\mathrm m}$ plane (inner 68\%~CL, outer 95\%~CL) when restricted to cosmic shear and galaxy-galaxy lensing. We illustrate the improvement in constraining power when fixing the galaxy bias of each lens sample ($b_{\rm 2dFLOZ}$, $b_{\rm 2dFHIZ}$, $b_{\rm LOWZ}$, $b_{\rm CMASS}$). In the panel to the left, we show $\xi_{\pm}$ in green, $\{\xi_{\pm}, \gamma_{\rm t}\}$ in pink, and $\{\xi_{\pm}, \gamma_{\rm t}\}$ where the galaxy bias parameters are fixed to their posterior means in purple (obtained from the $\{\xi_{\pm}, \gamma_{\rm t}\}$ run where they are varied). The contours in the right panel follow the same convention, except with conservative angular cuts. In both panels, we show the constraints from Planck 2015 in red.
}
\label{figxipmgtfixbias}
\end{figure}

Given the marginal improvement in the cosmological parameter constraints when including galaxy-galaxy lensing measurements in Section~\ref{xipmgtlab} (i.e.~considering $\{\xi_{\pm}, \gamma_{\rm t}\}$), we examined the constraints in the $\sigma_8$ -- $\Omega_{\rm m}$ plane when fixing the galaxy bias of the different samples to their best-fit values in Fig~\ref{figxipmgtfixbias}. Here, we find a significant improvement along the lensing degeneracy direction (with stronger marginalized constraints on $\sigma_8$ and $\Omega_{\rm m}$), but also a small improvement perpendicular to the degeneracy direction (as manifested in $S_8$). While this is true for both cuts to the $\gamma_{\rm t}$ measurements, the most significant improvement is when we retain the most angular scales in $\gamma_{\rm t}$ as expected.

\subsection{Galaxy bias posterior distributions}

\begin{figure*}
\begin{center}
\resizebox{8.53cm}{!}
{\includegraphics{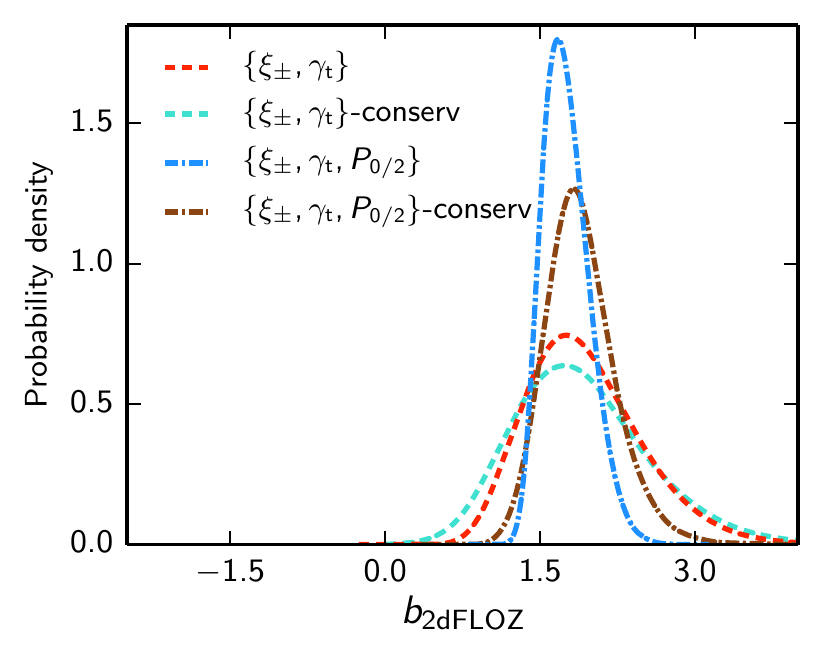}}
\resizebox{8.53cm}{!}
{\includegraphics{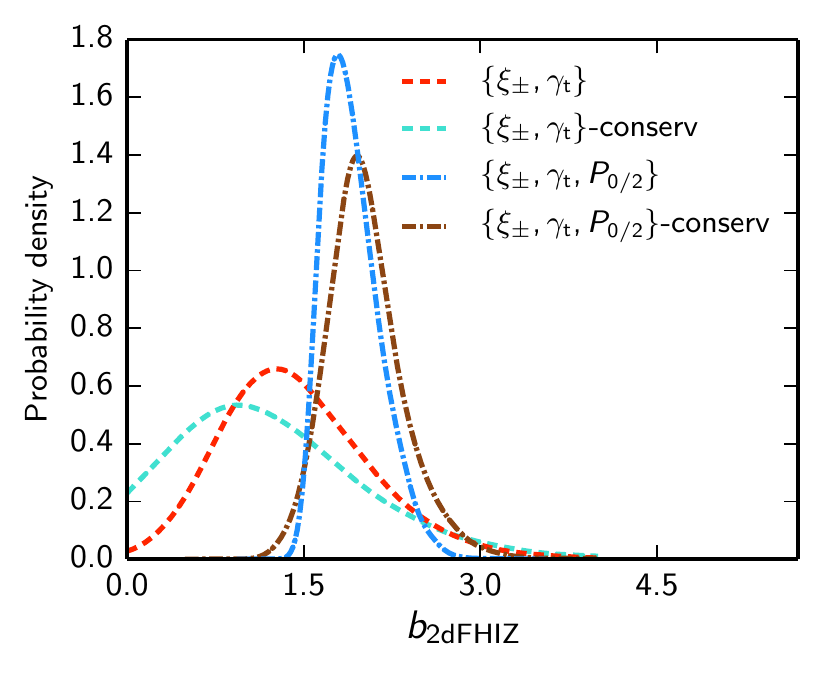}}
\resizebox{8.53cm}{!}
{\includegraphics{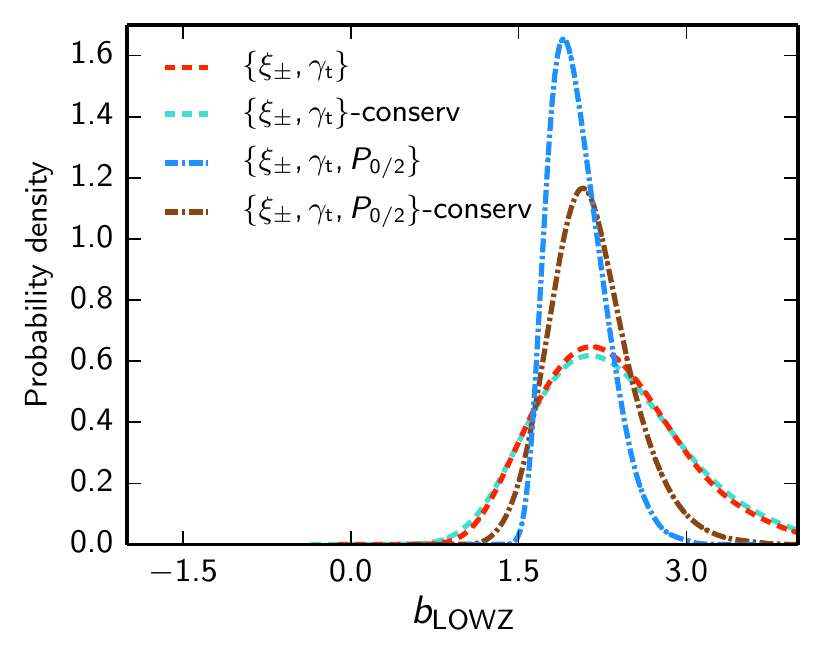}}
\resizebox{8.53cm}{!}
{\includegraphics{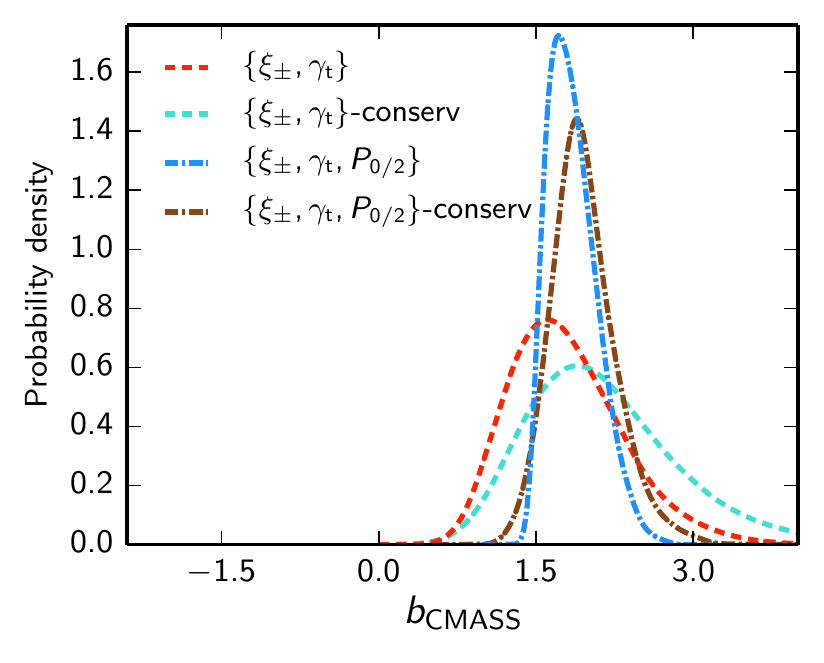}}
\end{center}
\vspace{-2.0em}
\caption{
Marginalized posterior distributions for the galaxy bias (2dFLOZ upper left, 2dFHIZ upper right, LOWZ lower right, and CMASS lower right) from $\{\xi_{\pm}, \gamma_{\rm t}\}$ in dashed red, $\{\xi_{\pm}, \gamma_{\rm t}\}$ with conservative data cuts in dashed cyan, $\{\xi_{\pm}, \gamma_{\rm t}, P_{0/2}\}$ in dot-dashed blue, and $\{\xi_{\pm}, \gamma_{\rm t}, P_{0/2}\}$ with conservative cuts in dot-dashed brown. All of the primary cosmological and astrophysical parameters are simultaneously varied in the analysis. For visual clarity, we do not show the posterior distributions for $\{\xi_{\pm}, P_{0/2}\}$, but note that they are similar to the distributions for the full data vector.
}
\label{figbias}
\end{figure*}

In Fig.~\ref{figbias}, we show the marginalized posterior distributions for the linear bias relating the galaxy and matter overdensities of each galaxy sample (2dFLOZ, 2dFHIZ, LOWZ, CMASS) from measurements of cosmic shear, galaxy-galaxy lensing, and redshift-space multipole power spectra. The underlying cosmological model assumed to generate these posterior distributions is $\Lambda$CDM.

As presented in Section~\ref{xipmgtpllab}, the galaxy biases are in agreement between the $\{\xi_{\pm}, \gamma_{\rm t}\}$ and $\{\xi_{\pm}, \gamma_{\rm t}, P_{0/2}\}$ data combinations, and between fiducial and conservative cuts to the data. Given $\{\xi_{\pm}, \gamma_{\rm t}\}$, the marginalized posterior distributions for the galaxy bias of 2dFHIZ (covering $0.43 < z < 0.70$) show a slight shift towards zero. This can be understood by the negative $\gamma_{\rm t}$ measurements in Fig.~\ref{figgt} for tomographic bins $0.1 < z_{\rm B} < 0.3$ and $0.3 < z_{\rm B} < 0.5$ that are counteracting the positive measurements in higher-redshift tomographic bins ($0.5 < z_{\rm B} < 0.7$ and $0.7 < z_{\rm B} < 0.9$). One alternative approach would be to exclude tomographic bins in which we do not expect much of a signal. We have checked that our cosmological parameter constraints are not particularly affected by the choice between the applied and alternative approaches (via full MCMC runs). 

\section{Extended astrophysical systematics} 
\label{extsystlab}

\begin{figure*}
\hspace{-0.2cm}
\resizebox{8.7cm}{!}{\includegraphics{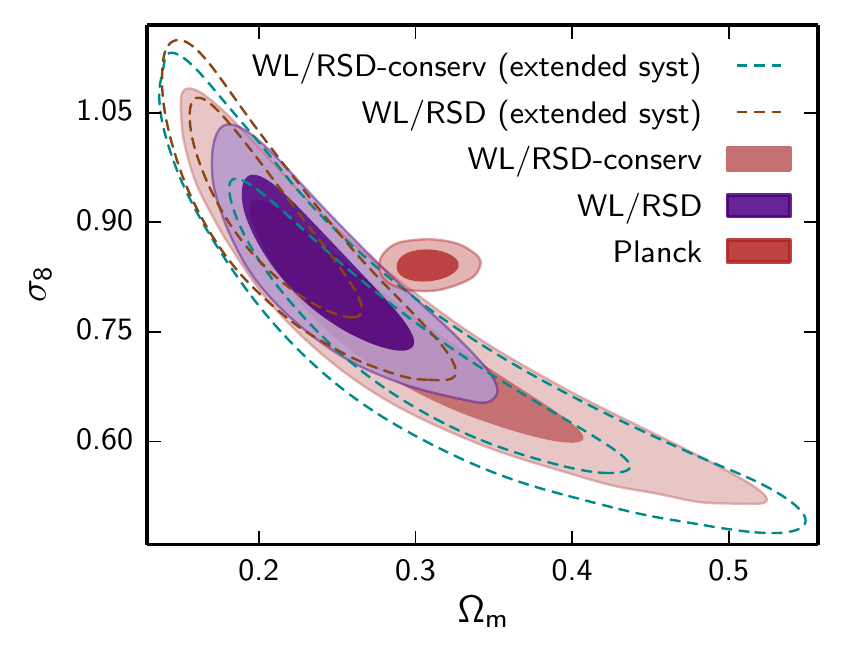}}
\resizebox{8.6cm}{!}{\includegraphics{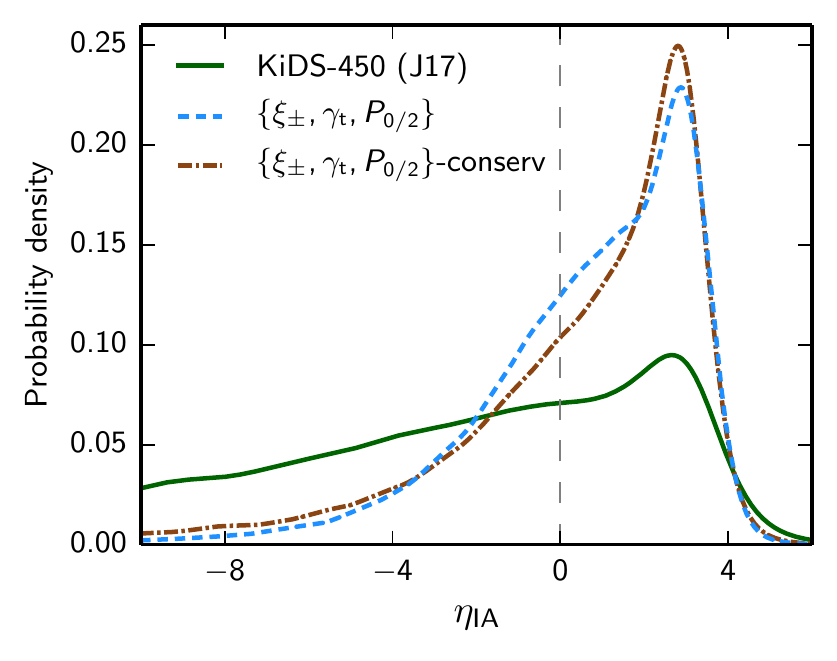}}
\vspace{-1.9em}
\caption{Left: Marginalized posterior contours in the $\sigma_8$ -- $\Omega_{\mathrm m}$ plane (inner 68\%~CL, outer 95\%~CL) when considering an extended treatment of the systematic uncertainties associated with observations of cosmic shear, galaxy-galaxy lensing, and redshift-space multipole power spectra. We show the contours for $\{\xi_{\pm}, \gamma_{\rm t}, P_{0/2}\}$ in solid purple, with an extended treatment of the systematic uncertainties in dashed brown, with conservative data cuts in solid pink, and with both an extended treatment of the systematic uncertainties and conservative data cuts in dashed cyan. For comparison, we show the constraints from Planck 2015 in solid red. Right: Given the extended systematics treatment, we show the marginalized posterior distributions for the power of the IA redshift dependence $\eta_{\rm IA}$ from $\{\xi_{\pm}, \gamma_{\rm t}, P_{0/2}\}$ in dashed blue, and with conservative data cuts in dot-dashed brown. For comparison, we show the distribution when restricted to $\xi_{\pm}$ in solid green (from \citealt{joudaki17}).
}
\label{figextsyst}
\end{figure*}

\subsection{Cosmological constraints}

In addition to the fiducial treatment of systematic uncertainties in the fully joint analysis of our measurements, we also consider an extended treatment to test the robustness of the cosmological constraints. This consists of simultaneously wider priors on all astrophysical parameters (detailed in~Table~\ref{tabpri}), and allows for the IA redshift dependence $\eta_{\rm IA}$ to vary as a free parameter in the MCMC.

We show our extended constraints in the $\sigma_8$ -- $\Omega_{\rm m}$ plane for both data cuts in Fig.~\ref{figextsyst}. While the area of our contours expand, perpendicular to the lensing degeneracy direction they do so mainly away from Planck. As a result, the tension with Planck is still significant at $T(S_8) = 2.4\sigma$ for fiducial data cuts and $T(S_8) = 2.9\sigma$ with conservative cuts. The contours also move along the lensing degeneracy direction, largely as a result of increasing the prior on the shot noise (as discussed in Section~\ref{snsec}).

\subsection{Astrophysical constraints}

We show the marginalized posterior distributions for $\eta_{\rm IA}$ in Fig.~\ref{figextsyst}. While the posteriors peak at $\eta_{\rm IA} \sim 3$, they are consistent with zero as $\eta_{\rm IA} = 0.6^{+3.1}_{-1.1}$ for fiducial data cuts, and $\eta_{\rm IA} = 0.2^{+3.6}_{-0.9}$ with conservative cuts. Our constraints are substantially stronger than from cosmic shear alone, which favors a similar positive peak, and where $\eta_{\rm IA} = -3.6^{+7.6}_{-2.7}$ \citep{joudaki17}. The distributions are asymmetric in that they favor a wider negative tail as increasingly negative values of $\eta_{\rm IA}$ drive the IA signal towards zero (such that the the shear signal dominates), while increasingly positive values of $\eta_{\rm IA}$ rapidly drive the intrinsic alignments towards values that are too large.

\begin{figure*}
\includegraphics[width=0.725\hsize]{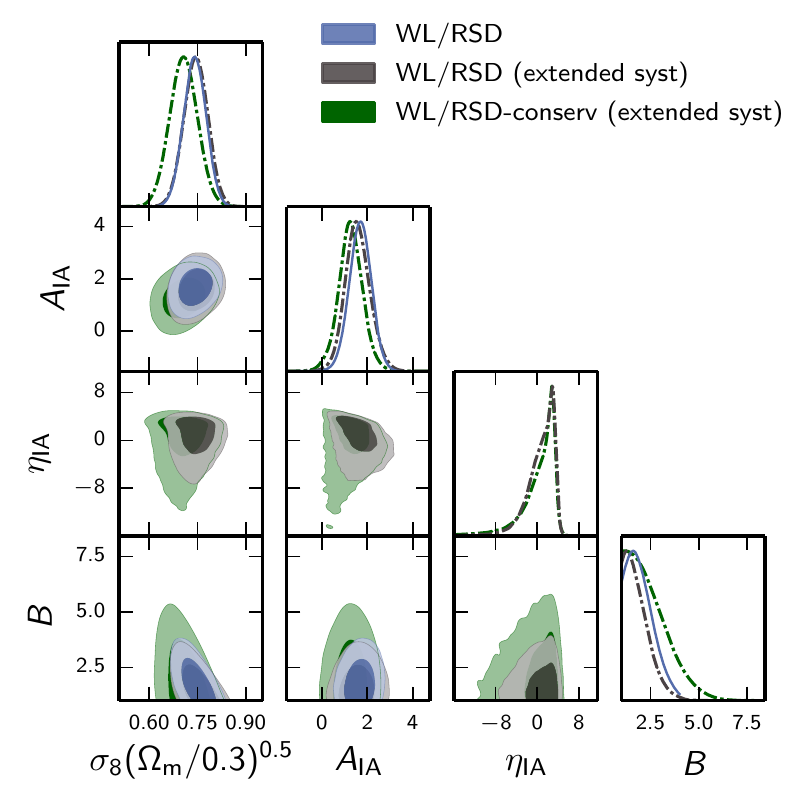}
\vspace{-1.6em}
\caption{\label{figsub} Marginalized posterior distributions of $S_8 = \sigma_8 \sqrt{\Omega_{\mathrm m}/0.3}$, IA amplitude ($A_{\rm IA}$), IA redshift dependence ($\eta_{\rm IA}$), baryon feedback amplitude ($B$), and their correlation, from the $\{\xi_{\pm}, \gamma_{\rm t}, P_{0/2}\}$ data vector (labelled `WL/RSD') with fiducial treatment of systematic uncertainties in blue, extended treatment of the astrophysical systematics in grey, and extended systematics with conservative data cuts in green. The vanilla cosmological parameters, and additional astrophysical parameters such as the galaxy bias, shot noise, and velocity dispersion (of each lens sample) are simultaneously varied in the analyses.}
\end{figure*}

In Fig.~\ref{figsub}, we illustrate the robustness of our results to the different treatments of the astrophysical systematics in the subspace occupied by $\{S_8, A_{\rm IA}, \eta_{\rm IA}, B\}$, along with their marginalized posterior distributions. The largest change in $S_8$ is found for conservative data cuts with extended systematics, where the posterior expands towards lower values. Given extended systematics, the IA amplitude experiences a minor shift and increase in the uncertainty, such that the preference for being positive decreases by $\sim0.5\sigma$. The feedback amplitude marginally shifts to lower values, such that $B < 3.1$ for fiducial data cuts, and $B < 4.4$ with conservative cuts (with posterior peaks at $B$ of 1.2 and 1.6, respectively, again indicating a preference for strong feedback). 

We do not quote our constraints on the shot noise, pairwise velocity dispersion, and galaxy bias in the extended systematics scenario, but note that they are consistent with the fiducial systematics constraints.

\section{Best-fit model parameters}
\label{sublab2}

\begin{figure*}
\includegraphics[width=0.74\hsize]
{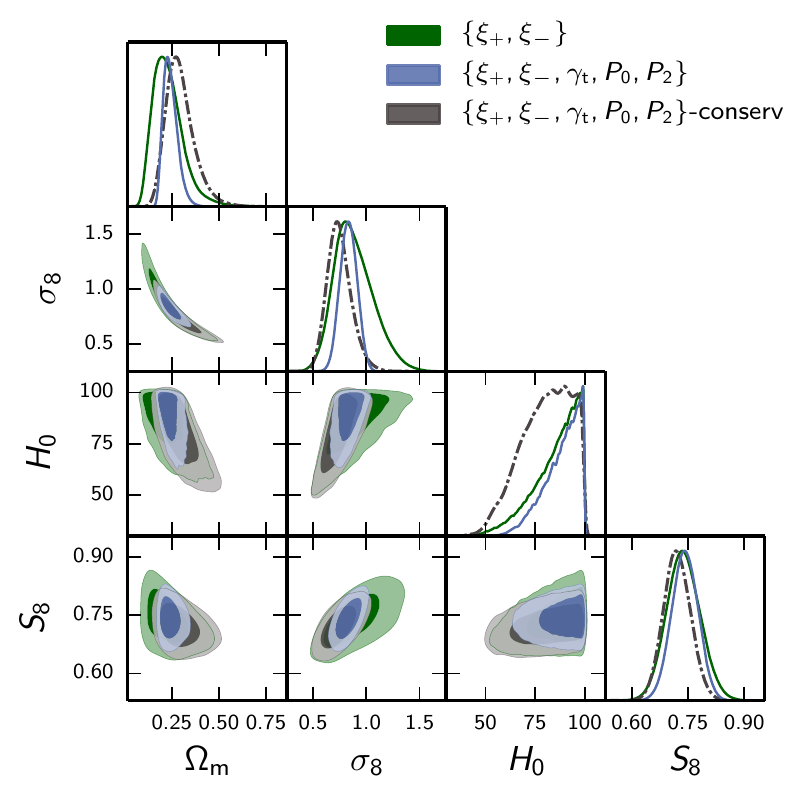}
\vspace{-2.0em}
\caption{\label{figbf} 
Marginalized posterior distributions of derived cosmological parameters $\{\Omega_{\mathrm m}, \sigma_8, H_0, S_8\}$, and their correlation, from $\{\xi_+, \xi_-\}$ in green, $\{\xi_+, \xi_-, \gamma_{\rm t}, P_0, P_2\}$ in blue, and $\{\xi_+, \xi_-, \gamma_{\rm t}, P_0, P_2\}$ with conservative data cuts in grey. The cosmic shear constraints are obtained using the same priors as in \citet{joudaki17}, with the exception of a wider prior on the baryon feedback amplitude (as described in Section~\ref{basesec}). 
}
\end{figure*}

\begin{figure*}
\includegraphics[width=1.00\hsize]{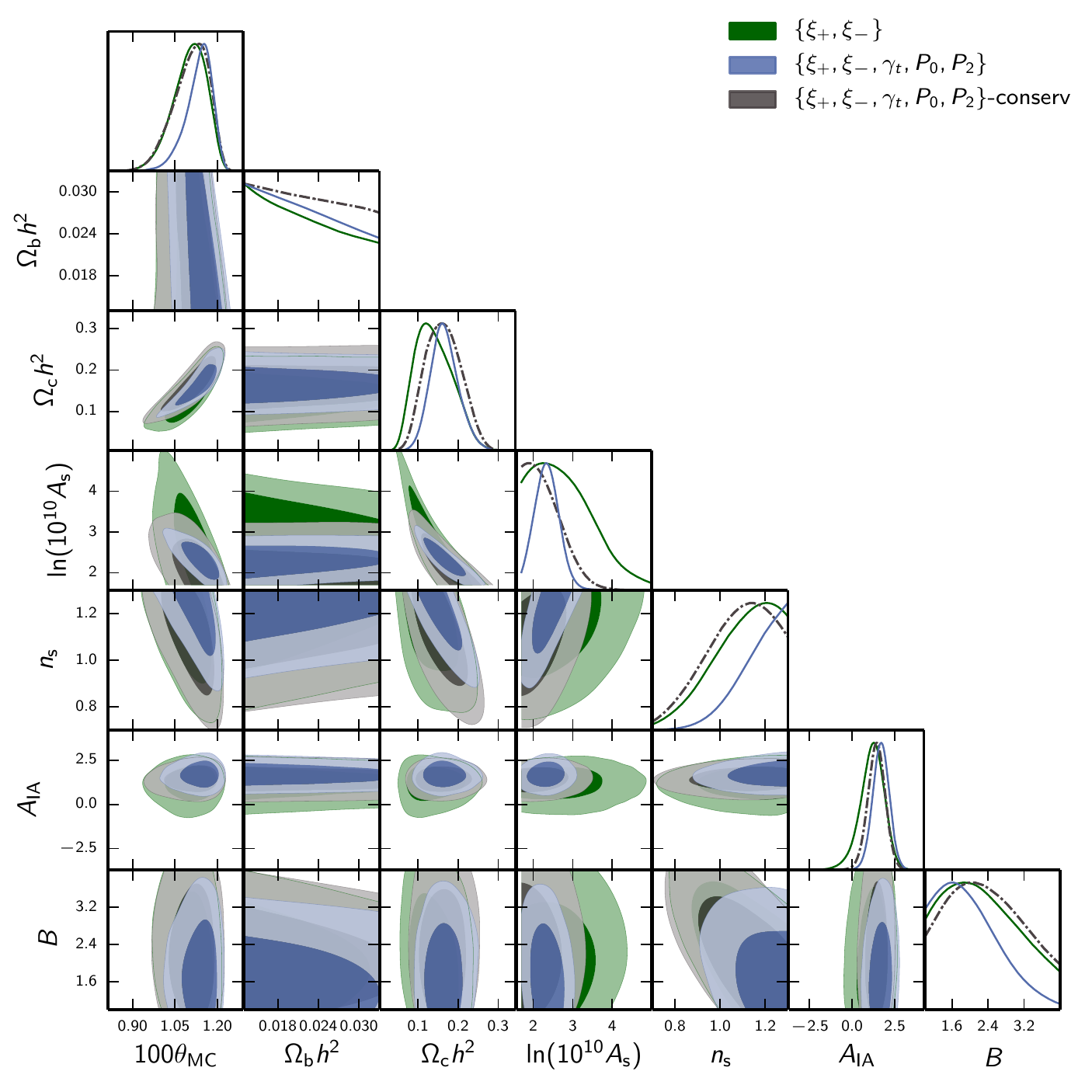}
\vspace{-3.0em}
\caption{\label{figprim} 
Marginalized posterior distributions of the primary model parameters $\{100 \theta_{\rm MC},\Omega_{\mathrm b} h^2, \Omega_{\mathrm c} h^2, \ln{(10^{10} A_{\mathrm s})}, n_s, A_{\rm IA}, B\}$ and their correlation from 
$\{\xi_+, \xi_-\}$ in green, $\{\xi_+, \xi_-, \gamma_{\rm t}, P_0, P_2\}$ in blue, and $\{\xi_+, \xi_-, \gamma_{\rm t}, P_0, P_2\}$ with conservative data cuts in grey. The cosmic shear constraints are obtained using the same priors as in \citet{joudaki17}, with the exception of a wider prior on the baryon feedback amplitude (as described in Section~\ref{basesec}). The 12 baseline astrophysical parameters associated with the galaxy bias, shot noise, and velocity dispersion are simultaneously varied in the MCMC runs. 
}
\end{figure*}

\begin{figure*}
\includegraphics[width=1.01\hsize]{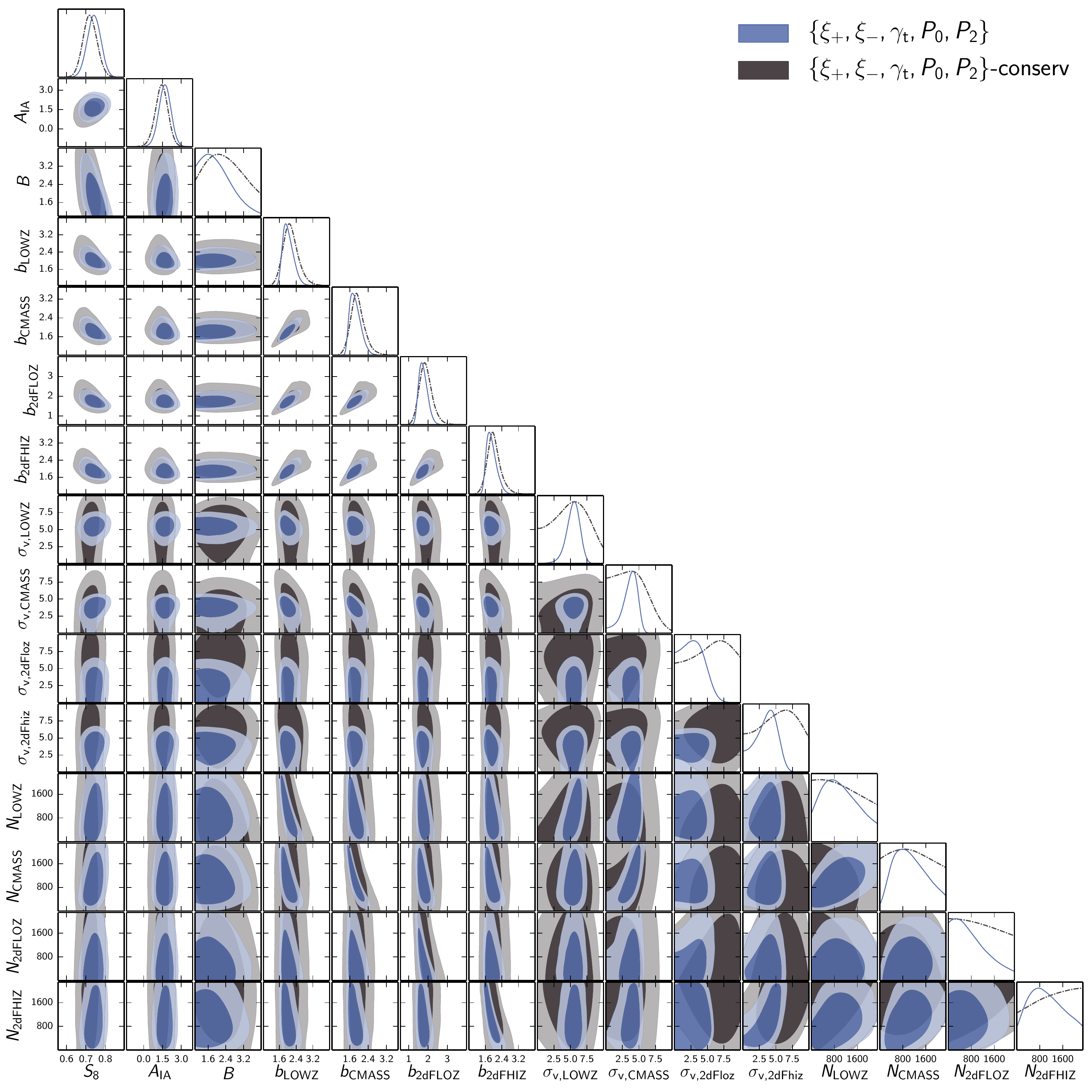}
\vspace{-2.4em}
\caption{\label{figsubmore} 
Marginalized posterior distributions of the derived $S_8 = \sigma_8 (\Omega_{\mathrm m}/0.3)^{0.5}$ parameter combination along with the primary astrophysical parameters $\{A_{\rm IA}, B, b_{\rm LOWZ}, b_{\rm CMASS}, b_{\rm 2dFLOZ}, b_{\rm 2dFHIZ}, \sigma_{\rm v, LOWZ}, \sigma_{\rm v, CMASS}, \sigma_{\rm v, 2dFLOZ}, \sigma_{\rm v, 2dFHIZ}, N_{\rm LOWZ}, N_{\rm CMASS}, N_{\rm 2dFLOZ}, N_{\rm 2dFHIZ}\}$, and their correlation. The constraints are obtained from $\{\xi_{\pm}, \gamma_{\rm t}, P_{0/2}\}$ in blue, and $\{\xi_{\pm}, \gamma_{\rm t}, P_{0/2}\}$ with conservative data cuts in grey. The vanilla cosmological parameters are simultaneously varied in the MCMC runs. 
}
\end{figure*}

Assuming a $\Lambda$CDM cosmological model, in Figure~\ref{figbf}, we show marginalized posterior distributions of derived cosmological parameters $\{\Omega_{\mathrm m}, \sigma_8, H_0, S_8\}$, and their correlation, from measurements of cosmic shear, galaxy-galaxy lensing, and redshift-space multipole power spectra for KiDS overlapping with 2dFLenS and BOSS. In Table~\ref{tabbf}, we list the marginalized posterior means and confidence intervals of these cosmological parameters. We also list $T(S_8)$ denoting the tension with Planck, the best-fit $\chi^2_{\rm eff}$, number of degrees of freedom, and DIC for each setup. In Figure~\ref{figprim}, we show marginalized posterior distributions of the primary model parameters $\{100 \theta_{\rm MC},\Omega_{\mathrm b} h^2, \Omega_{\mathrm c} h^2, \ln{(10^{10} A_{\mathrm s})}, n_s, A_{\rm IA}, B\}$ and their correlation (as that is the subspace shared with cosmic shear alone). Lastly, Figure~\ref{figsubmore} shows the marginalized posterior distributions of the derived $S_8$ parameter along with the primary astrophysical parameters and their correlation.

\begin{table*}
\begin{minipage}[t]{\textwidth}
\caption{\label{tabbf} Marginalized posterior means and 68\%\ confidence intervals of derived cosmological parameters $\{\Omega_{\mathrm m}, \sigma_8, H_0, S_8\}$ from observations of cosmic shear, galaxy-galaxy lensing, and redshift-space multipole power spectra for KiDS overlapping with 2dFLenS and BOSS. The cosmic shear $\{\xi_+, \xi_-\}$ constraints differ somewhat from the results in \citet{Hildebrandt16} and \citet{joudaki17} due to wider priors and use of covariance matrix constructed from $N$-body simulations. The Hubble constant constraints are dominated by the prior rather than the data. We further show $T(S_8)$ denoting the tension with Planck, the best-fit $\chi^2_{\rm eff}$, number of degrees of freedom (dof), and DIC for each setup. It is coincidental that the dof for $\{\xi_+, \xi_-, \gamma_{\rm t}\}$-conserv and $\{\xi_+, \xi_-, \gamma_{\rm t}, P_0, P_2\}$-conserv are the same, as the size of the data vector and number of free parameters are different. In the MCMC computations, we vary 7~free parameters for the data vector composed of $\{\xi_+, \xi_-\}$, 11 free parameters for $\{\xi_+, \xi_-, \gamma_{\rm t}\}$, and 19 free parameters for $\{\xi_+, \xi_-, \gamma_{\rm t}, P_0, P_2\}$.}
\renewcommand{\footnoterule}{} 
\begin{tabular}{l|rrrr|rrrr}
\hline
 & $ \Omega_{\rm m}$ & $ \sigma_8$ & $ S_8=\sigma_8\sqrt{\Omega_{\rm m}/0.3}$ & $ H_0$ [km\,s$ ^{-1}$ \,Mpc$ ^{-1}$ ] & $T(S_8)$ & $ \chi^2_{\rm eff}$ & dof & DIC \\ 
\hline
$\{\xi_+, \xi_-\}$ & $ 0.231^{+0.050}_{-0.095} $  & $ 0.884^{+0.144}_{-0.211} $  & $ 0.738^{+0.042}_{-0.046} $  & $ 85.5^{+14.4}_{-3.9} $  & 2.2 & 142.3 & 122 & 155.9 \\
$\{\xi_+, \xi_-, \gamma_{\rm t}\}$ & $ 0.238^{+0.048}_{-0.086} $  & $ 0.853^{+0.126}_{-0.177} $  & $ 0.731^{+0.037}_{-0.042} $  & $ 83.9^{+15.7}_{-5.1} $  & 2.6 & 199.0 & 182 & 223.0 \\
$\{\xi_+, \xi_-, \gamma_{\rm t}\}$-conserv & $ 0.311^{+0.061}_{-0.121} $  & $ 0.734^{+0.109}_{-0.166} $  & $ 0.715^{+0.037}_{-0.042} $  & $ 78.2^{+21.0}_{-7.6} $  & 2.9 & 180.5 & 166 & 202.6 \\
$\{\xi_+, \xi_-, P_0, P_2\}$ & $ 0.263^{+0.033}_{-0.054} $  & $ 0.780^{+0.075}_{-0.087} $  & $ 0.722^{+0.038}_{-0.037} $  & $ 86.0^{+13.7}_{-4.4} $  & 2.9 & 147.4 & 126 & 175.0 \\
$\{\xi_+, \xi_-, P_0, P_2\}$-conserv & $ 0.288^{+0.048}_{-0.085} $  & $ 0.753^{+0.100}_{-0.121} $  & $ 0.717^{+0.039}_{-0.039} $  & $ 81.7^{+17.6}_{-6.1} $  & 2.9 & 144.4 & 118 & 167.1 \\
$\{\xi_+, \xi_-, \gamma_{\rm t}, P_0, P_2\}$ & $ 0.243^{+0.026}_{-0.045} $  & $ 0.832^{+0.080}_{-0.079} $  & $ 0.742^{+0.035}_{-0.035} $  & $ 88.3^{+11.6}_{-4.5} $  & 2.6 & 206.9 & 190 & 236.1 \\
$\{\xi_+, \xi_-, \gamma_{\rm t}, P_0, P_2\}$-conserv & $ 0.295^{+0.052}_{-0.087} $  & $ 0.747^{+0.093}_{-0.125} $  & $ 0.721^{+0.036}_{-0.036} $  & $ 80.5^{+18.1}_{-6.9} $  & 3.0 & 184.3 & 166 & 207.4 \\
$\{\xi_+, \xi_-, \gamma_{\rm t}, P_0, P_2\}$-large scales & $ 0.286^{+0.055}_{-0.103} $  & $ 0.627^{+0.102}_{-0.132} $  & $ 0.593^{+0.081}_{-0.081} $  & $ 75.1^{+20.3}_{-11.1} $  & 3.0 & 65.6 & 50 & 83.9 \\
\hline
\end{tabular}
\end{minipage}
\end{table*}

\end{document}